\documentclass[twoside,12pt,a4paper]{report}
\usepackage{epsf}
\usepackage{graphics, graphicx}
\usepackage{amsmath}
\usepackage{amssymb}
\usepackage{mathtools}
\usepackage[dvipsnames]{xcolor}
\usepackage{latexsym}
\usepackage[margin=10pt,font=small,labelfont=bf]{caption}
\usepackage[utf8]{inputenc}
\usepackage[toc,page]{appendix}
\usepackage[T1]{fontenc}
\usepackage{stmaryrd}
\usepackage{color}
\usepackage{ulem}
\usepackage{hyperref}
\usepackage{fancyvrb}
\usepackage{tikz}
\usepackage{pgf}
\usepackage{scalefnt}
\usepackage{float}

\usetikzlibrary{calc,positioning,shapes.misc}

\hypersetup{
    colorlinks,
    linktoc=all,
    citecolor=black,
    filecolor=black,
    linkcolor=black,
    urlcolor=black
}

\allowdisplaybreaks

\begin{document}

\begin{titlepage}
 \begin{center}
  {\LARGE Eberhard Karls Universit\"at T\"ubingen}\\
  {\large Mathematisch-Naturwissenschaftliche Fakult\"at \\
Wilhelm-Schickard-Institut f\"ur Informatik\\[2cm]}
  {\huge Master Thesis\\[2cm]}
  {\Large\bf Context-Sensitive Abstract Interpretation of Dynamic Languages \\[1.5cm]}
 {\large Franciszek Piszcz}\\[0.5cm]
May 5, 2023 \\[2cm]
{\small\bf Reviewers}\\[0.5cm]
  \parbox{7cm}{\begin{center}{\large Jun. Prof. Jonathan Immanuel Brachthäuser}\\
    Software Engineering\\
  {\footnotesize Wilhelm-Schickard-Institut f\"ur Informatik\\
	Universit\"at T\"ubingen}\end{center}}\hfill\parbox{7cm}{\begin{center}
  {\large Prof. Klaus Ostermann}\\
   Programming Languages\\
  {\footnotesize Wilhelm-Schickard-Institut f\"ur Informatik\\
	Universit\"at T\"ubingen}\end{center}
 }
 {\small\bf Advisors}\\[0.5cm]
  \parbox{7cm}{\begin{center}{\large Dr. Philipp Schuster}\\
   Programming Languages\\
  {\footnotesize Wilhelm-Schickard-Institut f\"ur Informatik\\
	Universit\"at T\"ubingen}\end{center}}\hfill\parbox{7cm}{\begin{center}
  {\large Jun. Prof. Jonathan Immanuel Brachthäuser}\\
   Software Engineering\\
   {\footnotesize Wilhelm-Schickard-Institut f\"ur Informatik\\
   Universit\"at T\"ubingen}\end{center}
 }
  \end{center}
\end{titlepage}

\thispagestyle{empty}
\vspace*{\fill}
\begin{minipage}{11.2cm}
\textbf{Piszcz, Franciszek:}\\
\emph{Context-Sensitive Abstract Interpretation of Dynamic Languages}\\ Master Thesis Machine Learning\\
Eberhard Karls Universit\"at T\"ubingen\\
Thesis period: 15.11.2022 - 15.05.2023
\end{minipage}
\newpage

\pagenumbering{roman}
\setcounter{page}{1}

\section*{Abstract}
There is a vast gap in the quality of IDE tooling between static languages like
Java and dynamic languages like Python or JavaScript. Modern frameworks and libraries in these
languages heavily use their dynamic capabilities to achieve the best ergonomics and readability.
This has a side effect of making the current generation of IDEs blind to control flow and data flow,
which often breaks navigation, autocompletion and refactoring.

In this thesis we propose an algorithm that can bridge this gap between tooling for dynamic and
static languages by statically analyzing dynamic metaprogramming and runtime reflection in programs.
We use a technique called abstract interpretation to partially execute programs and extract
information that is usually only available at runtime. Our algorithm has been implemented in a
prototype analyzer that can analyze programs written in a subset of JavaScript.

\newpage
\section*{Acknowledgements}

I would like to thank Dr. Philipp Schuster and Prof. Jonathan Brachth\"auser
for giving me ideas and encouraging me to pursue them.

\cleardoublepage

\renewcommand{\baselinestretch}{1.3}
\small\normalsize

\tableofcontents

\renewcommand{\baselinestretch}{1}
\small\normalsize

\cleardoublepage

\pagenumbering{arabic}
\setcounter{page}{1}

\chapter{Introduction}

Dynamic languages, such as Python, JavaScript or R, offer a lot of flexibility
in the way programs are structured. Modern libraries and frameworks leverage it
to introduce declarative ways of writing code, tailored to their domain.
The result is concise, readable code, ergonomic APIs and little boilerplate.
This style of writing software is prevalent in the ecosystems of dynamic languages.
As a side effect, it is common that program components are wired up at runtime
according to some declarative description. This in turn makes the current generation
of IDEs and related tools completely blind to control flow and data flow. There is a vast
gap in tooling quality between dynamic languages like Python and static languages like Java.

Most efforts to bridge this gap concentrate on designing static type systems to model
common dynamic patterns. However, retrofitting static typing to languages whose main
strength lies in their flexibility feels wrong. To benefit from this approach, programmers
need to adapt their coding style to the type system, and write some type annotations.
This is not why they had chosen a dynamically-typed language in the first place.
Many instances of dynamic metaprogramming are hard to express in the language of types,
so these type systems make exceptions for the most popular frameworks and treat them in a
special way. For example, in MyPy it is achieved by extending the type checker with plugins.

In this thesis we propose an algorithm that can lift the quality of IDE support
for dynamic languages to the level commonly seen in statically-typed ones. Our approach
is general and does not rely on type annotations or special treatment of common patterns.
We use a technique called abstract interpretation to partially execute programs and extract
information that is usually only available at runtime. This enables our analysis to see
all control flow and data flow, even when it is buried deep in the internals of
some framework and obfuscated by the use of metaprogramming or runtime reflection.
Our algorithm has been implemented in a prototype analyzer that can analyze programs written
in a subset of JavaScript.

\section{Contents}

In Chapter 2 we formally describe a minimal subset of JavaScript. Illustrating our algorithms
in terms of a small language allows us to focus on principles and abstract away from technical details.
Chapter 3 describes static analysis of programs as a specific
computational problem and introduces the theory of abstract interpretation. In Chapter 4 we get familiar
with the framework of abstract interpretation by defining a naive static analysis. Chapter 5 contains our
main contribution, where we develop a context-sensitive analysis with heap cloning, along with underlying
theory and abstract domains. Finally, in Chapter 6 we talk about our prototype implementation.

\cleardoublepage

\chapter{TinyScript - a subset of JavaScript}

Our goal is to discuss and implement algorithms that can analyze code written in dynamic languages.
In particular, we are interested in methods that can accurately model dynamic behaviour, such as:
\begin{itemize}
\item runtime reflection,
\item dynamic class creation,
\item computed properties,
\item overwriting methods of a class at runtime.
\end{itemize}
Languages such as JavaScript or Python, in addition to these dynamic capabilities, also have complex object
models and offer a rich variety of builtin types and operations. For example, Python has:
\begin{itemize}
\item metaclasses - ability to customize class creation
\item \verb|__init_class__| - another, more lightweight way to customize class creation
\item \verb|__slots__| - a way for classes to define their memory layout to optimize space usage
\item multiple inheritance with complex rules for determining method resolution order
\item a mechanism to define class attributes that run custom code on read and write (property getters and setters)
\item another mechanism to intercept all attribute reads and writes at the level of a class (\verb|__getattr__| and \verb|__setattr__|)
\item a way to intercept all attribute reads and writes of a module (global \verb|__getattr__| and \verb|__setattr__|)
\item overloading operators for user-defined types
\item inheriting from builtin types such as int or list
\item etc.
\end{itemize}
On the other hand, JavaScript also has a quite elaborate data model:
\begin{itemize}
\item getset descriptors run arbitrary code on property reads and writes, like property getters and setters in Python
\item properties have a set of flags that control whether they can be enumerated, modified or redefined
\item a few different ways to define classes and methods, depending on language version (ES5 / ES6 / ES7)
\item 2 ways to define functions, with subtly different semantics when used as a method
\item special proxy objects that intercept all property accesses (like \verb|__getattr__| and \verb|__setattr__| in Python)
\item a few different ways to define and import modules (commonjs, AMD, UMD, ES6, <script> tags)
\item etc.
\end{itemize}
Therefore, the task of analyzing code written in these languages naturally leads to two distinct subproblems:
\begin{enumerate}
 \item Design of general methods to simulate dynamic behaviour during analysis
 \item Encoding all intricacies of a particular language's data model
\end{enumerate}
In this thesis we want to focus on the first one - on the essence - and we want to abstract away from technical details. To do this,
we will pick a "minimum viable subset" of JavaScript. This subset is capable of expressing most of commonly occurring dynamic patterns.
\section{Syntax}
{
\newcommand{\cat}[1]{{\textbf{#1}}}
\newcommand{\syn}[1]{\texttt{#1}}
\renewcommand{\|}{$|$}

\subsection*{Intermediate representation}

To further simplify the design of analysis algorithms, we define two variants of TinyScript. A high-level surface language and a
simpler intermediate language. There are two considerations that together make this split a desirable choice:
\begin{enumerate}
 \item First-class functions, and in particular their ability to capture variables in a lexical closure, are a central element of many dynamic patterns.
    Therefore, it is important to support these constructs.
 \item On the other hand, these lexical closures can be nested arbitrarily many times. Keeping track of which variable belongs to which
    lexical scope would be a great source of accidental complexity in an analysis algorithm. As we will see, program analysis is
    already a complex problem, so it's worth disentangling it from other aspects that can be handled separately.
\end{enumerate}
Therefore, we introduce a preprocessing step, \textit{closure conversion}, that converts the surface language TinyScript$^+$ into the intermediate language TinyScript.
This step rewrites all variable-capturing functions to explicitly take the outer environment (a closure) as an argument, and changes their runtime 
representation into a pair of closure and code. This results in programs that are easier to analyze: all data flow is explicit because reified
environments are first-class citizens.

\subsection*{Grammar}
We define the grammar of TinyScript$^+$ and TinyScript in a BNF-like notation:

\begin{align*}
    int &\in \cat{Int} & ::=& && \syn{0 \| 1 \| -1 \| 2 \| -2 \| ...} &\quad&\text{integer literals} \\
    str &\in \cat{Str} & ::=& && \syn{"" \| "abc" \| "123" \| ...} &\quad&\text{string literals} \\
    bool &\in \cat{Bool} & ::=& && \syn{true \| false} &\quad&\text{boolean literals} \\
    id &\in \cat{Id} & ::=& && \syn{a \| b \| foo \| bar \| ...} &\quad&\text{identifiers} \\
    \\
    e &\in \cat{Expr} & ::=& && \syn{$int$ \| $str$ \| $bool$ \| null} &\quad&\text{constants} \\
    &&|&&& \syn{$id$} &\quad&\text{variables} \\
    &&|&&& \syn{$e$.$id$} &\quad&\text{properties} \\
    &&|&&& \syn{$e_1$[$e_2$]} &\quad&\text{array elements} \\
    &&|&&& \syn{$id$ = $e$} &\quad&\text{variable assignments} \\
    &&|&&& \syn{$e_1$.$id$ = $e_2$} &\quad&\text{property assignments} \\
    &&|&&& \syn{$e_1$[$e_2$] = $e_3$} &\quad&\text{array element assignments} \\
    &&|&&& \syn{$e_1$ + $e_2$ \| $e_1$ - $e_2$} &\quad& \\
    &&|&&& \syn{$e_1$ * $e_2$ \| $e_1$ / $e_2$} &\quad&\text{binary operators} \\
    &&|&&& \syn{$e_0$($e_1$, $e_2$, ...)} &\quad& \text{function calls} \\
    &&|&&& \syn{\{$id_1$:\;$e_1$, $id_2$:\;$e_2$, ...\}} &\quad&\text{object literals} \\
    &&|&&& \syn{[$e_1$, $e_2$, ...]} &\quad&\text{array literals} \\
    &&|&&& \syn{bind-closure $id$} &\quad&\text{reify current environment as a lexical} \\
    &&&&& &\quad&\text{closure $cl$, create bound function} \\
    &&&&& &\quad&\text{object $(cl, id)$, where $id$ is a name of} \\
    &&&&& &\quad&\text{some global function declaration} \\
    \\
    e &\in \cat{Expr}^+ & ::=& && e \in \;(\cat{Expr} \setminus \{\syn{bind-closure $id$}\}) &\quad&\text{basic expressions} \\
    &&&&& &\quad&\text{excluding reified closures} \\
    &&|&&& \syn{function($id_1$, $id_2$, ...)\;\{} && \\
    &&&&& \syn{    $s$; return $e$} && \\
    &&&&& \syn{\}} &\quad&\text{lambda functions} \\
    \\
    s &\in \cat{Stmt} & ::=& && \syn{$\langle empty \rangle$} &\quad&\text{no-ops} \\
    &&|&&& \syn{$e$} &\quad&\text{expression statements} \\
    &&|&&& \syn{$s_1$; $s_2$} &\quad&\text{compound statements} \\
    &&|&&& \syn{var $id$ = $e$} &\quad&\text{variable declarations} \\
    &&|&&& \syn{if ($e$) \{ $s_1$ \} else \{ $s_2$ \}} &\quad&\text{conditionals} \\
    &&|&&& \syn{for (var $id$ in $e$) \{ $s$ \}} &\quad&\text{for loops} \\
    &&|&&& \syn{while ($e$) \{ $s$ \}} &\quad&\text{while loops} \\
    \\
    s &\in \cat{Stmt}^+ & ::=& && s \in \cat{Stmt} &\quad&\text{basic statements} \\
    &&|&&& \syn{function $id_0$($id_1$, $id_2$, ...)\;\{} && \\
    &&&&& \syn{    $s$; return $e$} && \\
    &&&&& \syn{\}} &\quad&\text{function declarations} \\
    \\
    d &\in \cat{Decl} & ::=& && \syn{function $id_0$($id_1$, $id_2$, ...)\;\{} && \\
    &&&&& \syn{    $s$; return $e$} && \\
    &&&&& \syn{\}} &\quad&\text{global function declarations} \\
    \\
    p &\in \cat{Prog} & ::=& && (s, d_1 d_2 ... d_n) &\quad&\text{programs in the intermediate} \\
    &&&&& &\quad&\text{representation consist of an entry} \\
    &&&&& &\quad&\text{point, which is a statement, and} \\
    &&&&& &\quad&\text{a sequence of global declarations} \\
    \\
    p &\in \cat{Prog}^+ & ::=& && s &\quad&\text{surface-level programs are simply} \\
    &&&&& &\quad&\text{(sequences of) statements}
\end{align*}
Note that in the above we made a small notational shortcut. By
\begin{align*}
    s &\in \cat{Stmt}^+ & ::=& && s \in \cat{Stmt} \quad | \quad \langle additions \rangle \\
    e &\in \cat{Expr}^+ & ::=& && e \in \;(\cat{Expr} \setminus \{\syn{bind-closure $id$}\}) \quad | \quad \langle additions \rangle
\end{align*}
we mean that we are extending the set of \textit{productions} that generate $\cat{Stmt}$ and $\cat{Expr}$ with
new $\langle additions \rangle$. This means that $\cat{Stmt}^+$ is \textit{not} defined in terms of $\cat{Stmt}$.
The two definitions are independent and both are recursive, and the set of productions that
generate $\cat{Stmt}^+$ is equal to the set of productions that generate $\cat{Stmt}$ extended with $\langle additions \rangle$.
}

{
\newcommand{\cat}[1]{{\textbf{#1}}}
\newcommand{\syn}[1]{\texttt{#1}}
\newcommand{\E}[1]{\mathcal{E} \llbracket \syn{#1} \rrbracket}
\renewcommand{\H}[1]{\mathcal{H} \llbracket \syn{#1} \rrbracket}
\renewcommand{\S}[1]{\mathcal{S} \llbracket \syn{#1} \rrbracket}
\renewcommand{\P}[1]{\mathcal{PROG} \llbracket \syn{#1} \rrbracket}

\section{Semantics of TinyScript$^{+}$}

\subsection*{Semantic domains}

\begin{align*}
\shortintertext{A running TinyScript program will be operating on some area of memory, which in particular
will include stack and heap-allocated data. For the purposes of defining the semantics of TinyScript, we will call this memory a
$\cat{Store}$. A store is a function that assigns a value to each memory location that belongs to the program,
and a special value $\syn{null}$ to all other (invalid) memory locations:}
\cat{Store} \quad&=\quad \cat{Loc} \rightarrow \cat{Val} \\
\shortintertext{Because usually primitive types are passed by value and composite types are passed by reference (they are sometimes called \textit{boxed types}),
it is beneficial to distinguish between 2 types of locations in the store:}
\cat{Loc} \quad&=\quad \langle\text{locations to store values of primitive types} \\
&\quad\quad \;\;\text{ and references to composite types}\rangle \\
\cat{Box} \quad&=\quad \langle\text{locations to store payloads of composite types}\rangle \\
\cat{Store} \quad&=\quad (\cat{Loc} \rightarrow \cat{Val}) \text{ or } (\cat{Box} \rightarrow \cat{Payload}) \text{, depending on the context} \\
\shortintertext{Special functions $alloc$ and $allocbox$ generate a fresh memory location and make it accessible for the program to manipulate:}
alloc \quad&:\quad \cat{Store} \rightarrow \cat{Loc} \\
allocbox \quad&:\quad \cat{Store} \rightarrow \cat{Box} \\
\shortintertext{Values manipulated by the program can be either primitive - integers, strings, booleans, functions and nulls, or they can be references to \textit{boxed} composite types - objects and arrays:}
\cat{Val} \quad&=\quad \cat{Int} \cup \cat{Str} \cup \cat{Bool} \cup \cat{Fun} \cup \{\syn{null}\} \cup \cat{Box} \\
\quad \cat{Payload} \quad&=\quad \cat{Obj} \cup \cat{Arr} \cup \{\syn{null}\} \\ \\
\shortintertext{Objects in TinyScript are functions that assign values to the names of their properties, and nulls to all other identifiers:}
\cat{Obj} \quad&=\quad \cat{Id} \rightarrow \cat{Val} \\
\shortintertext{Arrays are finite sequences of values:}
\cat{Arr} \quad&=\quad \cat{Val}^{*}
\shortintertext{Functions in TinyScript can have side effects, so they map tuples of (arguments, memory state before the call) into tuples of (result, memory state after the call):}
\cat{Fun} \quad&=\quad \cat{Val}^{*} \rightarrow \cat{Store} \rightarrow \cat{Val} \times \cat{Store} \\ \\
\shortintertext{There is one more semantic domain that we will need: lexical environments. They map variable names into memory locations in the store:}
\cat{Env} \quad&=\quad \cat{Id} \rightarrow \cat{Loc} \cup \{\syn{null}\}
\end{align*}

\subsection*{Semantic functions}

\begin{align*}
\shortintertext{
We will describe semantics of TinyScript in a denotational style. A set of mutually recursive partial functions map program and its input into
memory state at the end of successful execution. We need partial functions because some program runs will never terminate, and some
will encounter an unrecoverable runtime error. We define these partial functions using a system of equations. We define these equations
only for well-behaved programs. Whenever one of these partial functions is not defined for a given combination of arguments (there is no matching equation),
this means that a program has encountered an unrecoverable runtime error. Whenever an attempt to determine a value of a function using these
equations leads to an infinite number of reasoning steps, or a circular dependency on itself, this means that a corresponding execution
does not terminate. A program input is a single string, which is placed in a global variable called \syn{input} before the start of a program.
For simplicity we don't define any builtin operations for printing outputs and we treat memory state at the end as the output.
}
\P{$\cdot$} \;&:\; \cat{Prog}^+ \,\longrightarrow\, \cat{Str} \rightarrow \cat{Env} \times \cat{Store} \\
\shortintertext{
To allow for mutually recursive definitions in our programs, we split semantics of statements into 2 phases. In the first phase, $\mathcal{H}$,
we scan the code for declarations of functions and variables (in JavaScript this is known as \textit{hoisting}) and
gather them into a lexical environment. A second phase, $\mathcal{S}$, executes code inside this lexical environment.
}
\H{$\cdot$} \;&:\; \cat{Stmt}^+ \,\longrightarrow\, \cat{Env} \rightarrow \cat{Store} \rightarrow \cat{Env} \\
\S{$\cdot$} \;&:\; \cat{Stmt}^+ \,\longrightarrow\, \cat{Env} \rightarrow \cat{Store} \rightarrow \cat{Store} \\
\shortintertext{
Finally, we define semantics of expressions (which can have side effects) as functions that map lexical environment and memory
state before evaluating the expression into memory state afterwards and the computed value:
}
\E{$\cdot$} \;&:\; \cat{Expr}^+ \,\longrightarrow\, \cat{Env} \rightarrow \cat{Store} \rightarrow \cat{Val} \times \cat{Store} \\
& 
\end{align*}

\subsection*{Denotations of programs and statements}
\newcommand{\denotation}[1]{\\[-0.8\baselineskip] \shortintertext{$#1$}}

\begin{alignat*}{3}
    \\[-1.35\baselineskip]
    \denotation{\P{$s$} \, in \;=\; (\rho_2, \sigma_2) \:\text{ where}}
        \rho_{\varnothing} &:\; \cat{Env} & \;\;=\;\; & \lambda x.\syn{null} &\qquad \qquad \qquad \qquad \qquad \qquad & \\
        \sigma_{\varnothing} &:\; \cat{Store} & \;\;=\;\; & \lambda x.\syn{null} && \\
        loc &:\; \cat{Loc} & \;\;=\;\; & alloc \, \sigma_{\varnothing} && \\
        \rho_1 &:\; \cat{Env} & \;\;=\;\; & \rho_{\varnothing}[\syn{input} \, \mapsto \, loc] && \\
        \sigma_1 &:\; \cat{Store} & \;\;=\;\; & \sigma_{\varnothing}[loc \, \mapsto \, in] && \\
        \rho_2 &:\; \cat{Env} & \;\;=\;\; & \H{$s$} \, \rho_1 \, \sigma_1 && \\
        \sigma_2 &:\; \cat{Store} & \;\;=\;\; & \S{$s$} \, \rho_2 \, \sigma_1 && \\
    \\
    \denotation{\H{$\langle empty \rangle$} \, \rho \, \sigma \;=\; \rho}
    \denotation{\H{$e$} \, \rho \, \sigma \;=\; \rho}
    \denotation{\H{$s_1$; $s_2$} \, \rho \, \sigma \;=\; \rho_2 \:\text{ where} }
        \rho_1 &:\; \cat{Env} & \;\;=\;\; & \H{$s_1$} \, \rho \, \sigma && \\
        \rho_2 &:\; \cat{Env} & \;\;=\;\; & \H{$s_2$} \, \rho_1 \, \sigma && \\
    \denotation{\H{var $id$ = $e$} \, \rho \, \sigma \;=\; \rho[id \, \mapsto \, alloc \, \sigma]}
    \denotation{\H{if ($e$) \{ $s_1$ \} else \{ $s_2$ \}} \, \rho \, \sigma \;=\; \rho_2 \:\text{ where} }
        \rho_1 &:\; \cat{Env} & \;\;=\;\; & \H{$s_1$} \, \rho \, \sigma && \\
        \rho_2 &:\; \cat{Env} & \;\;=\;\; & \H{$s_2$} \, \rho_1 \, \sigma && \\
    \denotation{\H{for (var $id$ in $e$) \{ $s$ \}} \, \rho \, \sigma \;=\; (\H{$s$} \, \rho \, \sigma)[id \, \mapsto \, alloc \, \sigma]}
    \denotation{\H{while ($e$) \{ $s$ \}} \, \rho \, \sigma \;=\; \H{$s$} \, \rho \, \sigma}
    \denotation{\H{function $id_0$($id_1$, $id_2$, ...)\;\{ $s$; return $e$ \}} \, \rho \, \sigma \;=\; \rho[id_0 \, \mapsto \, alloc \, \sigma]}
\end{alignat*}
\\[-3.2\baselineskip]
\begin{alignat*}{2}
    \denotation{\S{$\langle empty \rangle$} \, \rho \, \sigma \;=\; \sigma}
    \denotation{\S{$e$} \, \rho \, \sigma \;=\; \sigma_1 \:\text{ where}}
        (\_, \sigma_1) &:\; \cat{Val} \times \cat{Store} \;\;=\;\; \E{$e$} \, \rho \, \sigma && \\
    \denotation{\S{$s_1$; $s_2$} \, \rho \, \sigma \;=\; \sigma_2 \:\text{ where}}
        \sigma_1 &:\; \cat{Store} \;\;=\;\; \S{$s_1$} \, \rho \, \sigma && \\
        \sigma_2 &:\; \cat{Store} \;\;=\;\; \S{$s_2$} \, \rho \, \sigma_1 && \\
    \denotation{\S{var $id$ = $e$} \, \rho \, \sigma \;=\; \sigma_1[\rho \, id \, \mapsto \, v] \:\text{ where}}
        (v, \sigma_1) &:\; \cat{Val} \times \cat{Store} \;\;=\;\; \E{$e$} \, \rho \, \sigma && \\
    \denotation{\S{if ($e$) \{ $s_1$ \} else \{ $s_2$ \}} \, \rho \, \sigma \;=\; \sigma_2 \:\text{ where}}
        (v, \sigma_1) &:\; \cat{Val} \times \cat{Store} \;\;=\;\; \E{$e$} \, \rho \, \sigma && \\
        \sigma_2 &:\; \cat{Store} \;\;=\;\;
            \begin{cases}
                \S{$s_2$} \, \rho \, \sigma_1 \; \text{ when } \; v \in \{\syn{false}, \syn{null}, \syn{0}, \syn{""}\} \\
                \S{$s_1$} \, \rho \, \sigma_1 \; \text{ otherwise}
            \end{cases} && \\
    \denotation{\S{for (var $id$ in $e$) \{ $s$ \}} \, \rho \, \sigma \;=\; f (arr) \, \sigma_1 \:\text{ where}}
        (v, \sigma_1) &:\; \cat{Val} \times \cat{Store} \;\;=\;\; \E{$e$} \, \rho \, \sigma && \\
        arr &:\; \cat{Val}^* \;\;=\;\; \sigma_1 v \; \text{ when } \; v \in \cat{Box} \; \text{ and } \; \sigma_1 v \in \cat{Arr} && \\
        f &:\; \cat{Val}^* \rightarrow \cat{Store} \rightarrow \cat{Store} && \\
        f&(v_1 \, ... \, v_n) \, \sigma \;\;=\;\;
            \begin{cases}
                \sigma \; \text{ when } \; n = 0 \\
                f(v_2 \, ... \, v_n)(\S{s} \, \rho \, (\sigma[\rho \, id \, \mapsto v_1]))  \; \text{ otherwise}
            \end{cases} && \\
    \denotation{\S{while ($e$) \{ $s$ \}} \, \rho \, \sigma \;=\; \sigma_2 \:\text{ where}}
        (v, \sigma_1) &:\; \cat{Val} \times \cat{Store} \;\;=\;\; \E{$e$} \, \rho \, \sigma && \\
        \sigma_2 &:\; \cat{Store} \;\;=\;\;
            \begin{cases}
                \sigma_1 \; \text{ when } \; v \in \{\syn{false}, \syn{null}, \syn{0}, \syn{""}\} \\
                \S{while ($e$) \{ $s$ \}} \, \rho \, (\S{$s$} \, \rho \, \sigma_1) \; \text{ otherwise}
            \end{cases} && \\
    \denotation{\S{function $id_0$($id_1$, $...$, $id_n$)\,\{ $s$; return $e$ \}} \, \rho \, \sigma_0 \; = \; \sigma_0[\rho \, id_0 \, \mapsto \, f]}
        \shortintertext{\quad with $f:\; \cat{Fun}, \; \; f(v_1 ... v_n) \, \sigma \; = \; \E{$e$} \, \rho_2 \, \sigma_2 \:\text{ where}$}
        \rho_1 &:\; \cat{Env} \;\;=\;\; \rho [id_1 \, \mapsto \, alloc \, \sigma] ... [id_n \, \mapsto \, alloc \, \sigma] && \\
        \rho_2 &:\; \cat{Env} \;\;=\;\; \H{$s$} \, \rho_1 \, \sigma && \\
        \sigma_1 &:\; \cat{Store} \;\;=\;\; \sigma [\rho_2 \, id_1 \, \mapsto \, v_1] ... [\rho_2 \, id_n \, \mapsto \, v_n] && \\
        \sigma_2 &:\; \cat{Store} \;\;=\;\; \S{$s$} \, \rho_2 \, \sigma_1
\end{alignat*}

\subsection*{Denotations of expressions}

\begin{alignat*}{2}
    \\[-1.35\baselineskip]
    \denotation{\E{$v$} \, \rho \, \sigma \;=\; (v, \sigma) \:\text{ when }\: v \in \cat{Int} \cup \cat{Str} \cup \cat{Bool} \cup \{\syn{null}\}}
    \denotation{\E{$id$} \, \rho \, \sigma \;=\; (\sigma (\rho \, id), \sigma)}
    \denotation{\E{$e$.$id$} \, \rho \, \sigma \;=\; (o \, id, \sigma_1) \:\text{ where}}
        (v, \sigma_1) &:\; \cat{Val} \times \cat{Store} \;\;=\;\; \E{$e$} \, \rho \, \sigma && \\
        o &:\; \cat{Obj} \;\;=\;\; \sigma_1 \, v \:\text{ when }\: v \in \cat{Box} \:\text{ and }\: \sigma_1 \, v \in \cat{Obj} && \\
    \denotation{\E{$e_1$[$e_2$]} \, \rho \, \sigma \;=\; (a_i, \sigma_2) \:\text{ where}}
        (v_1, \sigma_1) &:\; \cat{Val} \times \cat{Store} \;\;=\;\; \E{$e_1$} \, \rho \, \sigma && \\
        (v_2, \sigma_2) &:\; \cat{Val} \times \cat{Store} \;\;=\;\; \E{$e_2$} \, \rho \, \sigma_1 && \\
        a &:\; \cat{Arr} \;\;=\;\; \sigma_2 \, v_1 \:\text{ when }\: v_1 \in \cat{Box} \:\text{ and }\: \sigma_2 \, v_1 \in \cat{Arr} && \\
        i &:\; \cat{Int} \;\;=\;\; v_2 \:\text{ when }\: v_2 \in \cat{Int} && \\
    \denotation{\E{$id$ = $e$} \, \rho \, \sigma \;=\; (v, \sigma_1[\rho \, id \, \mapsto \, v]) \:\text{ where}}
        (v, \sigma_1) &:\; \cat{Val} \times \cat{Store} \;\;=\;\; \E{$e$} \, \rho \, \sigma && \\
    \denotation{\E{$e_1$.$id$ = $e_2$} \, \rho \, \sigma \;=\; (v_2, \sigma_3) \:\text{ where}}
        (v_1, \sigma_1) &:\; \cat{Val} \times \cat{Store} \;\;=\;\; \E{$e_1$} \, \rho \, \sigma && \\
        (v_2, \sigma_2) &:\; \cat{Val} \times \cat{Store} \;\;=\;\; \E{$e_2$} \, \rho \, \sigma_1 && \\
        b &:\; \cat{Box} \;\;=\;\; v_1 \:\text{ when }\: v_1 \in \cat{Box} && \\
        o &:\; \cat{Obj} \;\;=\;\; \sigma_2 \, b \:\text{ when }\: \sigma_2 \, b \in \cat{Obj} && \\
        \sigma_3 &:\; \cat{Store} \;\;=\;\; \sigma_2[b \, \mapsto \, o[id \, \mapsto \, v_2]] && \\
    \denotation{\E{$e_1$[$e_2$] = $e_3$} \, \rho \, \sigma \;=\; (v_3, \sigma_4) \:\text{ where}}
        (v_1, \sigma_1) &:\; \cat{Val} \times \cat{Store} \;\;=\;\; \E{$e_1$} \, \rho \, \sigma && \\
        (v_2, \sigma_2) &:\; \cat{Val} \times \cat{Store} \;\;=\;\; \E{$e_2$} \, \rho \, \sigma_1 && \\
        (v_3, \sigma_3) &:\; \cat{Val} \times \cat{Store} \;\;=\;\; \E{$e_3$} \, \rho \, \sigma_2 && \\
        b &:\; \cat{Box} \;\;=\;\; v_1 \:\text{ when }\: v_1 \in \cat{Box} && \\
        i &:\; \cat{Int} \;\;=\;\; v_2 \:\text{ when }\: v_2 \in \cat{Int} && \\
        a_1...a_n &:\; \cat{Arr} \;\;=\;\; \sigma_3 \, b \:\text{ when }\: \sigma_3 \, b \in \cat{Arr} && \\
        \sigma_4 &:\; \cat{Store} \;\;=\;\; \sigma_3[b \, \mapsto \, a_1...a_{i-1} v_3 a_{i+1}...a_n] && \\
    \denotation{\E{$e_1$ + $e_2$} \, \rho \, \sigma \;=\; (i, \sigma_2) \:\text{ where}}
        (v_1, \sigma_1) &:\; \cat{Val} \times \cat{Store} \;\;=\;\; \E{$e_1$} \, \rho \, \sigma && \\
        (v_2, \sigma_2) &:\; \cat{Val} \times \cat{Store} \;\;=\;\; \E{$e_2$} \, \rho \, \sigma_1 && \\
        i &:\; \cat{Int} \;\;=\;\; v_1 + v_2 \:\text{ when }\: v_1, v_2 \in \cat{Int} && \\
    \denotation{\E{$e_1$ - $e_2$} \, \rho \, \sigma \;=\; (i, \sigma_2) \:\text{ where}}
        (v_1, \sigma_1) &:\; \cat{Val} \times \cat{Store} \;\;=\;\; \E{$e_1$} \, \rho \, \sigma && \\
        (v_2, \sigma_2) &:\; \cat{Val} \times \cat{Store} \;\;=\;\; \E{$e_2$} \, \rho \, \sigma_1 && \\
        i &:\; \cat{Int} \;\;=\;\; v_1 - v_2 \:\text{ when }\: v_1, v_2 \in \cat{Int} && \\
    \denotation{\E{$e_1$ * $e_2$} \, \rho \, \sigma \;=\; (i, \sigma_2) \:\text{ where}}
        (v_1, \sigma_1) &:\; \cat{Val} \times \cat{Store} \;\;=\;\; \E{$e_1$} \, \rho \, \sigma && \\
        (v_2, \sigma_2) &:\; \cat{Val} \times \cat{Store} \;\;=\;\; \E{$e_2$} \, \rho \, \sigma_1 && \\
        i &:\; \cat{Int} \;\;=\;\; v_1 \cdot v_2 \:\text{ when }\: v_1, v_2 \in \cat{Int} && \\
    \denotation{\E{$e_1$ / $e_2$} \, \rho \, \sigma \;=\; (i, \sigma_2) \:\text{ where}}
        (v_1, \sigma_1) &:\; \cat{Val} \times \cat{Store} \;\;=\;\; \E{$e_1$} \, \rho \, \sigma && \\
        (v_2, \sigma_2) &:\; \cat{Val} \times \cat{Store} \;\;=\;\; \E{$e_2$} \, \rho \, \sigma_1 && \\
        i &:\; \cat{Int} \;\;=\;\; \left\lfloor \frac{v_1}{v_2} \right\rfloor \:\text{ when }\: v_1, v_2 \in \cat{Int} \:\text{ and }\: v_2 \neq 0 && \\
    \denotation{\E{$e_0$($e_1$, $...$, $e_n$)} \, \rho \, \sigma \;=\; f(v_1 ... v_n) \, \sigma_n \:\text{ where}}
        (v_0, \sigma_0) &:\; \cat{Val} \times \cat{Store} \;\;=\;\; \E{$e_0$} \, \rho \, \sigma && \\
        (v_1, \sigma_1) &:\; \cat{Val} \times \cat{Store} \;\;=\;\; \E{$e_1$} \, \rho \, \sigma_0 && \\
        ... & \qquad \qquad \qquad \quad \:\; ... && \\
        (v_n, \sigma_n) &:\; \cat{Val} \times \cat{Store} \;\;=\;\; \E{$e_n$} \, \rho \, \sigma_{n-1} && \\
        f &:\; \cat{Fun} \;\;=\;\; v_0 \:\text{ when }\: v_0 \in \cat{Fun} && \\
    \denotation{\E{\{$id_1$:\;$e_1$, $...$, $id_n$:\;$e_n$\}} \, \rho \, \sigma_0 \;=\; (b, \sigma_{n+1}) \:\text{ where}}
        (v_1, \sigma_1) &:\; \cat{Val} \times \cat{Store} \;\;=\;\; \E{$e_1$} \, \rho \, \sigma_0 && \\
        ... & \qquad \qquad \qquad \quad \:\; ... && \\
        (v_n, \sigma_n) &:\; \cat{Val} \times \cat{Store} \;\;=\;\; \E{$e_n$} \, \rho \, \sigma_{n-1} && \\
        o_{\varnothing} &:\; \cat{Obj} \;\;=\;\; \lambda id . \syn{null} && \\
        o &:\; \cat{Obj} \;\;=\;\; o_{\varnothing}[id_1 \, \mapsto \, v_1]...[id_n \, \mapsto \, v_n] && \\
        b &:\; \cat{Box} \;\;=\;\; allocbox \, \sigma_n && \\
        \sigma_{n+1} &:\; \cat{Store} \;\;=\;\; \sigma_n[b \, \mapsto \, o] && \\
    \denotation{\E{[$e_1$, $...$, $e_n$]} \, \rho \, \sigma_0 \;=\; (b, \sigma_{n+1}) \:\text{ where}}
        (v_1, \sigma_1) &:\; \cat{Val} \times \cat{Store} \;\;=\;\; \E{$e_1$} \, \rho \, \sigma_0 && \\
        ... & \qquad \qquad \qquad \quad \:\; ... && \\
        (v_n, \sigma_n) &:\; \cat{Val} \times \cat{Store} \;\;=\;\; \E{$e_n$} \, \rho \, \sigma_{n-1} && \\
        a &:\; \cat{Arr} \;\;=\;\; v_1 ... v_n && \\
        b &:\; \cat{Box} \;\;=\;\; allocbox \, \sigma_n && \\
        \sigma_{n+1} &:\; \cat{Store} \;\;=\;\; \sigma_n[b \, \mapsto \, a] && \\
    \denotation{\E{function($id_1$, $...$, $id_n$)\,\{ $s$; return $e$ \}} \, \rho \, \sigma_0 \;=\; (f, \sigma_0) \:\text{ where}}
        \shortintertext{\quad $f:\; \cat{Fun}, \; \; f(v_1 ... v_n) \, \sigma \; = \; \E{$e$} \, \rho_2 \, \sigma_2 \:\text{ where}$}
        \rho_1 &:\; \cat{Env} \;\;=\;\; \rho [id_1 \, \mapsto \, alloc \, \sigma] ... [id_n \, \mapsto \, alloc \, \sigma] && \\
        \rho_2 &:\; \cat{Env} \;\;=\;\; \H{$s$} \, \rho_1 \, \sigma && \\
        \sigma_1 &:\; \cat{Store} \;\;=\;\; \sigma [\rho_2 \, id_1 \, \mapsto \, v_1] ... [\rho_2 \, id_n \, \mapsto \, v_n] && \\
        \sigma_2 &:\; \cat{Store} \;\;=\;\; \S{$s$} \, \rho_2 \, \sigma_1
\end{alignat*}
}

\section{Closure conversion}
{
\newcommand{\syn}[1]{\texttt{#1}}
Closure conversion is a process that moves all nested functions to global scope, and replaces their previous definitions
with \syn{bind-closure $id$}, where $id$ is the name of the function in the global scope. These \syn{bind-closure} expressions
save current lexical environment into a closure, and pair it with an identifier of the global function. Later when these closures
are called, the global function receives saved outer lexical environment as the first argument, customarily called \syn{closure}.
In addition to moving nested functions to global scope, closure conversion also adds \syn{closure} as a first argument, and rewrites
all occurrences $a$ of variables from outer scope into $\syn{closure.}a$. If a function is nested more than one level, references to
variables in outer scopes are rewritten to $\syn{closure.}a$, $\syn{closure.closure.}a$, $\syn{closure.closure.closure.}a$, etc.,
depending on the level of nesting. A formal definition of closure conversion can be found in appendix.
}
\section{Semantics of TinyScript}

Here we describe the semantics of programs resulting from performing closure conversion on surface-level programs.
Notable differences from the semantics of the surface langauge are marked with \uline{underline}.

{
\newcommand{\cat}[1]{{\textbf{#1}}}
\newcommand{\syn}[1]{\texttt{#1}}
\newcommand{\E}[1]{\mathcal{E} \llbracket \syn{#1} \rrbracket}
\renewcommand{\H}[1]{\mathcal{H} \llbracket \syn{#1} \rrbracket}
\renewcommand{\S}[1]{\mathcal{S} \llbracket \syn{#1} \rrbracket}
\newcommand{\F}[1]{\mathcal{F} \llbracket \syn{#1} \rrbracket}
\newcommand{\PROG}[1]{\mathcal{PROG} \llbracket \syn{#1} \rrbracket}

\makeatletter
\newcommand*{\saved@uline}{}
\let\saved@uline\uline
\newcommand*{\mathuline}{%
  \mathpalette{\math@uline\saved@uline}%
}
\newcommand*{\math@uline}[3]{%
  \mbox{#1{$#2#3\m@th$}}%
}
\renewcommand*{\uline}{%
  \relax  
  \ifmmode
    \expandafter\mathuline
  \else
    \expandafter\saved@uline
  \fi
}
\makeatother

\subsection*{Semantic domains and functions}

\begin{align*}
    \shortintertext{Working memory of a program is represented the same way as in TinyScript$^+$:}
    \cat{Loc} \quad&=\quad \langle\text{locations to store values of primitive types} \\
    &\quad\quad \;\;\text{ and references to composite types}\rangle \\
    \cat{Box} \quad&=\quad \langle\text{locations to store payloads of composite types}\rangle \\
    \cat{Store} \quad&=\quad (\cat{Loc} \rightarrow \cat{Val}) \text{ or } (\cat{Box} \rightarrow \cat{Payload}) \text{, depending on the context} \\
    alloc \quad&:\quad \cat{Store} \rightarrow \cat{Loc} \\
    allocbox \quad&:\quad \cat{Store} \rightarrow \cat{Box} \\
    \shortintertext{Representation of values is extended with lexical environments, which become first-class citizens in the intermediate langauge:}
    \cat{Val} \quad&=\quad \cat{Int} \cup \cat{Str} \cup \cat{Bool} \cup \cat{Fun} \cup \{\syn{null}\} \cup \cat{Box} \uline{ \, \cup \, \cat{Env}} \\
    \quad \cat{Payload} \quad&=\quad \cat{Obj} \cup \cat{Arr} \cup \{\syn{null}\} \\
    \cat{Obj} \quad&=\quad \cat{Id} \rightarrow \cat{Val} \\
    \cat{Arr} \quad&=\quad \cat{Val}^{*} \\ \\
    \shortintertext{Representation of functions additionally stores the outer environment. Previously, this environment has been
        implicitly captured in the definition of (mathematical) function that represents a TinyScript$^+$ function. In other words, we have used
        closures in the meta-language (mathematical notation) to implement closures in the object language (TinyScript$^+$). Now, we represent functions
        as pairs of (closure, code pointer). The code pointer is an identifier that points to some global declaration. We additionally
        introduce a new type of environment, \cat{FEnv}, which resolves code pointers to actual functions.
        }
    \cat{RawFun} \quad&=\quad \cat{Val}^{*} \, \uline{\rightarrow \, \cat{FEnv}} \, \rightarrow \cat{Store} \rightarrow \cat{Val} \times \cat{Store} \\
    \cat{FEnv} \quad&=\quad \cat{Id} \rightarrow \cat{RawFun} \\
    \cat{Fun} \quad&=\quad \uline{ \cat{Env} \, \times \, \cat{Id} } \\ \\
    \shortintertext{The shape of lexical environments is unchanged - they still map identifiers to locations - but they no longer
        mix variables from different lexical scopes. Instead, variables from outer scopes are accessible through a special variable called
        \texttt{closure}, which contains reified outer environment.
    }
    \cat{Env} \quad&=\quad \cat{Id} \rightarrow \cat{Loc} \cup \{\syn{null}\} \\ \\
    \shortintertext{Semantic functions have almost the same signature as before, but denotations of statements and expressions
    additionally take \cat{FEnv} as an argument:}
    \PROG{$\cdot$} \;&:\; \cat{Prog} \,\longrightarrow\, \cat{Str} \rightarrow \cat{Env} \times \cat{Store} \\
    \H{$\cdot$} \;&:\; \cat{Stmt} \,\longrightarrow\, \cat{Env} \rightarrow \cat{Store} \rightarrow \cat{Env} \\
    \S{$\cdot$} \;&:\; \cat{Stmt} \,\longrightarrow\, \cat{Env} \; \uline{ \rightarrow \cat{FEnv}} \rightarrow \cat{Store} \rightarrow \cat{Store} \\
    \E{$\cdot$} \;&:\; \cat{Expr} \,\longrightarrow\, \cat{Env} \; \uline{ \rightarrow \cat{FEnv}} \rightarrow \cat{Store} \rightarrow \cat{Val} \times \cat{Store} \\
    \shortintertext{We additionally introduce one more semantic function, $\F{$\cdot$}$, that interprets global function declarations:}
    \F{$\cdot$} \;&:\; \cat{Decl} \,\longrightarrow\, \cat{RawFun} \\ \\
\end{align*}

\subsection*{Denotations}

\newcommand{\denotation}[1]{\\[-0.8\baselineskip] \shortintertext{$#1$}}

\newcommand{\ulinex}[2]{{\renewcommand{\ULdepth}{#1 pt} \uline{#2}}}

\begin{alignat*}{3}
    \\[-1.35\baselineskip]
    \denotation{\PROG{$(s, d_1 ... d_n)$} \, in \;=\; (\rho_2, \sigma_2) \:\text{ where}}
        \rho^\varnothing_{f} &:\; \cat{FEnv} & \;\;=\;\; & \lambda x. \bot && \\
        f_1 ... f_n &:\; \cat{Id}^* & \;\;=\;\; & \langle \text{function names corresponding to} && \\
            &&& \;\;\text{global declarations $d_1 ... d_n$} \rangle && \\
        \ulinex{5}{\rho_f} &:\; \cat{FEnv} & \;\;=\;\; & \ulinex{6}{\rho^\varnothing_{f} [f_1 \, \mapsto \, \F{$d_1$}] ... [f_n \, \mapsto \, \F{$d_n$}]} && \\
        \rho_{\varnothing} &:\; \cat{Env} & \;\;=\;\; & \lambda x.\syn{null} &\qquad \qquad \qquad & \\
        \sigma_{\varnothing} &:\; \cat{Store} & \;\;=\;\; & \lambda x.\syn{null} && \\
        loc &:\; \cat{Loc} & \;\;=\;\; & alloc \, \sigma_{\varnothing} && \\
        \rho_1 &:\; \cat{Env} & \;\;=\;\; & \rho_{\varnothing}[\syn{input} \, \mapsto \, loc] && \\
        \sigma_1 &:\; \cat{Store} & \;\;=\;\; & \sigma_{\varnothing}[loc \, \mapsto \, in] && \\
        \rho_2 &:\; \cat{Env} & \;\;=\;\; & \H{$s$} \, \rho_1 \, \sigma_1 && \\
        \sigma_2 &:\; \cat{Store} & \;\;=\;\; & \S{$s$} \, \rho_2 \, \ulinex{5}{\rho_f} \, \sigma_1 && \\
\end{alignat*}
\begin{alignat*}{2}
    \\[-2.4\baselineskip]
    \denotation{\F{function $id_0$($id_1$, $...$, $id_n$)\,\{ $s$; return $e$ \}} \; = \; \lambda \, (v_1 ... v_n) \, \rho_f \, \sigma \, . \, (v, \sigma_3)}
        \shortintertext{\quad where}
        \ulinex{4}{\rho_0} &:\; \cat{Env} \;\;=\;\; \ulinex{4}{\lambda x . \syn{null}} \quad \leftarrow \text{ previously here we were} && \\
        & \qquad \qquad \qquad \quad \text{using environment coming from the outer} && \\
        & \qquad \qquad \qquad \quad \text{lexical scope, now we use an empty one} && \\
        \rho_1 &:\; \cat{Env} \;\;=\;\; \ulinex{4}{\rho_0} [id_1 \, \mapsto \, alloc \, \sigma] ... [id_n \, \mapsto \, alloc \, \sigma] && \\
        \rho_2 &:\; \cat{Env} \;\;=\;\; \H{$s$} \, \rho_1 \, \sigma && \\
        \sigma_1 &:\; \cat{Store} \;\;=\;\; \sigma [\rho_2 \, id_1 \, \mapsto \, v_1] ... [\rho_2 \, id_n \, \mapsto \, v_n] && \\
        \sigma_2 &:\; \cat{Store} \;\;=\;\; \S{$s$} \, \rho_2 \, \rho_f \, \sigma_1 && \\
        (v, \sigma_3) &:\; \cat{Val} \times \cat{Store} \;\;=\;\; \E{$e$} \, \rho_2 \, \rho_f \, \sigma_2 \, && \\ \\
    \shortintertext{Denotations of statements are essentially the same as in the surface langauge - the only difference is that they gain one
    more argument, $\rho_f$, and they pass it through to all recursive invocations of $\S{$\cdot$}$ and $\E{$\cdot$}$:}
    \S{$\cdot$} \;&:\; \cat{Stmt} \,\longrightarrow\, \cat{Env} \; \uline{ \rightarrow \cat{FEnv}} \rightarrow \cat{Store} \rightarrow \cat{Store} && \\ \\
    \shortintertext{Denotations of most expressions are also almost the same as in the surface language. Similarly, they only gain one additional argument that they
    pass through. So here we will describe only these expressions where the difference in semantics is non-trivial.}
    \E{$\cdot$} \;&:\; \cat{Expr} \,\longrightarrow\, \cat{Env} \; \uline{ \rightarrow \cat{FEnv}} \rightarrow \cat{Store} \rightarrow \cat{Val} \times \cat{Store}  && \\ \\
    \denotation{\E{bind-closure $id$} \, \rho \, \rho_f \, \sigma \;=\; ((\rho, id), \sigma)}
    \denotation{\E{$e_0$($e_1$, $...$, $e_n$)} \, \rho \, \rho_f \, \sigma \;=\; f(cl \, v_1 ... v_n) \, \rho_f \, \sigma_n \:\text{ where}}
        (v_0, \sigma_0) &:\; \cat{Val} \times \cat{Store} \;\;=\;\; \E{$e_0$} \, \rho \, \sigma && \\
        (v_1, \sigma_1) &:\; \cat{Val} \times \cat{Store} \;\;=\;\; \E{$e_1$} \, \rho \, \sigma_0 && \\
        ... & \qquad \qquad \qquad \quad \:\; ... && \\
        (v_n, \sigma_n) &:\; \cat{Val} \times \cat{Store} \;\;=\;\; \E{$e_n$} \, \rho \, \sigma_{n-1} && \\
        \uline{(cl, id_f)} &:\; \uline{\cat{Env} \times \cat{Id}} \;\;=\;\; v_0 \:\text{ when }\: v_0 \in \cat{Fun} && \\
        f &:\; \cat{RawFun} \;\;=\;\; \ulinex{4}{ \rho_f \, {id_f}} && \\
    \denotation{\E{$e$.$id$} \, \rho \, \sigma \;=\; (o \, id, \sigma_1) \:\text{ where}}
        (v, \sigma_1) &:\; \cat{Val} \times \cat{Store} \;\;=\;\; \E{$e$} \, \rho \, \sigma && \\
        res &:\; \cat{Obj} \;\;=\;\; \begin{cases}
            (\sigma_1 \, v) \, id \\
                \qquad \quad \text{when }\: v \in \cat{Box} \\
                \qquad \quad \text{and }\: o = \sigma_1 \, v \in \cat{Obj} \\
            \ulinex{5}{\sigma_1 \, (v \, id)} \\
            \qquad \quad \text{ when}\: v \in \cat{Env}
        \end{cases} && \\
    \denotation{\E{$e_1$.$id$ = $e_2$} \, \rho \, \sigma \;=\; (v_2, \sigma_3) \:\text{ where}}
        (v_1, \sigma_1) &:\; \cat{Val} \times \cat{Store} \;\;=\;\; \E{$e_1$} \, \rho \, \sigma && \\
        (v_2, \sigma_2) &:\; \cat{Val} \times \cat{Store} \;\;=\;\; \E{$e_2$} \, \rho \, \sigma_1 && \\
        \sigma_3 &:\; \cat{Store} \;\;=\;\; \begin{cases}
            \sigma_2[v_1 \, \mapsto \, o[id \, \mapsto \, v_2]] \\
                \qquad \quad \text{when }\: v_1 \in \cat{Box} \\
                \qquad \quad \text{and }\: o = \sigma_2 \, v_1 \in \cat{Obj} \\
            \ulinex{5}{\sigma_2[v_1 id \, \mapsto \, v_2]} \\
                \qquad \quad \text{ when}\: v_1 \in \cat{Env}
        \end{cases} && \\
\end{alignat*}
\vspace*{\fill}
}

\cleardoublepage

\chapter{Program analysis}

{
\newcommand{\cat}[1]{{\textbf{#1}}}
\newcommand{\syn}[1]{\texttt{#1}}
\renewcommand{\P}{\mathcal{P}}

\newcommand{\E}[1]{\mathcal{E} \llbracket \syn{#1} \rrbracket}
\renewcommand{\H}[1]{\mathcal{H} \llbracket \syn{#1} \rrbracket}
\renewcommand{\S}[1]{\mathcal{S} \llbracket \syn{#1} \rrbracket}
\newcommand{\PROG}[1]{\mathcal{PROG} \llbracket \syn{#1} \rrbracket}
\newcommand{\COL}[1]{\mathcal{COL} \llbracket \syn{#1} \rrbracket}

\section{Problem statement}

Now that we have picked a concrete language with a well-defined semantics, we can
formulate the problem of program analysis. In this thesis we will focus on algorithms
that can (approximately) answer the following question: for each expression occurring in the
program, what are all possible values to which this expression can evaluate during program execution?

\section{Collecting semantics of TinyScript}

Answering this question can be thought of as computing some property of a program.
\textit{Collecting semantics} is a method to precisely describe properties of programs.
We first observe that our question starts with "for each expression occurring in the program, what
are all possible values". This suggests that our property, which we will call $\cat{ExprValues}$,
is a mapping from expressions to sets of values:
$$
    \cat{ExprValues} \;=\; \cat{Expr} \rightarrow \P(\cat{Val})
$$
For simplicity, we assume that each expression in our domain $\cat{Expr}$ has a unique label - if the same expression occurs in different
places of the program, we treat it as different expressions and we assign separate sets of values to them.
Let $\PROG{$\cdot$}$ be an instrumented TinyScript semantics that gives our property for a particular program execution over a single input string:
$$
    \PROG{$\cdot$} \;:\; \cat{Prog} \rightarrow \cat{Str} \rightarrow \cat{ExprValues}
$$
It is easy to imagine how this instrumented semantics can be defined by modifying standard semantics to additionally pass around and update
the $\cat{ExprValues}$ mapping while evaluating expressions. Please refer to the appendix for details. We treat this instrumented semantics as
a total function, and define its value for non-terminating and erroneous executions as an empty $\cat{ExprValues}$, equal to $\lambda \, e. \, \varnothing$.

However, we are interested in the property of a program and not of a particular execution. Our question starts with "what are all possible X?",
which means that we are interested in taking a union over all possible executions. Therefore, we define the \textit{collecting semantics}
of a TinyScript program $p$ as follows:
\begin{align*}
    \COL{$\cdot$} \;&:\; \cat{Prog} \rightarrow \cat{ExprValues} \\
    \COL{$p$} \;&=\; \bigcup_{in \,\in\, \cat{Str}} \, \PROG{$p$} \, in \\[0.3\baselineskip]
    \shortintertext{Where the union of two $v_1, v_2 \in \cat{ExprValues}$ is defined as:}
    v_1 \cup v_2 \;&=\; \lambda \, e . \: v_1 (e) \cup v_2 (e) \\
\end{align*}
We must observe that directly computing collecting semantics of most programs is intractable.
It is a union over an infinite set of possible inputs. Even computing one term of this union requires
running the program, which can take an unbounded amount of time or not terminate at all. Therefore, all
program analyses focus on computing some over-approximation of it, with varying precision and computational cost.

\section{Abstract interpretation as a framework for building program analyzers}

Abstract interpretation is a method of systematically deriving program analysis algorithms, pioneered by Cousot and Cousot in \cite{cousot1,cousot2,cousot3},
and described very well in a static analysis textbook \cite{spa}.
Given a program and a property of interest, one defines a system of equations that, when solved, would give us some over-approximation of the
desired property. Variables in this system approximate the semantics of relevant parts of the program, and their choice depends on the nature of
the property that we are interested in. The most common setup is that for each point in the program, there is a variable that approximates
program state at this point, and each statement generates an equation that relates program states before and after executing this statement.
An attentive reader might notice that the structure of these equations should closely resemble the denotational semantics of a language. Indeed,
once we decide on a suitable abstract domain to approximate program states, writing a system of equations for a given program becomes a mere formality.
This connection between denotational semantics and abstract interpretation has been explored in depth in \cite{abstr}.

\subsubsection{Lattices and fixed points}
We must note that the framework of abstract interpretation imposes some restrictions on the way we construct these systems of equations.
Every variable must range over some lattice. Left-hand sides of equations should be variables. Right-hand sides of equations, when viewed
as functions of the variables that occur in them, should be monotone wrt. to the lattice order. These conditions ensure that
there exists a unique smallest solution to this system and it is equal to the least fixed point of combined right-hand sides of these
equations. These systems are usually solved by iteratively finding this least fixed point. For this iteration to converge, either all
lattices must be of finite height, or additional techniques must be used to avoid infinite loops. \cite{spa} contains a good
introduction to the theory of lattices and fixed points.
}

\cleardoublepage

\chapter{A naive approach}

{
\newcommand{\cat}[1]{{\textbf{#1}}}
\newcommand{\syn}[1]{\texttt{#1}}
\renewcommand{\P}{\mathcal{P}}
In this chapter we make a first attempt at designing a static analyzer of TinyScript. Our goal now is to establish a
baseline that is easy to understand, and later will serve as a foundation for more advanced approaches.
We formulate a simple, context-insensitive analysis in the framework of abstract interpretation.
It will iteratively compute the following pieces of information:
\begin{itemize}
    \item An abstract state of execution at each point in the program: an approximation of the $\textbf{Store}$ semantic domain
    defined in Chapter 2.
    \item A call graph: a mapping from each function call expression in the program to a set of possible functions that this call might refer to.
    \item An abstract equivalent of the $\textbf{ExprValues}$ from the previous chapter: approximate sets of values that expressions in the
    program might evaluate to.
\end{itemize}

\subsection*{Interdependent call graph and data flow}

\DefineVerbatimEnvironment{TsCode}{Verbatim}{
    gobble=0,numbers=left,numbersep=2mm,
    frame=lines,framerule=0.8mm,
}

It is important to note that computing the call graph and other information about a program must be done simultaneously. Since functions are first-class values, programs
are free to manipulate them in complex ways. This means that it is entirely possible to have multiple rounds of the following feedback loop:
\begin{enumerate}
\item resolve some function calls
\item propagate data flow through these function calls
\item resolve more function calls based on newly-propagated values of callees in call expressions
\item propagate more dataflow through these newly-resolved function calls
\item ...
\end{enumerate}
For example, consider the following TinyScript$^+$ program:
\begin{TsCode}
function curried_add(a) {
    return function(b) {
        return function(c) {
            return a + b + c
        }
    }
}
function wrapper(x) {
    var array = []
    array.push(x)
    return array[0]
}
var adder1 = wrapper(curried_add(2))
var adder2 = wrapper(adder1(3))
var result = wrapper(adder2(5))
\end{TsCode}
Here we see that resolving the call to \texttt{adder2} at line 10 requires that we know the value of \texttt{adder2}. This in turn requires resolving
the call to \texttt{adder1} at line 9 and then propagating data flow through array operations in the body of \texttt{wrapper}. Resolving the call
to \texttt{adder1} in turn requires performing similar operations for \texttt{curried\_add}.

While this example is a bit contrived, similar patterns are frequently seen in practice. Frameworks and libraries tend to introduce
seemingly superfluous layers of indirection to accomodate complex real-world use cases. Examples include plugin systems, dependency injection in OOP, state
management in frontend frameworks and extensive use of decorators in web servers written in Python. It is essential that these indirections do not confuse
code analysis tools.

\section{Abstract domains}

\subsection*{Operating memory}
As we have already mentioned, we will represent operating memory at each point of the program using some abstract domain called \cat{State}, which is
an approximation of the concrete domain \cat{Store} defined in Chapter 2. We might be tempted to define it like this:
\begin{align*}
\cat{State} \quad&=\quad (\cat{Loc} \rightarrow \cat{AbsVal}) \text{ or } (\cat{Box} \rightarrow \cat{AbsPayload}) \text{, depending} \\
& \qquad \qquad \qquad \qquad \qquad \qquad \qquad \qquad \qquad  \qquad \;\;\; \text{on the context} \\
\cat{AbsVal} \quad&=\quad \langle \text{some abstract domain approximating } \P(\cat{Val}) \rangle \\
\cat{AbsPayload} \quad&=\quad \langle \text{some abstract domain approximating } \P(\cat{Payload}) \rangle \\
\cat{Loc} \text{ and } \cat{Box} \quad&=\quad \langle \text{same as in Chapter 2} \rangle
\end{align*}
However, before we do it, let us observe that a program during its execution might use an arbitrarily large number of memory locations.
Even in a small program, a loop or recursion might cause a function to be called an unbounded number of times, and each
call might need separate memory locations for its local variables and for objects or arrays created in this function.
This means that the number of \cat{Loc}s and \cat{Box}es required to analyze a program can be very large.
For a static analysis to be practical, the total number of
variables and equations should be bounded by some quantity related to the size of a program, and this
quantity should not depend too much on its runtime behaviour. Therefore, we need to somehow compress the
potentially-very-large number of memory locations. We will do it by keeping only one set of local variables for each
function -- it will over-approximate local variables in all possible calls to this function.
We will treat global variables as local variables of a special function called \syn{main}, and
later on we will use the same logic for handling both local and global variables.
For heap-allocated data (objects and arrays), we will only keep one copy per each place in the program
where an array or an object is created. This is known as an \textit{allocation-site abstraction}.
\\ \\
We define the abstract equivalent of the concrete domain \cat{Store} as:
\begin{align*}
\cat{State} \quad&=\quad (\cat{Loc} \rightarrow \cat{Val}) \text{ or } (\cat{Box} \rightarrow \cat{Payload}) \text{, depending} \\
& \qquad \qquad \qquad \qquad \qquad \qquad \qquad \qquad \qquad \;\;\; \text{on the context} \\
\cat{Loc} \quad&=\quad \cat{FunctionId} \times \cat{Id} \\
\cat{Box} \quad&=\quad \{e \in \cat{Expr} \;|\; e \text{ is an object or array literal}\} \\
\cat{FunctionId} \quad&=\quad \{id \in \cat{Id} \;|\; id \text{ is a function name}\} \; \cup \; \{\syn{main}\} \\ \\
\cat{Val} \quad&=\quad \langle \text{abstract domain approximating sets of values} \rangle \\
\cat{Payload} \quad&=\quad \langle \text{abstract domain approximating objects and arrays} \rangle \\
\end{align*}
For convenience, for abstract domains \cat{Loc}, \cat{Box}, \cat{Val}, \cat{Payload}, etc. we reuse the names from concrete domains.
From now on, we will be dealing only with abstract domains most of the time. In cases when we need to distinguish abstract and concrete
variants, we will write $\cat{Loc}^A$, $\cat{Box}^A$, $\cat{Val}^A$, $\cat{Payload}^A$ and $\cat{Loc}^C$, $\cat{Box}^C$, $\cat{Val}^C$, $\cat{Payload}^C$.

\subsection*{Value lattice}

Now we will design abstract domains to approximate sets of values, objects and arrays. Since this is a baseline that aims to be easy
to understand, we will choose a simple representation. The simplest possible choice would be
$\cat{Val}^A = \P(\cat{Val}^C)$ and $\cat{Payload}^A = \P(\cat{Payload}^C)$. This creates similar problems as the
set of memory locations discussed in previous section. $\cat{Val}^C$ and $\cat{Payload}^C$ are infinite, and therefore:
\begin{itemize}
\item The lattices $\P(\cat{Val}^C)$ and
$\P(\cat{Payload}^C)$ have infinite height. This means that it's not feasible to solve equations over this lattice by iteratively
computing the least fixed point, since this process might not converge for lattices of infinite height. In fact, most programs that
humans write do contain some parts that would cause the fixpoint iteration to diverge.
\item Representing elements of $\P(\cat{Val}^C)$ and
$\P(\cat{Payload}^C)$ can consume arbitrarily large amount of memory.
\end{itemize}
To address these two problems, we introduce a \textit{restricted powerset lattice}, $\P_{\leq N}(\cdot)$. It consists
of all subsets containing at most $N$ elements, and a top element representing the whole set:
$$
    \P_{\leq N}(A) \quad=\quad \{s \subseteq A \;:\; |s| \leq N \} \; \cup \; \{\top_{A}\} \quad \text{ where } \quad \top_A = A
$$
We introduce a distinct symbol $\top_A$ in place of $A$ to emphasize its role as an element of the lattice. It also aligns well
with the fact that implementations will use a special placeholder value to represent $\top_A$, and they will represent
other subsets of $A$ by directly enumerating their elements. Similarly, we will often indicate the bottom element of this lattice as
$\bot_A$ or $\bot$ instead of $\varnothing$.
\\ \\
Since $\P_{\leq N}(A)$ is a sub-lattice of the powerset lattice $\P(A)$, it inherits its structure from it. The \textit{join} operation
(a.k.a. the least upper bound) is fully determined by the partial order inherited from $\P(A)$, and consists of handling two simple cases:
$$
    \text{for } X, Y \in \P_{\leq N}(A), \qquad X \sqcup Y \;=\; \begin{cases}
        X \cup Y \:\text{ when }\: | X \cup Y | \leq N \\ 
        \top_A  \quad \:\:\:\text{ otherwise }
    \end{cases}
$$
We can also take the least upper bound of more than 2 elements -- it is well-defined for any subset of $\P_{\leq N}(A)$:
$$
    \text{for } S \subseteq \P_{\leq N}(A), \qquad \bigsqcup S \;=\; \begin{cases}
        \bigcup S \:\text{ when }\: | \bigcup S | \leq N \\ 
        \top_A  \,\:\:\text{ otherwise }
    \end{cases}
$$
Equipped with a new tool, we can proceed to define the lattice to compactly approximate sets of values. This lattice is parameterized by
some number $N \in \mathbb{N}$ that controls the trade-off between analysis precision and computational cost.
$$
\cat{Val} \quad=\quad \P_{\leq N}(\cat{Int}) \, \times \, \P_{\leq N}(\cat{Str})
        \, \times \, \P(\cat{Bool} \cup \{\syn{null}\} \cup \cat{Box} \cup \cat{Fun} \cup \cat{Env})
$$
We define three auxiliary functions, that select the first, second and third component of a \cat{Val}:
\begin{align*}
    \text{int} \; &\,: \; \cat{Val} \rightarrow \P_{\leq N}(\cat{Int}) \\
    \text{str} \; &\,: \; \cat{Val} \rightarrow \P_{\leq N}(\cat{Str}) \\
    \text{other} \; &\,: \; \cat{Val} \rightarrow \P(\cat{Bool} \cup \{\syn{null}\} \cup \cat{Box} \cup \cat{Fun} \cup \cat{Env})
\intertext{
We also define auxiliary functions to select various subsets of the third component:
}
    \text{bool} \; &\,: \; \cat{Val} \rightarrow \P(\cat{Bool}) \\
    \text{null} \; &\,: \; \cat{Val} \rightarrow \P(\{\syn{null}\}) \\
    \text{box} \; &\,: \; \cat{Val} \rightarrow \P(\cat{Box}) \\
    \text{fun} \; &\,: \; \cat{Val} \rightarrow \P(\cat{Fun}) \\
    \text{env} \; &\,: \; \cat{Val} \rightarrow \P(\cat{Env})
\intertext{
We represent functional values as pairs of an outer environment (a closure) and a function name, as in Chapter 2:
}
    \cat{Fun} \quad&=\quad \cat{Env} \, \times \, \cat{FunctionId} \\
\intertext{
We note that this analysis does not distinguish
between different calls to the same function: it computes only a single environment that over-approximates environments
of all possible calls to a function. This means that we can represent the outer environment simply as the name of the enclosing function,
since there is only one environment per function anyway:
}
    \cat{Env} \quad&=\quad \cat{FunctionId} \\
\intertext{
The value lattice is of finite height because it is a product of three finite-height lattices. The first two are restricted powerset lattices,
which by construction are of finite height. The last one is a powerset of a finite set. The set
$\cat{Bool} \cup \{\syn{null}\} \cup \cat{Box} \cup \cat{Fun} \cup \cat{Env}$ is finite:
}
    |\cat{bool}| \quad&=\quad 2 & \text{ -- finite} \\
    |\{\syn{null}\}| \quad&=\quad 1 & \text{ -- finite} \\
    |\cat{Box}| \quad&=\quad \langle \text{number of object and array} \\
     & \qquad \;\;\; \text{literals occurring in the program} \rangle & \text{ -- finite} \\
    |\cat{Fun}| \quad&=\quad |\cat{FunctionId}|^2 \\
     &=\quad \langle \text{number of functions in} \\
     & \qquad \;\;\; \text{the program, squared} \rangle & \text{ -- finite} \\
    |\cat{Env}| \quad&=\quad |\cat{FunctionId}| & \text{ -- finite}
\intertext{
We define the lattices of object and array payloads as follows:
}
    \cat{Payload} \quad&=\quad \cat{Obj} \, \times \, \cat{Arr} \\
    \cat{Obj} \quad&=\quad \cat{Id} \, \rightarrow \, \cat{Val} \\
    \cat{Arr} \quad&=\quad \cat{Val}
\intertext{
As with \cat{Val}, we define auxiliary functions that select first and second component of \cat{Payload}:
}
    \text{obj} \; &\,: \; \cat{Payload} \rightarrow \cat{Obj} \\
    \text{arr} \; &\,: \; \cat{Payload} \rightarrow \cat{Arr}
\end{align*}
We represent objects as mappings from property names to elements of the value lattice approximating values of these properties.
Since the total number of object properties in a given program is finite, and the value lattice is of finite height, the height of
the object lattice is also finite. Our representation of arrays is crude: we lump all array elements together irrespective of their position
and maintain only a single approximation of the set of possible values any element of the array can take. We also do not maintain any information
about the length of the array.

\subsection*{Call graph lattice}

We define the lattice of possible call graphs of a given program as follows:
\begin{align*}
    \cat{CallGraph} \quad&=\quad \cat{CallSite} \rightarrow \P(\cat{FunctionId}) \\
    \cat{CallSite} \quad&=\quad \{e \in \cat{Expr} \;|\; e \text{ is a call expression}\}
\end{align*}
The height of this lattice is finite, since all programs have a finite number of call expressions and function definitions.

\section{Analysis algorithm}
We compute an approximation of the \cat{ExprValues} property of an input program $p$ in the following steps:
\begin{enumerate}
    \item Translate program $p$ into a set of lattice-valued variables and a system of equations relating these variables,
        such that a solution of this system will approximate the property of interest.
    \item Solve this system of equations by iteratively computing the least fixed point of functions that
        relate left and right hand sides of equations. Start from a point where all variables are assigned the bottom
        element of their lattice.
    \item Retrieve the property of interest from the solution to this constraint system. This usually means just
        selecting the right subset of variables.
\end{enumerate}

\subsection*{Prerequisites}
We start by enumerating various syntactic elements of the input program, such that later we can easily refer to them when defining
corresponding variables and equations:
\begin{alignat*}{2}
    \{e_1, ..., e_n\} \;&=\; \cat{Expr} & \quad - \quad & \text{the set of expressions occurring in $p$} \\
    \{s_1, ..., s_m\} \;&=\; \cat{Stmt} & \quad - \quad & \text{the set of statements occurring in $p$} \\
    \{d_1, ..., d_k\} \;&=\; \cat{Decl} & \quad - \quad & \text{the set of function declarations in $p$} \\
    \{f_1, ..., f_k\} \;&=\; \cat{FunctionId} & \quad - \quad & \text{the set of function names in $p$} \\
    \{\alpha_1, ..., \alpha_l\} \;&=\; \cat{CallSite} & \quad - \quad & \text{the set of call expressions in $p$} \\
    \{b_1, ..., b_t\} \;&=\; \cat{Box} & \quad - \quad & \text{the set of array and object } \\
        &&& \text{literal expressions in $p$}
\end{alignat*}

\subsection*{Variables}

\newcommand{\sigmain}{\sigma^{\text{\tiny in}}}

\begin{alignat*}{2}
    \sigmain_{e_1}, \, ... \, , \, \sigmain_{e_n} \: &\in \: \textbf{State} & \quad \;\; \text{---} \quad \;\; & \text{execution states before} \\
        &&& \text{evaluating expressions $e_1, ..., e_n$} \\
    \sigma_{e_1}, \, ... \, , \, \sigma_{e_n} \: &\in \: \textbf{State} & \quad \;\; \text{---} \quad \;\; & \text{execution states after} \\
        &&& \text{evaluating expressions $e_1, ..., e_n$} \\
    v_{e_1}, \, ... \, , \, v_{e_n} \: &\in \: \textbf{Val} & \quad \;\; \text{---} \quad \;\; & \text{approximate sets of values that} \\
        &&& \text{expressions $e_1, ..., e_n$ might evaluate to} \\ \\
    \sigmain_{s_1}, \, ... \, , \, \sigmain_{s_m} \: &\in \: \textbf{State} & \quad \;\; \text{---} \quad \;\; & \text{execution states before} \\
        &&& \text{executing statements  $s_1, ..., s_m$} \\
    \sigma_{s_1}, \, ... \, , \, \sigma_{s_m} \: &\in \: \textbf{State} & \quad \;\; \text{---} \quad \;\; & \text{execution states after} \\
        &&& \text{executing statements  $s_1, ..., s_m$} \\ \\
        \sigmain_{f_1}, \, ... \, , \, \sigmain_{f_k} \: &\in \: \textbf{State} & \quad \;\; \text{---} \quad \;\; & \text{execution states before} \\
        &&& \text{entering functions  $f_1, ..., f_k$} \\
    \sigma_{f_1}, \, ... \, , \, \sigma_{f_k} \: &\in \: \textbf{State} & \quad \;\; \text{---} \quad \;\; & \text{execution states after} \\
        &&& \text{exiting functions  $f_1, ..., f_k$} \\
    v_{f_1}, \, ... \, , \, v_{f_k} \: &\in \: \textbf{Val} & \quad \;\; \text{---} \quad \;\; & \text{approximate sets of values that} \\
        &&& \text{functions $f_1, ..., f_k$ might return} \\ \\
    \shortintertext{$\qquad \; c_{\alpha_1}, \, ... \, , \, c_{\alpha_l} \: \in \: \P(\textbf{FunctionId})$}
    \\[-0.8\baselineskip]
     & & \quad \;\; \text{---} \quad \;\; & \text{sets of functions that} \\
        &&& \text{call sites $\alpha_1, ..., \alpha_l$ might refer to}
\end{alignat*}

\subsection*{Equations}

\newcommand{\case}[2]{\\[-0.5\baselineskip] \shortintertext{$\llbracket \syn{#1} \rrbracket$ \hfill}}

Let $f$ be the name of some function in the input program. For each $f$, we generate the following equations:
\begin{alignat*}{2}
    \\[-1.5\baselineskip]
    \case{function $f$($...$) \{ $s_0$; return $e_0$ \}}{function}
        &\Longrightarrow& \quad \sigmain_{s_0} &= \sigmain_f \quad \land \quad \sigmain_{e_0} = \sigma_{s_0} \quad \land \quad
        v_f = v_{e_0} \quad \land \quad \sigma_f = \sigma_{e_0} \qquad \qquad \qquad \qquad \qquad \qquad \qquad
\end{alignat*}
Let $e$ be an expression occurring inside the body of $f$, including the
result expression after the \syn{return} keyword. We also include here all expressions occurring in the main program,
and for them we set $f = \syn{main}$. For each pair of $(f, e)$, we generate the following equations, depending on the type of $e$:
\begin{alignat*}{2}
    \\[-1.5\baselineskip]
    \case{$id$}{variable}
        &\Longrightarrow& \quad \sigma_e &= \sigmain_e \quad \land \quad v_e = \sigmain_e (f, id) \qquad \qquad \qquad \qquad \qquad \\
    \case{$e_1$.$id$}{property}
        &\Longrightarrow& \quad \sigmain_{e_1} &= \sigmain_e \quad \land \quad \sigma_e = \sigma_{e_1} \quad \land \\
        && v_e &= \bigsqcup \, \big\lbrace \text{obj}(\sigma_e \, b) \, id \;|\; b \in \text{box} (v_{e_1}) \big\rbrace \\
        && & \quad \sqcup \, \bigsqcup \, \big\lbrace \sigma_e(env, id) \;|\; env \in \text{env} (v_{e_1}) \big\rbrace \\
    \case{$e_1$[$e_2$]}{array element}
        &\Longrightarrow& \quad \sigmain_{e_1} &= \sigmain_e \quad \land \quad \sigmain_{e_2} = \sigma_{e_1} \quad \land \quad \sigma_e = \sigma_{e_2} \\
        && v_e &= \bigsqcup \, \big\lbrace \text{arr}(\sigma_e \, b) \;|\; b \in \text{box} (v_{e_1}) \big\rbrace \\
    \case{$id$ = $e_1$}{variable assignment}
        &\Longrightarrow& \quad \sigmain_{e_1} &= \sigmain_e \quad \land \quad \sigma_e = \sigma_{e_1} [(f, id) \, \mapsto \, v_{e_1}] \\
    \case{$e_1$.$id$ = $e_2$}{property assignment}
        &\Longrightarrow& \quad \sigmain_{e_1} &= \sigmain_e \quad \land \quad \sigmain_{e_2} = \sigma_{e_1} \quad \land \\
        && \sigma_e &= \bigsqcup \big\lbrace \sigma_{e_2}[b \, \mapsto \, o] \;|\; b \in \text{box} (v_{e_1}) \:\text{ and} \\
        &&& \qquad \qquad \qquad \qquad \quad o = (\text{obj}(\sigma_{e_2} \, b)[id \, \mapsto \, v_{e_2}] , \bot) \: \sqcup \: \sigma_{e_2} \, b \big\rbrace \\
        && & \quad \sqcup \, \bigsqcup \, \big\lbrace \sigma_{e_2} [(env, id) \, \mapsto \, v_{e_2}] \;|\; env \in \text{env} (v_{e_1}) \big\rbrace \\
    \case{$e_1$[$e_2$] = $e_3$}{array element assignment}
        &\Longrightarrow& \quad \sigmain_{e_1} &= \sigmain_e \quad \land \quad \sigmain_{e_2} = \sigma_{e_1} \quad \land \quad \sigmain_{e_3} = \sigma_{e_2} \quad \land \\
        && \sigma_e &= \bigsqcup \big\lbrace \sigma_{e_3}[b \, \mapsto \, a] \;|\; b \in \text{box} (v_{e_1}) \:\text{ and} \\
        &&& \qquad \qquad \qquad \qquad \quad a = (\bot, \text{arr}(\sigma_{e_3} \, b) \, \sqcup \, v_{e_3}) \big\rbrace \\
    \\[-0.5\baselineskip] \shortintertext{$
    \llbracket \syn{$e_1$ + $e_2$} \rrbracket, \; \llbracket \syn{$e_1$ - $e_2$} \rrbracket, \;
    \llbracket \syn{$e_1$ * $e_2$} \rrbracket, \; \llbracket \syn{$e_1$ / $e_2$} \rrbracket$ \hfill}
        &\Longrightarrow& \quad \sigmain_{e_1} &= \sigmain_e \quad \land \quad \sigmain_{e_2} = \sigma_{e_1} \quad \land \quad \sigma_e = \sigma_{e_2} \\
        && v_e &= \bigsqcup \, \big\lbrace (v_1 \, op \, v_2, \bot, \bot) \;|\; v_1 \in \text{int} (v_{e_1}) \:\text{ and }\: v_2 \in \text{int}(v_{e_2}) \big\rbrace \\
        && & \quad \text{where } op \, = \, + \, \text{or} \, - \, \text{or} \, \cdot \, \text{or} \, \lfloor / \rfloor \\
    \case{$e_0$($e_1$, $...$, $e_d$)}{function call}
    &\Longrightarrow& \quad \sigmain_{e_0} &= \sigmain_e \quad \land \quad \sigmain_{e_1} = \sigma_{e_0} \quad \land \quad ... \quad \land \quad \sigmain_{e_d} = \sigma_{e_{d-1}} \quad \land \\
    && \sigmain_{f_1} &= \sigmain_{f_1} \sqcup call_{f_1} \quad \land \quad ... \quad \land \quad \sigmain_{f_k} = \sigmain_{f_k} \sqcup call_{f_k} \quad \land \\
    && & \; \text{where} \\[-0.7\baselineskip]
    && & \qquad call_{f_i} = \begin{cases}
        \sigma_{e_d}[(f_i, id_0^{\,i}) \, \mapsto \, \text{cl}(v_{e_0}, f_i)] \\
        \quad \; [(f_i, id_1^{\,i}) \, \mapsto \, v_{e_1}] ... [(f_i, id_d^{\,i}) \, \mapsto \, v_{e_d}] \\
        \qquad \,\text{when }\: f_i \in c_e \:\text{ and }\: \#f_i = d \\
        \bot \quad \text{ otherwise} \end{cases} \\
    && & \qquad id_0^{\,i}, id_1^{\,i}, ..., id_{\#f_i}^{\,i} \quad \text{--} \quad \text{formal parameters of function $f_i$} \\
    && & \qquad \text{note: in programs generated by closure conversion,}\\
    && & \qquad \text{$id_0^{\,i}$ is always equal to \syn{closure}} \\
    && & \qquad \#f_i \quad \text{--} \quad \text{number of formal parameters of function $f_i$} \\
    && & \qquad \text{cl}(v, f_i) = (\bot, \bot, \{ cl \;|\; (cl, f) \in \text{fun}(v) \:\text{ and }\: f = f_i\}) \\
    && \sigma_e &= \bigsqcup \, \big\lbrace \sigma_{f} \;|\; f \in c_e \big\rbrace \quad \land \quad v_e = \bigsqcup \, \big\lbrace v_{f} \;|\; f \in c_e \big\rbrace \quad \land \\
    && c_e &= \{ f \;|\; (cl, f) \in \text{fun}(v_{e_0})\} \\
    \\[-0.5\baselineskip] \shortintertext{$
    \llbracket \syn{\{$id_1$:\;$e_1$, $...$, $id_d$:\;$e_d$\}} \rrbracket$ \hfill}
        &\Longrightarrow& \quad \sigmain_{e_1} &= \sigmain_e \quad \land \quad \sigmain_{e_2} = \sigma_{e_1} \quad \land \quad ... \quad \land \quad \sigmain_{e_d} = \sigma_{e_{d-1}} \quad \land \\
        && v_{e} &= e \quad \land \quad \sigma_e = \sigma_{e_d}[e \, \mapsto \, ((\lambda id. \, \bot)[
            id_1 \, \mapsto \, v_{e_1}, ..., id_d \, \mapsto \, v_{e_d}
        ], \bot)] \\
    \\[-0.5\baselineskip] \shortintertext{$
    \llbracket \syn{[$e_1$, $...$, $e_d$]} \rrbracket$ \hfill}
        &\Longrightarrow& \quad \sigmain_{e_1} &= \sigmain_e \quad \land \quad \sigmain_{e_2} = \sigma_{e_1} \quad \land \quad ... \quad \land \quad \sigmain_{e_d} = \sigma_{e_{d-1}} \quad \land \\
        && v_{e} &= e \quad \land \quad \sigma_e = \sigma_{e_d}[e \, \mapsto \, (\bot, v_{e_1} \, \sqcup \, ... \, \sqcup \, v_{e_d})] \\
    \case{bind-closure $id$}{closure capture}
    &\Longrightarrow& \quad \sigma_e &= \sigmain_e \quad \land \quad v_e = (f, id)
\end{alignat*}
Let $s$ be a statement occurring inside the body of $f$.
For each pair of $(f, s)$, we generate the following equations, depending on the type of $s$:
\begin{alignat*}{2}
    \\[-1.5\baselineskip]
    \case{$\langle empty \rangle$}{}
        &\Longrightarrow& \quad \sigma_s &= \sigmain_s \qquad \qquad \qquad \qquad \qquad \qquad \qquad \qquad \qquad \qquad \qquad \qquad \qquad \\
    \case{$e$}{}
        &\Longrightarrow& \quad \sigmain_e &= \sigmain_s \quad \land \quad \sigma_s = \sigma_e \\
    \case{$s_1$; $s_2$}{}
        &\Longrightarrow& \quad \sigmain_{s_1} &= \sigmain_s \quad \land \quad \sigmain_{s_2} = \sigma_{s_1} \quad \land \quad \sigma_s = \sigma_{s_2} \\
    \case{var $id$ = $e$}{}
        &\Longrightarrow& \quad \sigmain_{e} &= \sigmain_s \quad \land \quad \sigma_s = \sigma_{e} [(f, id) \, \mapsto \, v_{e}] \\
    \case{if ($e$) \{ $s_1$ \} else \{ $s_2$ \}}{}
        &\Longrightarrow& \quad \sigmain_e &= \sigmain_s \quad \land \\
        && \sigmain_{s_1} &= \sigma_e \text{ if } \syn{true} \in \text{bool}(v) \text{ else } \bot_{\cat{State}} \quad \land \\
        && \sigmain_{s_2} &= \sigma_e \text{ if } \syn{false} \in \text{bool}(v) \text{ else } \bot_{\cat{State}} \quad \land \\
        && \sigma_s &= (\sigma_{s_1} \text{ if } \syn{true} \:\; \in \text{bool}(v) \text{ else } \bot_{\cat{State}}) \\
        &&& \: \sqcup (\sigma_{s_2} \text{ if } \syn{false} \in \text{bool}(v) \text{ else } \bot_{\cat{State}}) \\
    \case{for (var $id$ in $e$) \{ $s_0$ \}}{}
        &\Longrightarrow& \quad \sigmain_e &= \sigmain_s \quad \land \quad \sigma_s = \sigma_{s_0} \, \sqcup \, \sigma_e \quad \land \\
        && \sigmain_{s_0} &= (\sigma_e \, \sqcup \, \sigma_{s_0})\big[(f, id) \: \mapsto \: \bigsqcup \, \big\lbrace \text{arr}(\sigma_e \, b) \;|\; b \in \text{box} (v_e) \big\rbrace\big] \\
    \case{while ($e$) \{ $s_0$ \}}{}
        &\Longrightarrow& \quad \sigmain_e &= \sigmain_s \, \sqcup \, \sigma_{s_0} \quad \land \quad
        \sigmain_{s_0} = \sigma_e \quad \land \quad \sigma_s = \sigma_{s_0} \, \sqcup \, \sigma_e
\end{alignat*}

}

\cleardoublepage

\chapter{Context-sensitive analysis}

{
\newcommand{\cat}[1]{{\textbf{#1}}}
\newcommand{\syn}[1]{\texttt{#1}}
\renewcommand{\P}{\mathcal{P}}
\newcommand{\Pfin}{\mathcal{P}_{\text{fin}}}

\newcommand*\mcup{\mathbin{\mathpalette\mcupinn\relax}}
\newcommand*\mcupinn[2]{\vcenter{\hbox{$\mathsurround=0pt
  \ifx\displaystyle#1\textstyle\else#1\fi\bigcup$}}}
\newcommand*\mcap{\mathbin{\mathpalette\mcapinn\relax}}
\newcommand*\mcapinn[2]{\vcenter{\hbox{$\mathsurround=0pt
  \ifx\displaystyle#1\textstyle\else#1\fi\bigcap$}}}
\newcommand*\msqcup{\mathbin{\mathpalette\msqcupinn\relax}}
\newcommand*\msqcupinn[2]{\vcenter{\hbox{$\mathsurround=0pt
  \ifx\displaystyle#1\textstyle\else#1\fi\bigsqcup$}}}
\newcommand*\msqcap{\mathbin{\mathpalette\msqcapinn\relax}}
\newcommand*\msqcapinn[2]{\vcenter{\hbox{$\mathsurround=0pt
  \ifx\displaystyle#1\textstyle\else#1\fi\bigsqcap$}}}

\makeatletter
\newcommand*{\saved@uline}{}
\let\saved@uline\uline
\newcommand*{\mathuline}{%
  \mathpalette{\math@uline\saved@uline}%
}
\newcommand*{\math@uline}[3]{%
  \mbox{#1{$#2#3\m@th$}}%
}
\renewcommand*{\uline}{%
  \relax  
  \ifmmode
    \expandafter\mathuline
  \else
    \expandafter\saved@uline
  \fi
}
\makeatother

\newcommand{\addr}[1]{{\texttt{\#}}_{#1}}

In this chapter we will develop an improved analysis that overcomes the shortcomings
of the naive approach from Chapter 4. We will start by introducing the concept of
a calling context. Then, we will describe a way to lift the value lattice to be
context-sensitive. Let's suppose we have some memory location and we are at some
particular point in the program. We will no longer ask:
\textit{what values this memory location might hold?}
Instead, it will be: \textit{how is the set of possible
values of this memory location dependent on the current call stack?}
In other words, the lattice of context-sensitive values will consist of mappings from
call stacks into approximate sets of values.
\\ \\
Equipped with this more precise representation, we will define context-sensitive versions of other abstract
domains and derive our analysis in the framework of abstract interpretation.

\section{Calling contexts}

\subsection*{Frame of reference}

First, let us make a small observation. In the previous chapter,
when generating equations for the naive analysis, we always operated within a
\textit{frame of reference} corresponding to some point in the program. We were always located
at some point in the code inside the body of some function. We analyzed this function
irrespective of where this function had been called from and what chain of function
calls had led to this particular invocation. Our question was \textit{a program somehow
got to execute this function, what can we say about its execution?} and not \textit{
the entry point of our program is main(), what can we say about the execution
of all transitive callees?}
While the latter directly corresponds to our problem statement
defined in Chapter 3 (\textit{collecting semantics}), the former is closer to what programmers
think when trying to understand code. They read the code of a function and most of the time
their question can be answered by looking only at the local neighbourhood of callers and callees,
without going all the way to the main entry point. This perspective also has a tremendous user
experience benefit: all code will be analyzed no matter whether it is reachable from main
or not. While unreachable functions are rarely commited into code repositories, it is common
to write some subroutines and only later connect them with the rest of the program.
\\ \\
Therefore, when thinking about call stacks, our frame of reference will always be the function
that is currently being executed - the one at the top of the call stack. During analysis, we start with the
assumption that \textit{a program somehow got to execute this function}. At first we analyze its properties
that are independent of the caller, akin to constant propagation. Then we look at the direct
callers of this function, and propagate data flow coming from them. Later we look at data flow
originating from callers of callers of this function. And so on.

\subsection*{Call stacks}

\DefineVerbatimEnvironment{TsCode}{Verbatim}{
    gobble=0,numbers=left,numbersep=2mm,
    frame=lines,framerule=0.8mm,
    commandchars=\\\{\},codes={\catcode`$=3\catcode`_=8},
}

We represent call stacks as sequences of zero or more call sites. It follows naturally that we should anchor them
to our frame of reference - the function at the top. Therefore, we adopt a convention where call sites are enumerated
top-to-bottom, starting with the direct caller of the function that is currently being executed.
\begin{align*}
    \cat{CallSite} \quad&=\quad \{e \in \cat{Expr} \;|\; e \text{ is a call expression}\} \\
    \cat{Stack} \quad&=\quad \cat{CallSite}^*
\end{align*}
The space of call stacks admits a natural tree-like structure, where full stacks correspond to leaves
and common prefixes correspond to nodes. Consider the following program:
\begin{TsCode}
function f() \{
    $\longleftarrow$ we are here
\}

function bar() \{
    f$_\alpha$()
\}

function foo1() \{
    bar$_\beta$()
\}

function foo2() \{
    bar$_\gamma$()
\}

function baz() \{
    f$_\delta$()
\}

function main() \{
    foo1$_{m1.1}$()
    foo1$_{m1.2}$()
    foo2$_{m2}$()
    baz$_{m3}$()
\}
\end{TsCode}
Possible call stacks during execution of \syn{f()} can be organized into a tree of common prefixes like this:
\begin{align*}
    \\[-3\baselineskip]
    \begin{split}
    & \text{Call stacks:} \\
    & \bullet \; \alpha \: \beta \: m_{1.1} \\
    & \bullet \; \alpha \: \beta \: m_{1.2} \\
    & \bullet \; \alpha \: \gamma \: m_2 \\
    & \bullet \; \delta \: m_3
    \end{split}
    \quad
    \begin{split}
        \\ \\
        &\text{Tree:} \\
        &\begin{tikzpicture}[thick]
            \node (f) at ( 0,2) [circle,draw=blue!50,fill=blue!20] {};
            \node (bar) at ( -2,1) [circle,draw=blue!50,fill=blue!20] {};
            \node (baz) at ( 2,1) [circle,draw=blue!50,fill=blue!20] {};
            \node (main3) at ( 2.7,0) [circle,draw=blue!50,fill=blue!20] {};
            \node (foo1) at ( -3,0) [circle,draw=blue!50,fill=blue!20] {};
            \node (foo2) at ( -1,0) [circle,draw=blue!50,fill=blue!20] {};
            \node (main11) at ( -3.7,-1) [circle,draw=blue!50,fill=blue!20] {};
            \node (main12) at ( -2.3,-1) [circle,draw=blue!50,fill=blue!20] {};
            \node (main2) at ( -0.3,-1) [circle,draw=blue!50,fill=blue!20] {};
            \draw[-] (bar) -- node[auto, inner sep=1.5pt] {$\alpha$} (f);
            \draw[-] (baz) -- node[auto, inner sep=1.5pt] {$\delta$} (f);
            \draw[-] (foo1) -- node[auto, inner sep=1.5pt] {$\beta$} (bar);
            \draw[-] (foo2) -- node[auto, inner sep=1.5pt] {$\gamma$} (bar);
            \draw[-] (main11) -- node[auto, inner sep=1.5pt] {$m_{1.1}$} (foo1);
            \draw[-] (main12) -- node[auto, inner sep=5pt, right] {$m_{1.2}$} (foo1);
            \draw[-] (main2) -- node[auto, inner sep=1.5pt] {$m_{2}$} (foo2);
            \draw[-] (main3) -- node[auto, inner sep=1.5pt] {$m_{3}$} (baz);
          \end{tikzpicture}
    \end{split}
\end{align*}

\subsection*{Abstracting over call stacks}

So now we have a nice tool to distinguish between different call stacks at a given point in the
program. In principle, we could use it as a foundation for building our context-sensitive analysis.
However, it turns out that it's much better to operate on subtrees that share a common prefix instead
of individual call stacks \cite{bdds,merging}. Let's consider an example with a simple question:
\begin{TsCode}
function \_do\_work(workspace, command) \{
    workspace.table $\longleftarrow$ \textit{what are possible values of this}
                         \textit{expression, depending on the call stack?}
    // ... some code doing the work ...
\}

function initialize() \{
    // Let's pretend that it is a heavy calculation to precompute
    // something using hardcoded data. Examples include preparing
    // configuration, building a search index, a lookup table, or
    // precomputing some statistics over a known dataset.
    return 'table'
\}

function library\_routine(workspace, command) \{
    if (!workspace.is\_initialized) \{
        workspace.table = initialize$_\beta$()
        workspace.is\_initialized = true
    \}
    \_do\_work$_\alpha$(workspace, command);
\}

function application\_routine\_1() \{
    workspace = \{\}
    library\_routine$_{\gamma_1}$(workspace, "command 1")
\}

...

function application\_routine\_n() \{
    workspace = \{\}
    library\_routine$_{\gamma_n}$(workspace, "command n")
\}

function main() \{
    application\_routine\_1$_{\delta_1}$()
    ...
    application\_routine\_n$_{\delta_n}$()

    other\_workspace = \{table: 'other\_table'\}
    \_do\_work$_\sigma$(other\_workspace, 'other\_command')
\}
\end{TsCode}
\pagebreak
Let's assume that we only have individual call stacks at our disposal. To answer our
question, we need to enumerate all of them and assign a set of values to each one:
\begin{align*}
    \alpha \: \gamma_1 \: \delta_1 \; &\mapsto \; \{ \syn{'table'} \} \\
     &\;\;\vdots \\
    \alpha \: \gamma_n \: \delta_n \; &\mapsto \; \{ \syn{'table'} \} \\
    \sigma \;\;\;\;\;\;\;\;\;\, &\mapsto \; \{ \syn{'other\_table'} \}
\end{align*}
Moreover, not only we have to store $n$ copies of \syn{'table'} instead of one. To
arrive at this result, we have to analyze the \syn{initialize()} function $n$ times, each
time performing the same computation, but with a different call stack. It is clearly a lot
of waste and we feel that it should be possible to share this computation.
\\ \\
We can do it by changing our representation to store generalized results that are valid for
groups of call stacks that share a common prefix. They make the answer to our question much
more compact and elegant:
\begin{align*}
    \alpha \: * \; &\mapsto \; \{ \syn{'table'} \} \\
    \sigma \: * \; &\mapsto \; \{ \syn{'other\_table'} \}
\end{align*}
The stars at the end are wildcards and mean that these answers are valid for any call stack
that starts with $\alpha$ or $\sigma$ respectively. There are two more benefits of this representation:
\begin{itemize}
    \item It is more friendly to running the analysis interactively with incremental updates.
        Editing or deleting any of \syn{application\_routine\_1}, ..., \syn{application\_routine\_n}
        or \syn{main} will not trigger recomputation of results for \syn{\_do\_work} and
        \syn{library\_routine}.
    \item It will compute correct results regardless of whether \syn{\_do\_work} and
        \syn{library\_routine} are reachable from main. As we already argued, this is a huge user experience win.
\end{itemize}

\subsection*{The lattice of calling contexts}

We will now develop this concept of grouping call stacks and operating on whole groups into a proper, lattice-theoretic
abstract domain. First, we define \mbox{\cat{SubCtx}} - the family of subtrees of the tree of all potential call stacks:
\begin{align*}
\cat{SubCtx} \quad &\subseteq \quad \P(\cat{Stack}) \\
\cat{SubCtx} \quad &= \quad \{ \text{subtree}(a) \;|\; a \in \cat{CallSite}^*\} \\
\text{subtree} \quad &\:: \quad \cat{CallSite}^* \; \rightarrow \; \P(\cat{Stack}) \\ 
\text{subtree}(a_1 ... a_n) \quad &= \quad \{ a_1...a_n b_1...b_m \;|\; b_1 ... b_m \in \cat{CallSite}^* \}
\end{align*}
We have used elements of this domain in the previous section when we referred to groups of call stacks that share
a common prefix. This domain is useful for representing context-sensitive values in a compact way, but it does not
allow taking unions, intersections and complements. We will later see that they are necessary to model
 data flow in programs. We need to design a domain that supports these operations.
\\ \\
We start by observing that \cat{SubCtx} is defined as a family of sets, but this family is not closed under these
set-theoretic operations. This means that we can use a standard technique and define \cat{Ctx} as the smallest family of subsets
of \cat{Stack} that contains all elements of \cat{SubCtx} and is closed under finite unions, intersections and complements:
\begin{align*}
    \cat{Ctx} \quad &\subseteq \quad \P(\cat{Stack}) \\
    \cat{Ctx} \quad &\supseteq \quad \cat{SubCtx} \\
    \forall_{c_1, c_2 \in \cat{Ctx}} \quad & \\
    & c_1 \cup c_2 \; \in \cat{Ctx} \\
    & c_1 \cap c_2 \; \in \cat{Ctx} \\
    & \cat{Stack} \setminus c_1 \; \in \cat{Ctx}
\end{align*}
It turns out that the domain of calling contexts defined this way is exactly what we need.

\subsection*{Representation of contexts}
We now have a domain that meets our requirements of supporting basic set-theoretic operations. However, the way we have
defined it does not tell us much about how its elements look like. It's not clear whether we can feasibly implement
it as a data structure:
\begin{itemize}
 \item can elements of \cat{Ctx} be represented using a finite amount of memory?
 \item can we efficiently compute unions, intersections and complements?
\end{itemize}
The answer to both of these questions is yes. First, we need some notation to talk about finite subsets of a set:
$$
\Pfin(X) \; = \; \{A \subseteq X \;|\; A \text{ is finite}\}
$$
We start by introducing the following lemma:
\begin{align*}
    &\forall_{\textit{ctx} \:\in\: \cat{Ctx}} \quad
    \exists_{\substack{t_1, ..., t_n \:\in\: \cat{SubCtx} \qquad \\ H_1, ..., H_n \:\in\: \Pfin(\cat{SubCtx})}}
    \quad \textit{ctx} \; = \; \mcup_{i=1}^n \, \left[ t_i \; \setminus \; \mcup_{h \,\in H_i} \, h \right]
\end{align*}
This means that every context can be expressed using a finite number of subtrees.
Moreover, there always exists a unique \textit{normal form} that uses the smallest number of subtrees.
We leave this fact and the above lemma without proof because the proof is elementary, technical and does
not carry any valuable insight.
\\ \\
Consequently, we can represent each context using a finite number of parts, where each part consists of
a subtree and a finite number of \textit{holes}:
\begin{align*}
    \cat{Ctx} \quad &\simeq \quad \Pfin(\cat{SubCtx} \: \times \: \Pfin(\cat{SubCtx}))
\intertext{We recall that each subtree is uniquely identified by the common prefix of all call stacks in that subtree:}
    \cat{SubCtx} \quad &\simeq \quad \cat{CallSite}^*
\intertext{Therefore, we can represent calling contexts using a finite number of finite sequences of call sites:}
    \cat{Ctx} \quad &\simeq \quad \Pfin(\cat{CallSite}^* \: \times \: \Pfin(\cat{CallSite}^*))
\end{align*}
This representation, coupled with a normalization procedure, allows for practical computation
of unions, intersections and complements in polynomial time.

\section{Value lattice}

\subsection*{Raw values}

We will use elements of a simple value lattice resembling the one from Chapter 4 as building blocks from which we will
assemble context-sensitive values:
$$
\cat{RawVal} \;\;=\;\; \P_{\leq N}(\cat{Int}) \, \times \, \P_{\leq N}(\cat{Str})
        \, \times \, \P(\cat{Bool} \cup \{\syn{null}\} \cup \cat{AllocPath} \cup \cat{Fun})
$$
Similarly to Chapter 4, we define auxiliary functions to conveniently select relevant components of this lattice:
\begin{align*}
    \text{int} \; &\,: \; \cat{RawVal} \rightarrow \P_{\leq N}(\cat{Int}) \\
    \text{str} \; &\,: \; \cat{RawVal} \rightarrow \P_{\leq N}(\cat{Str}) \\
    \text{other} \; &\,: \; \cat{RawVal} \rightarrow \P(\cat{Bool} \cup \{\syn{null}\} \cup \cat{Box} \cup \cat{Fun} \cup \cat{Env}) \\
    \text{bool} \; &\,: \; \cat{RawVal} \rightarrow \P(\cat{Bool}) \\
    \text{null} \; &\,: \; \cat{RawVal} \rightarrow \P(\{\syn{null}\}) \\
    \text{alloc\_path} \; &\,: \; \cat{RawVal} \rightarrow \P(\cat{AllocPath}) \\
    \text{fun} \; &\,: \; \cat{RawVal} \rightarrow \P(\cat{Fun})
\end{align*}
We will define \cat{AllocPath} and \cat{Fun} later in this chapter. For now we treat them as finite sets
of opaque identifiers.
\\ \\
We define primitive operations on \cat{RawVal} similarly as in Chapter 4:
\begin{align*}
    \text{for } a, b \:\in\: \cat{RawVal}, \quad a + b &= \msqcup \, \{ (\{a_i + b_i\}, \bot, \bot) \; | \; a_i \, \in \, \text{int}(a), \; b_i \, \in \, \text{int}(b) \} \\
        & \quad \sqcup \, \msqcup \, \{ (\bot, \{a_i b_i\}, \bot) \; | \; a_i \, \in \, \text{str}(a), \; b_i \, \in \, \text{str}(b) \} \\[0.3\baselineskip]
        a - b &= \msqcup \, \{ (\{a_i - b_i\}, \bot, \bot) \; | \; a_i \, \in \, \text{int}(a), \; b_i \, \in \, \text{int}(b) \} \\[0.3\baselineskip]
        a * b &= \msqcup \, \{ (\{a_i \cdot b_i\}, \bot, \bot) \; | \; a_i \, \in \, \text{int}(a), \; b_i \, \in \, \text{int}(b) \} \\[0.3\baselineskip]
        a / b &= \msqcup \, \{ (\{\lfloor \frac{a_i}{b_i} \rfloor\}, \bot, \bot) \; | \; a_i \, \in \, \text{int}(a), \; b_i \, \in \, \text{int}(b) \}
\end{align*}

\subsection*{Context-sensitive values}

\newcommand{\stackval}{\cat{Stack} \; \rightarrow \; \cat{RawVal}}
\newcommand{\stackfinval}{\cat{Stack} \; \overset{\text{fin}} \longrightarrow \; \cat{RawVal}}

Guided by the question from the beginning of this chapter, \textit{how is the set of possible values of this memory location dependent
on the current call stack?}, we define the value lattice as the space of functions from call stacks to raw values:
$$
    \cat{Val} \quad = \quad \stackval
$$
However, as we already discussed in 5.1 Calling contexts, the best way of representing such functions is to abstract
over call stacks and encode generalized information that holds for whole subtrees of the space of call stacks. We can
achieve it by imposing two restrictions on every function $f \in \cat{Val}$:
\begin{itemize}
    \item The image of $f$ can contain only a finite number of distinct values from the base lattice \cat{RawVal}.
    \item For each value $v \in \cat{RawVal}$, the set of corresponding call stacks $f^{-1}(\{v\})$ must be expressible as a calling context \cat{Ctx}.
\end{itemize}
More formally, we define the space of \textit{finitely-representable functions} from call stacks to raw values as:
\begin{align*}
    & \stackfinval \quad \subseteq \quad \stackval \\[0.5\baselineskip]
    &f \quad \in \quad \stackfinval \\[0.5\baselineskip]
    & \qquad \qquad \text{if and only if} \\[0.5\baselineskip]
    & f(\cat{Stack}) \text{ is finite } \quad \land \quad \forall_{v \in \cat{RawVal}} \quad f^{-1}(\{v\}) \; \in \; \cat{Ctx}
\end{align*}
We redefine the value lattice accordingly:
$$
    \cat{Val} \quad = \quad \stackfinval
$$

\subsection*{Representation}

It is clear that elements of the value lattice can be represented as finite sets of pairs of contexts and raw values:
$$
    \cat{Val} \quad \simeq \quad \Pfin(\cat{Ctx} \, \times \, \cat{RawVal})
$$
This enables practical implementation of a corresponding data structure.

\subsection*{Lattice structure}

A lattice is a partial order with top and bottom elements and where every finite subset
of elements has a unique least upper bound (join) and greatest lower bound (meet).
We inherit this structure from the pointwise lattice of functions $\stackval$ and observe
that $\stackfinval$ is a sublattice. Therefore, ordering, join, meet, top and bottom are
the same as in the base lattice:
\begin{align*}
    \text{for } f_1, f_2 \:\in\: \cat{Val}, \quad f_1 \, \sqsubseteq \, f_2 \quad &\Longleftrightarrow \quad \forall s. \, f_1(s) \, \sqsubseteq f_2(s) \\
    f_1 \, \sqcup \, f_2 \; &= \; \lambda s. \, f_1(s) \, \sqcup f_2(s) \\
    f_1 \, \sqcap \, f_2 \; &= \; \lambda s. \, f_1(s) \, \sqcap f_2(s) \\
    \top_{\cat{Val}} \; &= \; \lambda s. \, \top_{\cat{RawVal}} \\
    \bot_{\cat{Val}} \; &= \; \lambda s. \, \bot_{\cat{RawVal}}
\end{align*}
It turns out that \cat{Val} is closed under join and meet defined this way, and therefore it is a sublattice.

\subsection*{Pointwise operations}
What is more, \cat{Val} is closed under any n-ary operation on \cat{RawVal} lifted to operate pointwise on \cat{Val}:
\begin{align*}
    op \; &\:: \; \cat{RawVal}^n \; \rightarrow \; \cat{RawVal} \\
    \text{lift}^{op} \; &\:: \; (\stackfinval)^n \; \rightarrow \; (\stackval) \\
    \text{lift}^{op}(f_1, ..., f_n) \; &= \; \lambda s. \, op(f_1(s), ..., f_n(s))
    \intertext{We will show that}
    \text{lift}^{op} \; &\:: \; (\stackfinval)^n \; \rightarrow \; (\stackfinval)
    \intertext{Let}
    f_1, ..., f_n \; &\in \; \stackfinval \\
    f \; &= \; \lambda s. \, op(f_1(s), ..., f_n(s))
    \intertext{We first note that the image of $f$, $f(\cat{Stack})$, is finite. From the definition of $\stackfinval$ we
    have that $f_1(\cat{Stack}), ..., f_n(\cat{Stack})$ are finite. Consequently, $op(f_1(\cat{Stack}), ..., f_n(\cat{Stack}))$
    is finite, because it is an image of $op$ over a finite set $f_1(\cat{Stack}) \, \times \, ... \, \times \, f_n(\cat{Stack})$.
    }
    \intertext{Now we need to prove that $\forall_{v \in \cat{RawVal}} \;\; f^{-1}(\{v\}) \; \in \; \cat{Ctx}$ and we will be done.
    Let $v \in \cat{RawVal}$.}
    f^{-1}(\{v\}) \; &= \; \{ s \in \cat{Stack} \; | \; f(s) = v \} \\
    &= \; \{ s \in \cat{Stack} \; | \; op(f_1(s), ..., f_n(s)) = v \} \\
    &= \; \{ s \in \cat{Stack} \; | \; (f_1(s), ..., f_n(s)) \: \in \: op^{-1}(\{v\}) \} \\
    &= \; \{ s \in \cat{Stack} \; | \; (v_1, ..., v_n) \: \in \: op^{-1}(\{v\}), \\
    & \qquad \qquad \qquad \quad \: v_1 = f_1(s), \: ..., \: v_n = f_n(s) \} \\
    &= \; \{ s \in \cat{Stack} \; | \; (v_1, ..., v_n) \: \in \: op^{-1}(\{v\}), \\
    & \qquad \qquad \qquad \quad \: s \:\in\: f_1^{-1}(\{v_1\}), \: ..., \: s \:\in\: f_n^{-1}(\{v_n\}) \} \\
    &= \; \bigcup_{(v_1, ..., v_n) \:\in\: op^{-1}(\{v\})} \;\: \bigcap_{i = 1}^{n} \;\: f_i^{-1}(\{v_i\}) \;\; = \;\; \text{(1)}
    \intertext{Because images of $f_1, ..., f_n$ are finite, in the above union, only a finite number of terms
     ($\bigcap_{i = 1}^{n} \;\: f_i^{-1}(\{v_i\})$) will be non-empty sets. Therefore, (1) is a finite union,
     where each term is a finite intersection of preimages of $f_1, ..., f_n$. Since $f_1, ..., f_n \; \in \; \stackfinval$,
     these preimages are elements of \cat{Ctx}. And since \cat{Ctx} is closed under finite unions and intersections,
     (1) is an element of \cat{Ctx}. Consequently, $f^{-1}(\{v\}) \:\in\: \cat{Ctx}$.}
    \intertext{This means that $f \; : \; \stackfinval$ and therefore}
    \text{lift}^{op} \; &\:: \; (\stackfinval)^n \; \rightarrow \; (\stackfinval)
\end{align*}
Later, when writing an equation system for the analysis, we will be using lifted versions of
primitive operations: $\text{lift}^{+}, \; \text{lift}^{-}, \; \text{lift}^{*}, \; \text{lift}^{/}$.

\subsection*{Updating context on function calls and returns}

We define two operations, \textit{enter\_ctx} and \textit{exit\_ctx}, that update
calling contexts of a \cat{Val} when calling and returning from functions.
When a context-sensitive value enters a function call $\alpha$,
its mapping from call stacks to raw values should now expect one more stack frame
at the top of the stack, corresponding to the call site $\alpha$:
\begin{align*}
    \text{enter\_ctx} \quad &\:: \quad \cat{CallSite} \; \rightarrow \; \cat{Val} \; \rightarrow \; \cat{Val} \\
    \text{enter\_ctx}(\alpha)(v) \quad &= \quad \lambda (s_1 ... s_n). \, \begin{cases}
       v(s_2 ... s_n) \;\text{ when }\; s_1 = \alpha \\
       \bot_{\cat{RawVal}} \;\text{ otherwise}
    \end{cases}
\intertext{
Correspondingly, when a context-sensitive value returns from a function call $\alpha$,
its mapping from call stacks to raw values should now expect one less stack frame at the top,
and select only a subtree of the original value that corresponds to the call site $\alpha$:
}
    \text{exit\_ctx} \quad &\:: \quad \cat{CallSite} \; \rightarrow \; \cat{Val} \; \rightarrow \; \cat{Val} \\
    \text{exit\_ctx}(\alpha)(v) \quad &= \quad \lambda (s_1 ... s_n). \, v(\alpha s_1 ... s_n) 
\end{align*}

\subsection*{Notation}

In next sections we will use the following notation to describe elements of the value lattice:
\begin{align*}
    \text{for } \; \alpha_i, \, \beta_i^j \, &\in \, \cat{CallSite}^* \; \text{ and } \; v_i \, \in \, \cat{RawVal}, \\
    \text{by } \; f &= \left[ \substack{
        \alpha_1 \, * \, \setminus \{ \beta_1^1, ..., \beta_1^{k_1} \} \: \mapsto \: v_1 \\
        \alpha_2 \, * \, \setminus \{ \beta_2^1, ..., \beta_2^{k_2} \} \: \mapsto \: v_2 \\
        \vdots \\
        \alpha_n \, * \, \setminus \{ \beta_n^1, ..., \beta_n^{k_n} \} \: \mapsto \: v_n
        } \right] \\[0.6\baselineskip]
    \text{we mean that }& f \in \cat{Val} \text{ and for any} \; s \in \cat{Stack}, \\
    f(s) &= \begin{cases}
        v_1 \; \text{ if } \; s \in \text{subtree}(\alpha_1), \; s \notin \text{subtree}(\alpha_1 \beta_1^1), \; ..., \; s \notin \text{subtree}(\alpha_1 \beta_1^{k_1}) \\
        v_2 \; \text{ if } \; s \in \text{subtree}(\alpha_2), \; s \notin \text{subtree}(\alpha_2 \beta_2^1), \; ..., \; s \notin \text{subtree}(\alpha_2 \beta_2^{k_2}) \\
        \; \vdots \\
        v_n \; \text{ if } \; s \in \text{subtree}(\alpha_n), \; s \notin \text{subtree}(\alpha_n \beta_n^1), \; ..., \; s \notin \text{subtree}(\alpha_n \beta_n^{k_n}) \\
        \bot_{\cat{RawVal}} \; \text{ otherwise}
    \end{cases} %
    \intertext{where}
    \text{subtree} \;\; &\:: \;\; \cat{CallSite}^* \; \rightarrow \; \P(\cat{Stack}) \\ 
    \text{subtree}(a_1 ... a_n) \;\; &= \;\; \{ a_1...a_n b_1...b_m \;|\; b_1 ... b_m \in \cat{CallSite}^* \}
\end{align*}
$$
\text{additionally, when $k_i = 0$, we will write ``} \alpha_i * \text{'' instead of ``} \alpha_i \, * \, \setminus \{ \beta_i^1, ..., \beta_i^{k_i} \} \text{''}
$$

\vspace*{\fill}

\section{Examples of context-sensitivity}

Now we have the value lattice, we know how to perform
operations on it, and we know what happens to values when they enter and
exit function calls. Let's see it in action by analyzing a few short
programs. Figures 5.1 and 5.2 show source code annotated with data flow
of abstract values.

\vspace*{\fill}

\tikzset{
    node distance=0mm,
    code/.append style={
      rectangle,
      minimum size=0mm,
      font=\ttfamily,
      execute at begin node=\obeyspaces,
      anchor=west,
      inner sep=0mm,
    },
    underline/.style args={#1 by #2 and #3}{
        insert path={
            [draw] ($(#1.south west) + (0mm, 0.7mm)$) -- ($(#1.south west) + (0mm, -#2)$)
            -- ($(#1.south east) + (0mm, -#2)$) -- ($(#1.south east) + (0mm, 0.7mm)$)
        },
        insert path={
            [draw] ($(#1.south) + (0mm, -#2)$) -- ++(0mm, -#3)
        },
        thick,
    },
    underline2/.style args={#1 and #2 by #3 and #4}{
        insert path={
            [draw] ($(#1.south west) + (0mm, 0.7mm)$) -- ($(#1.south west) + (0mm, -#3)$)
            -- ($(#2.south east) + (0mm, -#3)$) -- ($(#2.south east) + (0mm, 0.7mm)$)
        },
        insert path={
            [draw] ($(#1.south)!0.5!(#2.south) + (0mm, -#3)$) -- ++(0mm, -#4)
        },
        thick,
    },
    overline/.style args={#1 by #2 and #3}{
        insert path={
            [draw] ($(#1.north west) + (0mm, -0.7mm)$) -- ($(#1.north west) + (0mm, #2)$)
            -- ($(#1.north east) + (0mm, #2)$) -- ($(#1.north east) + (0mm, -0.7mm)$)
        },
        insert path={
            [draw] ($(#1.north) + (0mm, #2)$) -- ++(0mm, #3)
        },
        thick,
    },
    overline2/.style args={#1 and #2 by #3 and #4}{
        insert path={
            [draw] ($(#1.north west) + (0mm, -0.7mm)$) -- ($(#1.north west) + (0mm, #3)$)
            -- ($(#2.north east) + (0mm, #3)$) -- ($(#2.north east) + (0mm, -0.7mm)$)
        },
        insert path={
            [draw] ($(#1.north)!0.5!(#2.north) + (0mm, #3)$) -- ++(0mm, #4)
        },
        thick,
    },
    colorbox/.style args={#1}{
        inner sep=2mm, thick,
        draw=#1!40,
        fill=#1!20,
        text=black,
        scale=0.8,
    },
    colorbox2/.style args={#1}{
        scale=1.1, inner sep=3.5mm,
        thick,
        draw=#1!40,
        fill=#1!20,
        text=black,
    },
    operation/.style={
        inner sep=2mm, thick,
        rounded corners=2.5mm,
        draw=gray!40,
        fill=gray!20,
        text=black,
        scale=0.8,
    },
    colorop/.style args={#1}{
        inner sep=2mm, thick,
        rounded corners=2.5mm,
        draw=#1!40,
        fill=#1!20,
        text=black,
        scale=0.8,
    },
    text height=1.5ex,
    text depth=0ex,
    x={(0.6cm,0cm)},y={(0cm, -0.6cm)},
}

\begin{figure}[H]
\begin{tikzpicture}

\node(i2) [code] at (-1,1) {{}  return  {}}; \node(x2) [code, right=of i2] {x}; \node(plus) [code, right=of x2] {{} +  {}}; \node(one) [code, right=of plus] {1};
\path[overline=x2 by 0mm and 10mm, color=teal!40] -- ++(6, 0);
\node(i1) [code] at (-1,0) {{}function increment(}; \node(x1) [code, right=of i1] {x}; \node(i2) [code, right=of x1] {) \{};
\path[overline=x1 by 0mm and 1mm, color=teal!40] -- ++(3.5, 0);
\path[overline=one by 0.5mm and 1mm, color=orange!40] -- ++(10.7, 0) node(one1) [pos=1.0, colorbox=orange] {$[* \, \mapsto \, \{1\}]$};
\path[underline2=x2 and one by 1mm and 4mm, color=violet!40] -- ++(15.9, 0) node(result) [pos=1.0, colorbox=violet, scale=1.4, inner sep=3.5mm]
    {$\left[\substack{ \alpha * \, \mapsto \, \{4\} \\ \beta * \, \mapsto \, \{6\} }\right]$};
\node[code] at (-1,2) {{}\}};

\node[code] at (-1,7) {{}function main() \{};
\node(m10) [code] at (-1,8) {{}   }; \node(m11) [code, right=of m10] {increment$_\alpha$(}; \node(m12) [code, right=of m11] {3}; \node(m13) [code, right=of m12] {)};
\path[overline=m12 by 0.5mm and 1.7mm, color=orange!40] -- ++(2.4, 0) node(n3) [pos=1.0, anchor=west, colorbox=orange] {$[* \, \mapsto \, \{3\}]$};
\path[underline2=m11 and m13 by 1mm and 3mm, color=blue!40] -- ++(12.7, 0) node(n4) [pos=1.0, anchor=west, colorbox=blue] {$[* \, \mapsto \, \{4\}]$};
\node(m20) [code] at (-1,9) {{}   }; \node(m21) [code, right=of m20] {increment$_\beta$(}; \node(m22) [code, right=of m21] {5}; \node(m23) [code, right=of m22] {)};
\path[overline=m22 by 0.5mm and 1mm, color=orange!40] -- ++(8.4, 0) -- ++(0, -0.66) node(n5) [pos=1.0, anchor=south, colorbox=orange] {$[* \, \mapsto \, \{5\}]$};
\path[underline2=m21 and m23 by 1mm and 2mm, color=blue!40] -- ++(18, 0) -- ++(0, -0.36) node(n6) [pos=1.0, anchor=south, colorbox=blue] {$[* \, \mapsto \, \{6\}]$};;
\node[code] at (-1,10) {{}\}};

\node(n3alpha) [colorbox=teal, above=-2 of n3] {$[\alpha * \, \mapsto \, \{3\}]$};
\node(n5beta) [colorbox=teal, above=-2 of n5] {$[\beta * \, \mapsto \, \{5\}]$};

\path[-stealth, thick, draw=teal!40] (n3.north) -> (n3alpha.south) node[pos=0.5, colorop=teal] {enter\_ctx$(\alpha)$}; 
\path[-stealth, thick, draw=teal!40] (n5.north) -> (n5beta.south) node[pos=0.5, colorop=teal] {enter\_ctx$(\beta)$};

\node(xunion) [colorop=teal] at ($(n3alpha.north)!0.5!(n5beta.north) + (0, -1)$) {$\sqcup$};

\path[-stealth, thick, draw=teal!40] (n3alpha.north) -> ++(0, -1) -> (xunion.west);
\path[-stealth, thick, draw=teal!40] (n5beta.north) -> ++(0, -1) -> (xunion.east);

\node(x) [colorbox=teal, above=-2.5 of xunion, scale=1.4, inner sep=3.5mm] {$\left[\substack{ \alpha * \, \mapsto \, \{3\} \\ \beta * \, \mapsto \, \{5\} }\right]$};

\path[-stealth, thick, draw=teal!40] (xunion.north) -> (x.south);

\node(lift) [colorop=violet] at (result |- x) {lift$^+$};

\path[-stealth, thick, draw=teal!40] ($(x.east) + (0mm, 0.7mm)$) -> ($(lift.west) + (0.3mm, 0.7mm)$);
\path[-stealth, thick, draw=violet!40] (lift.south) -> (result.north);
\path[stealth-, thick, draw=orange!40] ($(lift.west) + (0.3mm, -0.7mm)$) -- ++(-3.7, 0) node (lifthelper) [pos=1.0] {} -- (lifthelper |- one1.north);

\node(resulthelper) [below=-1.5 of result] {};
\path[thick, draw=violet!40] (result.south) -- (resulthelper);
\node(exitalpha) [colorop=violet, above=-1.5 of n4] {exit\_ctx$(\alpha)$};
\node(exitbeta) [colorop=violet, above=-1.5 of n6] {exit\_ctx$(\beta)$};
\path[-stealth, thick, draw=violet!40] (exitalpha.south) -> (n4.north);
\path[-stealth, thick, draw=violet!40] (exitbeta.south) -> (n6.north);
\path[thick, draw=violet!40] (resulthelper.north) -- (exitalpha |- resulthelper.north) -- (exitalpha.north);
\path[thick, draw=violet!40] (resulthelper.north) -- (exitbeta |- resulthelper.north) -- (exitbeta.north);
\end{tikzpicture}

\caption{
A program that demonstrates separating data from different calls to one function. It is also
an example of using pointwise operations on our lattice to combine context-specific and context-independent
data. Rectangles contain abstract values of expressions, rounded rectangles denote operations and arrows indicate
direction of computation.
}
\end{figure}
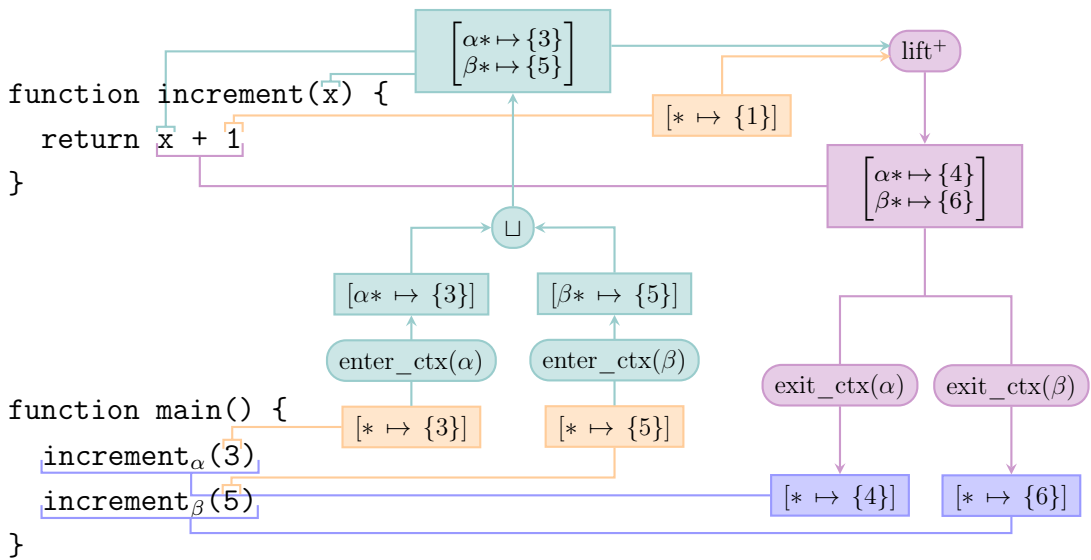

\begin{figure}[H]
\input{figures/imperative_init}
\end{figure}

\section{Revisiting allocation-site abstraction}

We remember from Chapter 4 that one way to handle heap-allocated data \cite{heapabstractions}
in a static analysis is to identify each object with the place in the
program where it has been created. This means that we keep one abstract
object for each allocation site, and it represents a group of objects that
have been allocated at this site. This crude approximation has
one very serious drawback: it does not allow strong updates. When we process
a write operation to an abstract memory location, we can perform a
strong update or a weak update. A strong update overwrites previous set
of values with a new one. On the other hand, a weak update is more conservative
and only adds the new values to the set of possibilities, leaving old values intact.
Figure 5.3 shows what can go wrong when using strong updates with
allocation-site abstraction. Weak updates must be used to retain soundness,
which leads to a significant loss of precision (Figure 5.4).

\vspace*{\fill}

\begin{figure}[H]
\begin{tikzpicture}

\node(c1) [code] at (0, 0) {function create(val) \{};
\node(c2) [code, below=-1 of d1.west, anchor=west] {{}  var res =  {}};
\node(c2_alloc) [code, right=of c2] {\{\}};
\node(c3) [code, below=-1 of c2.west, anchor=west] {{}   {}};
\node(c3_res) [code, right=of c3] {res};
\node(c3_dot) [code, right=of c3_res] {.};
\node(c3_x) [code, right=of c3_dot] {x};
\node(c3_eq) [code, right=of c3_x] {{} =  {}};
\node(c3_val) [code, right=of c3_eq] {val};
\node(c4) [code, below=-1 of c3.west, anchor=west] {{}  return  {}};
\node(c4_res) [code, right=of c4] {res};
\node(c5) [code, below=-1 of c4.west, anchor=west] {\}};

\node(m1) [code] at (0, 7) {function main() \{};
\node(m2) [code, below=-1 of m1.west, anchor=west] {{}  var a =  {}};
\node(m2_create) [code, right=of m2] {create$_\alpha$(1)};
\node(m3) [code, below=-1 of m2.west, anchor=west] {{}  var b =  {}};
\node(m3_create) [code, right=of m3] {create$_\beta$(2)};
\node(m4) [code, below=-1 of m3.west, anchor=west] {{}   {}};
\node(m4_ax) [code, right=of m4] {a.x};
\node(m5) [code, below=-1 of m4.west, anchor=west] {\}};

\path[underline=m2_create by 1mm and 1mm, draw=orange!40] -- ++(4.2, 0) node[colorbox=orange, inner sep=1.5mm] {$[* \, \mapsto \, \{ \addr{1} \}]$};
\path[underline=m3_create by 1mm and 1mm, draw=orange!40] -- ++(4.2, 0) node[colorbox=orange, inner sep=1.5mm] {$[* \, \mapsto \, \{ \addr{1} \}]$};

\path[underline=m4_ax by 1mm and 1mm, draw=blue!40] -- ++(3, 0) node[colorbox=blue, inner sep=1.5mm] {$[* \, \mapsto \, \{ 2 \}]$};

\draw[stealth-, teal!40, thick] ($(m2.north west) + (0.8, -0.15)$) -- ++(-1, 0) -- ++(0, -0.3) -- ++(-4.5, 0) -- ++(0, -0.3)
    node(menter) [pos=1.0, anchor=south, colorbox=teal] {$
        \addr{1}.x \: \mapsto \: [* \, \mapsto \, \bot]
    $};
\draw[stealth-, violet!40, thick] ($(m3.north west) + (0.8, -0.25)$) -- ++(-5.5, 0)
    node(mmiddle) [pos=1.0, anchor=center, colorbox=violet] {$
        \addr{1}.x \: \mapsto \: [* \, \mapsto \, \{1\}]
    $};
\draw[stealth-, blue!40, thick] ($(m3.south west) + (0.8, 0.15)$) -- ++(-1, 0) -- ++(0, 0.3) -- ++(-4.5, 0) -- ++(0, 0.3)
    node(mexit) [pos=1.0, anchor=north, colorbox=blue] {$
        \addr{1}.x \: \mapsto \: [* \, \mapsto \, \{2\}]
    $};

\draw[stealth-, teal!40, thick] ($(c3.north west) + (0.8, -0.15)$) -- ++(-5.5, 0) -- ++(0, -0.3)
    node(center) [pos=1.0, anchor=south, colorbox2=teal] {$
        \addr{1}.x \: \mapsto \: \left[\substack{\alpha * \, \mapsto \, \bot \\ \beta * \, \mapsto \, \{1\}}\right]
    $};
\draw[stealth-, blue!40, thick] ($(c3.south west) + (0.8, 0.15)$) -- ++(-5.5, 0) -- ++(0, 0.3)
    node(cexit) [pos=1.0, anchor=north, colorbox2=blue] {$
        \addr{1}.x \: \mapsto \: \left[\substack{\alpha * \, \mapsto \, \{1\} \\ \beta * \, \mapsto \, \{2\}}\right]
    $};

\draw[stealth-, teal!40, thick] ($(center.west)$) -- ++(-1.5, 0)
    node(center_union) [pos=1.0, anchor=center, colorop=teal] {$\sqcup$};
\draw[-stealth, violet!40, thick] (mmiddle.west) -- (center_union.south east |- mmiddle.west)
    -- (center_union.south east) node [pos=0.15, colorop=violet] {enter\_ctx$(\beta)$};
\draw[-stealth, teal!40, thick] (menter.west) -- (center_union.south west |- menter.west)
    -- (center_union.south west) node [pos=0.15, colorop=teal] {enter\_ctx$(\alpha)$};

\draw[-stealth, violet!40, thick] ($(cexit.south east) + (-0.4, 0)$) -- ++(0, 0) node(iexit_helper) [pos=1.0, anchor=center] {}
    -- (iexit_helper |- mmiddle.north east) node[pos=0.15, colorop=violet, inner sep=1.5mm] {exit\_ctx$(\alpha)$}
    -- ++(0, 0.27) node(iexit_helper2) [pos=1.0, anchor=center] {} -- (mmiddle.east |- iexit_helper2);
\draw[-stealth, blue!40, thick] ($(cexit.east) + (0.0, 0.4)$) -- ++(0.4, 0) node(iexit_helper3) [pos=1.0, anchor=center] {}
    -- (iexit_helper3 |- mexit.east) node[pos=0.33, colorop=blue, inner sep=1.5mm] {exit\_ctx$(\beta)$} -- (mexit.east);

\draw[-stealth, violet!40, thick] (center.east) -- ++(11, 0) node(setattr_helper) [pos=1.0, anchor=center] {}
    -- (setattr_helper |- cexit.east) node(setattr) [pos=0.5, colorop=violet] {setattr$(v_{\text{res}}, \text{``x''}, v_{\text{val}})$} -- (cexit.east);

\path[underline=c3_res by 1mm and 2mm, color=orange!40, thick] -- ++(3.15, 0) -- ++(0, 1.5)
    node(res) [pos=1.0, anchor=north, colorbox=orange] {$v_{\text{res}} = [* \, \mapsto \, \{ \addr{1} \}]$};
\path[underline=c3_val by 1mm and 1mm, color=teal!40, thick] -- ++(1.5, 0) -- ++(0, 0.5)
    -- ++(3.5, 0) -- ++(0, 2)
    node(val) [pos=1.0, anchor=north, colorbox2=teal] {$\left[\substack{
        \alpha * \, \mapsto \, \{ 1 \} \\
        \beta * \, \mapsto \, \{ 2 \}
    }\right]$};

\draw[-stealth, teal!40, thick] (val.east) -| ($(setattr.south) + (2, 0)$);
\draw[-stealth, orange!40, thick] (res.east) -| ($(setattr.south) + (1, 0)$);

\end{tikzpicture}

\caption{
An example of why strong updates cannot be used with allocation-site
abstraction. A single abstract object with address $\addr{1}$ represents both
\texttt{a} and \texttt{b}. When \texttt{b} is initialized, the field \texttt{x}
of both \texttt{a} and \texttt{b} is overwritten, resulting in an incorrect
value of expression \texttt{a.x} on the next line.
}
\end{figure}
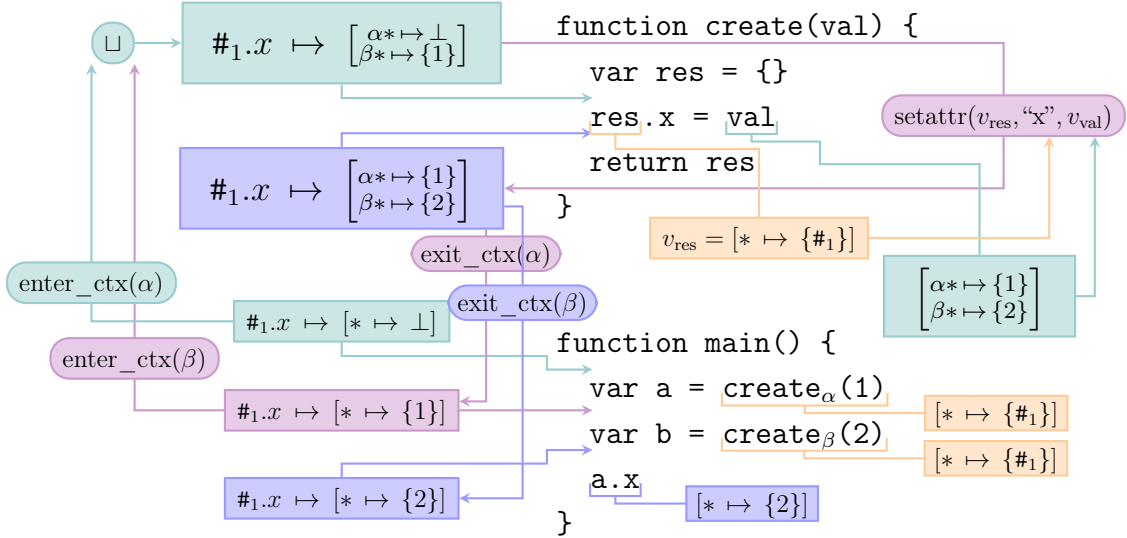

\begin{figure}[H]
\begin{tikzpicture}

\node(c1) [code] at (0, 0) {function create(val) \{};
\node(c2) [code, below=-1 of d1.west, anchor=west] {{}  var res =  {}};
\node(c2_alloc) [code, right=of c2] {\{\}};
\node(c3) [code, below=-1 of c2.west, anchor=west] {{}   {}};
\node(c3_res) [code, right=of c3] {res};
\node(c3_dot) [code, right=of c3_res] {.};
\node(c3_x) [code, right=of c3_dot] {x};
\node(c3_eq) [code, right=of c3_x] {{} =  {}};
\node(c3_val) [code, right=of c3_eq] {val};
\node(c4) [code, below=-1 of c3.west, anchor=west] {{}  return  {}};
\node(c4_res) [code, right=of c4] {res};
\node(c5) [code, below=-1 of c4.west, anchor=west] {\}};

\node(m1) [code] at (0, 7) {function main() \{};
\node(m2) [code, below=-1 of m1.west, anchor=west] {{}  var a =  {}};
\node(m2_create) [code, right=of m2] {create$_\alpha$(1)};
\node(m3) [code, below=-1 of m2.west, anchor=west] {{}  var b =  {}};
\node(m3_create) [code, right=of m3] {create$_\beta$(2)};
\node(m4) [code, below=-1 of m3.west, anchor=west] {{}   {}};
\node(m4_ax) [code, right=of m4] {a.x};
\node(m5) [code, below=-1 of m4.west, anchor=west] {\}};

\path[underline=m2_create by 1mm and 1mm, draw=orange!40] -- ++(4.2, 0) node[colorbox=orange, inner sep=1.5mm] {$[* \, \mapsto \, \{ \addr{1} \}]$};
\path[underline=m3_create by 1mm and 1mm, draw=orange!40] -- ++(4.2, 0) node[colorbox=orange, inner sep=1.5mm] {$[* \, \mapsto \, \{ \addr{1} \}]$};

\path[underline=m4_ax by 1mm and 1mm, draw=blue!40] -- ++(3, 0) node[colorbox=blue, inner sep=1.5mm] {$[* \, \mapsto \, \{1, 2\}]$};

\draw[stealth-, teal!40, thick] ($(m2.north west) + (0.8, -0.15)$) -- ++(-1, 0) -- ++(0, -0.3) -- ++(-4.5, 0) -- ++(0, -0.3)
    node(menter) [pos=1.0, anchor=south, colorbox=teal] {$
        \addr{1}.x \: \mapsto \: [* \, \mapsto \, \bot]
    $};
\draw[stealth-, violet!40, thick] ($(m3.north west) + (0.8, -0.25)$) -- ++(-5.5, 0)
    node(mmiddle) [pos=1.0, anchor=center, colorbox=violet] {$
        \addr{1}.x \: \mapsto \: [* \, \mapsto \, \{1\}]
    $};
\draw[stealth-, blue!40, thick] ($(m3.south west) + (0.8, 0.15)$) -- ++(-1, 0) -- ++(0, 0.3) -- ++(-4.5, 0) -- ++(0, 0.3)
    node(mexit) [pos=1.0, anchor=north, colorbox=blue] {$
        \addr{1}.x \: \mapsto \: [* \, \mapsto \, \{1, 2\}]
    $};

\draw[stealth-, teal!40, thick] ($(c3.north west) + (0.8, -0.15)$) -- ++(-5.5, 0) -- ++(0, -0.3)
    node(center) [pos=1.0, anchor=south, colorbox2=teal] {$
        \addr{1}.x \: \mapsto \: \left[\substack{\alpha * \, \mapsto \, \bot \\ \beta * \, \mapsto \, \{1\}}\right]
    $};
\draw[stealth-, blue!40, thick] ($(c3.south west) + (0.8, 0.15)$) -- ++(-5.5, 0) -- ++(0, 0.3)
    node(cexit) [pos=1.0, anchor=north, colorbox2=blue] {$
        \addr{1}.x \: \mapsto \: \left[\substack{\alpha * \, \mapsto \, \{1\} \\ \beta * \, \mapsto \, \{1, 2\}}\right]
    $};

\draw[stealth-, teal!40, thick] ($(center.west)$) -- ++(-1.5, 0)
    node(center_union) [pos=1.0, anchor=center, colorop=teal] {$\sqcup$};
\draw[-stealth, violet!40, thick] (mmiddle.west) -- (center_union.south east |- mmiddle.west)
    -- (center_union.south east) node [pos=0.15, colorop=violet] {enter\_ctx$(\beta)$};
\draw[-stealth, teal!40, thick] (menter.west) -- (center_union.south west |- menter.west)
    -- (center_union.south west) node [pos=0.15, colorop=teal] {enter\_ctx$(\alpha)$};

\draw[-stealth, violet!40, thick] ($(cexit.south east) + (-0.4, 0)$) -- ++(0, 0) node(iexit_helper) [pos=1.0, anchor=center] {}
    -- (iexit_helper |- mmiddle.north east) node[pos=0.15, colorop=violet, inner sep=1.5mm] {exit\_ctx$(\alpha)$}
    -- ++(0, 0.27) node(iexit_helper2) [pos=1.0, anchor=center] {} -- (mmiddle.east |- iexit_helper2);
\draw[-stealth, blue!40, thick] ($(cexit.east) + (0.0, 0.4)$) -- ++(0.4, 0) node(iexit_helper3) [pos=1.0, anchor=center] {}
    -- (iexit_helper3 |- mexit.east) node[pos=0.33, colorop=blue, inner sep=1.5mm] {exit\_ctx$(\beta)$} -- (mexit.east);

\draw[-stealth, violet!40, thick] (center.east) -- ++(11, 0) node(setattr_helper) [pos=1.0, anchor=center] {}
    -- (setattr_helper |- cexit.east) node(setattr) [pos=0.5, colorop=violet] {setattr$(v_{\text{res}}, \text{``x''}, v_{\text{val}})$} -- (cexit.east);

\path[underline=c3_res by 1mm and 2mm, color=orange!40, thick] -- ++(3.15, 0) -- ++(0, 1.5)
    node(res) [pos=1.0, anchor=north, colorbox=orange] {$v_{\text{res}} = [* \, \mapsto \, \{ \addr{1} \}]$};
\path[underline=c3_val by 1mm and 1mm, color=teal!40, thick] -- ++(1.5, 0) -- ++(0, 0.5)
    -- ++(3.5, 0) -- ++(0, 2)
    node(val) [pos=1.0, anchor=north, colorbox2=teal] {$\left[\substack{
        \alpha * \, \mapsto \, \{ 1 \} \\
        \beta * \, \mapsto \, \{ 2 \}
    }\right]$};

\draw[-stealth, teal!40, thick] (val.east) -| ($(setattr.south) + (2, 0)$);
\draw[-stealth, orange!40, thick] (res.east) -| ($(setattr.south) + (1, 0)$);

\end{tikzpicture}

\caption{
Weak updates allow us to retain soundness when using allocation-site abstraction.
Assigning $\{1, 2\}$ to \texttt{a.x} is a sound but imprecise solution.
}
\end{figure}
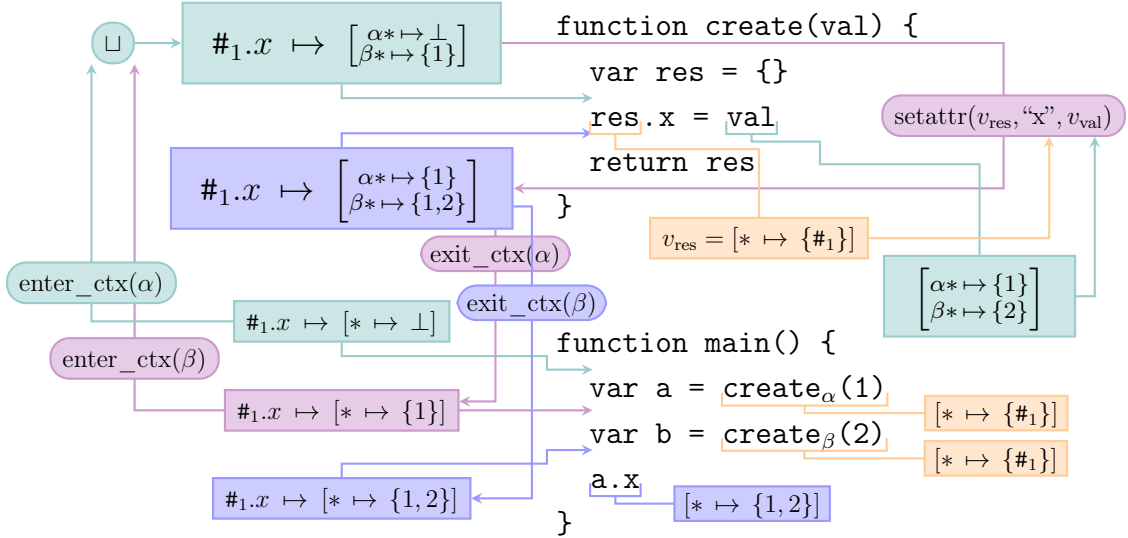

\vspace*{\fill}

\subsection*{Recency abstraction}
A technique called \textit{recency abstraction} has been introduced in \cite{recency} to enable strong updates
in some cases. The main idea is to use two abstract addresses for each
allocation site: \textit{r} for the most recently allocated object, and \textit{o} for all older objects.
While the term \textit{the most recently allocated object} can refer to different objects depending on
the context, we can perform strong updates on it. This is because at any given time during the execution of a program,
there is only one most recently allocated object. On the other hand, for the abstract object that summarizes all older objects,
\textit{o}, we still have to use weak updates. When analysis encounters an allocation site, it merges the previous
most recently allocated object \textit{r} into the summary of older objects \textit{o} using the join operation
of appropriate lattice. After this, it points \textit{r} into a newly-created, empty object. While this technique
definitely is an improvement over allocation-site abstraction, it has some limitations. The most important one is that
analysis precision is not robust to refactoring changes in the code. Let's assume there is some piece of code that creates
and initializes some data structure, and it is repeated in a few places. Extracting it into a common routine will cause the
analyzer to mix up most instances of this data structure into one summary representation, with one exception of the most
recently allocated instance. Changing the code to be cleaner results in a degraded precision of analysis.

\subsection*{Heap specialization}
Looking at the example in Figure 5.4, a natural idea comes to mind: if we inlined the \texttt{create}
function into \texttt{main}, we would have two allocation sites, corresponding to the call sites $\alpha$ and $\beta$. Since
these allocation sites are now in the main entry point, each of them can only refer to a single object. This means that we
can perform strong updates on them. Going further, we can imagine that transitively inlining all functions into \texttt{main}
would dramatically improve precision of our abstract heap. Obviously this would cause a lot of duplicate work
and would blow up memory usage. However, it turns out that we can do something similar but only inline allocation sites (Figure 5.5).
This technique is known as \textit{heap specialization} or \textit{heap cloning} \cite{cloning1,cloning2,merging}, and so far it has only been used in
flow-insensitive analyzers. Later in this chapter we will incorporate it into our flow-sensitive analysis. We will also show
that heap specialization is a generalization of recency abstraction, and allows performing strong updates more often.

\begin{figure}[H]
\begin{tikzpicture}

\node(c1) [code] at (0, 0) {function create(val) \{};
\node(c2) [code, below=-1 of d1.west, anchor=west] {{}  var res =  {}};
\node(c2_alloc) [code, right=of c2] {\{\}};
\node(c3) [code, below=-1 of c2.west, anchor=west] {{}   {}};
\node(c3_res) [code, right=of c3] {res};
\node(c3_dot) [code, right=of c3_res] {.};
\node(c3_x) [code, right=of c3_dot] {x};
\node(c3_eq) [code, right=of c3_x] {{} =  {}};
\node(c3_val) [code, right=of c3_eq] {val};
\node(c4) [code, below=-1 of c3.west, anchor=west] {{}  return  {}};
\node(c4_res) [code, right=of c4] {res};
\node(c5) [code, below=-1 of c4.west, anchor=west] {\}};

\node(m1) [code] at (0, 10) {function main() \{};
\node(m2) [code, below=-1 of m1.west, anchor=west] {{}  var a =  {}};
\node(m2_create) [code, right=of m2] {create$_\alpha$(1)};
\node(m3) [code, below=-1 of m2.west, anchor=west] {{}  var b =  {}};
\node(m3_create) [code, right=of m3] {create$_\beta$(2)};
\node(m4) [code, below=-1 of m3.west, anchor=west] {{}   {}};
\node(m4_ax) [code, right=of m4] {a.x};
\node(m5) [code, below=-1 of m4.west, anchor=west] {\}};

\path[underline=m2_create by 1mm and 1mm, draw=orange!40] -- ++(4.2, 0)
    node(create_alpha) [colorbox=orange, inner sep=1.5mm] {$[* \, \mapsto \, \{ \addr{1/\alpha} \}]$};
\path[underline=m3_create by 1mm and 1mm, draw=orange!40] -- ++(4.2, 0)
    node(create_beta) [colorbox=orange, inner sep=1.5mm] {$[* \, \mapsto \, \{ \addr{1/\beta} \}]$};

\path[underline=m4_ax by 1mm and 1mm, draw=blue!40] -- ++(3, 0) node[colorbox=blue, inner sep=1.5mm] {$[* \, \mapsto \, \{ 1 \}]$};

\draw[stealth-, teal!40, thick] ($(m2.north west) + (0.8, -0.15)$) -- ++(-1, 0) -- ++(0, -0.3) -- ++(-5, 0) -- ++(0, -0.3)
    node(menter) [pos=1.0, anchor=south, colorbox2=teal] {$
        \substack{
            \addr{1/\alpha}.x \: \mapsto \: [* \, \mapsto \, \bot] \\
            \addr{1/\beta}.x \: \mapsto \: [* \, \mapsto \, \bot]
        }
    $};
\draw[stealth-, violet!40, thick] ($(m3.north west) + (0.8, -0.25)$) -- ++(-6, 0)
    node(mmiddle) [pos=1.0, anchor=center, colorbox2=violet] {$
        \substack{
            \addr{1/\alpha}.x \: \mapsto \: [* \, \mapsto \, \{1\}] \\
            \addr{1/\beta}.x \: \mapsto \: [* \, \mapsto \, \bot] \;\;
        }
    $};
\draw[stealth-, blue!40, thick] ($(m3.south west) + (0.8, 0.15)$) -- ++(-1, 0) -- ++(0, 0.3) -- ++(-5, 0) -- ++(0, 0.3)
    node(mexit) [pos=1.0, anchor=north, colorbox2=blue] {$
        \substack{
            \addr{1/\alpha}.x \: \mapsto \: [* \, \mapsto \, \{1\}] \\
            \addr{1/\beta}.x \: \mapsto \: [* \, \mapsto \, \{2\}]
        }
    $};

\draw[stealth-, teal!40, thick] ($(c3.north west) + (0.8, -0.15)$) -- ++(-5.5, 0) -- ++(0, -0.3)
    node(center) [pos=1.0, anchor=south, colorbox=teal] {$
        \addr{1}.x \: \mapsto \: \left[* \, \mapsto \, \bot\right]
    $};
\draw[stealth-, blue!40, thick] ($(c3.south west) + (0.8, 0.15)$) -- ++(-5.5, 0) -- ++(0, 0.3)
    node(cexit) [pos=1.0, anchor=north, colorbox2=blue] {$
        \addr{1}.x \: \mapsto \: \left[\substack{\alpha * \, \mapsto \, \{1\} \\ \beta * \, \mapsto \, \{2\}}\right]
    $};

\draw[stealth-, teal!40, thick] ($(center.west)$) -- ++(-3, 0)
    node(center_union) [pos=1.0, anchor=center, colorop=teal] {$\sqcup$};
\draw[-stealth, violet!40, thick] (mmiddle.west) -- (center_union.south east |- mmiddle.west)
    -- (center_union.south east) node [pos=0.55, colorop=violet] {enter\_call$(\beta)$};
\draw[-stealth, teal!40, thick] (menter.west) -- (center_union.south west |- menter.west)
    -- (center_union.south west) node [pos=0.6, colorop=teal] {enter\_call$(\alpha)$};

\draw[-stealth, violet!40, thick] ($(cexit.south east) + (-0.4, 0)$) -- ++(0, 0) node(iexit_helper) [pos=1.0, anchor=center] {}
    -- (iexit_helper |- mmiddle.north east) node[pos=0.15, colorop=violet, inner sep=1.5mm] {exit\_call$(\alpha)$}
    -- ++(0, 0.27) node(iexit_helper2) [pos=1.0, anchor=center] {} -- (mmiddle.east |- iexit_helper2);
\draw[-stealth, blue!40, thick] ($(cexit.east) + (0.0, 0.4)$) -- ++(0.4, 0) node(iexit_helper3) [pos=1.0, anchor=center] {}
    -- (iexit_helper3 |- mexit.east) node[pos=0.28, colorop=blue, inner sep=1.5mm] {exit\_call$(\beta)$} -- (mexit.east);

\draw[-stealth, violet!40, thick] (center.east) -- ++(13, 0) node(setattr_helper) [pos=1.0, anchor=center] {}
    -- (setattr_helper |- cexit.east) node(setattr) [pos=0.5, colorop=violet] {setattr$(v_{\text{res}}, \text{``x''}, v_{\text{val}})$} -- (cexit.east);

\path[underline=c3_res by 1mm and 2mm, color=orange!40, thick] -- ++(3.15, 0) -- ++(0, 1.5)
    node(res) [pos=1.0, anchor=north, colorbox=orange] {$v_{\text{res}} = [* \, \mapsto \, \{ \addr{1} \}]$};

\draw[orange!40, thick, stealth-] (create_alpha) -- ++(0, -1.5)
    node[pos=1.0, anchor=south, colorop=orange] {exit\_call$(\alpha)$}
    -- ++(0, -2) -| (res.south);

\draw[orange!40, thick, stealth-] (create_beta) -- ++(4, 0) -- ++(0, -2.5)
    node[pos=1.0, anchor=south, colorop=orange] {exit\_call$(\beta)$}
    -- ++(0, -2) -| (res.south);

\path[underline=c3_val by 1mm and 1mm, color=teal!40, thick] -- ++(1.5, 0) -- ++(0, 0.5)
    -- ++(3.5, 0) -- ++(0, 2)
    node(val) [pos=1.0, anchor=north, colorbox2=teal] {$\left[\substack{
        \alpha * \, \mapsto \, \{ 1 \} \\
        \beta * \, \mapsto \, \{ 2 \}
    }\right]$};

\draw[-stealth, teal!40, thick] (val.east) -| ($(setattr.south) + (2, 0)$);
\draw[-stealth, orange!40, thick] (res.east) -| ($(setattr.south) + (1, 0)$);

\end{tikzpicture}

\caption{
An example of heap specialization. Inside the scope of \syn{create} there is
a single abstract object with address $\addr{1}$. However, upon exiting to
\syn{main}, it is specialized as $\addr{1/\alpha}$ or $\addr{1/\beta}$,
depending on the call site. This enables us to distinguish two objects that
allocation-site abstraction would normally mix up, and additionally we can
perform strong updates on them. Auxiliary functions enter\_call and exit\_call
are defined later in a section about abstract states. They are similar to
enter\_ctx and exit\_ctx, but additionally handle heap specialization.
}
\end{figure}
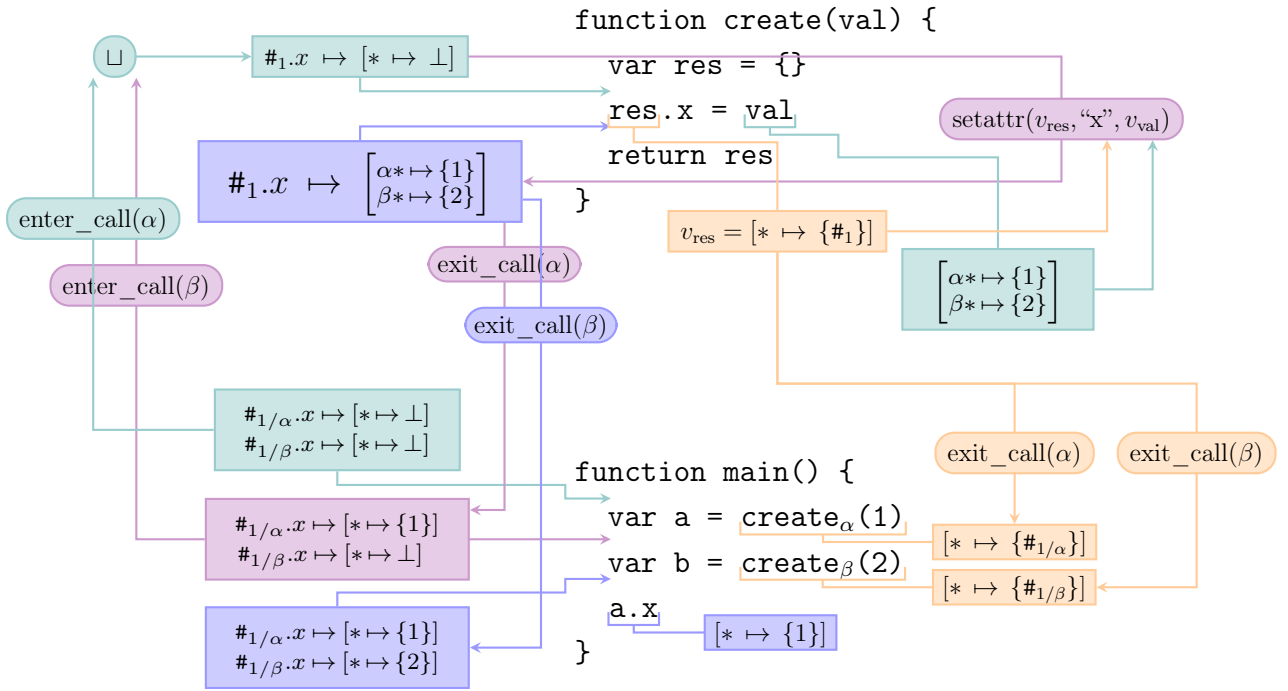

\subsection*{Allocation paths}

We define two domains, \textit{allocation sites} and \textit{allocation paths}, to represent addresses on the abstract heap:
\begin{align*}
    \cat{AllocSite} \quad &= \quad \{e \in \cat{Expr} \;|\; e \text{ is an object or array literal}\} \\
    \cat{AllocPath} \quad &= \quad \cat{AllocSite} \, \times \, \cat{CallSite}^* \, \times \, \mathbb{Z}_{\geq 0}
\end{align*}
We use $\addr{1}, \addr{2}, ...$ to identify allocation sites and $\addr{1}$, $\addr{1+1}$, $\addr{1+2}$, $\addr{1/\alpha}$, $\addr{1/\alpha+1}$,
$\addr{1/\alpha+2}$, $\addr{1/\alpha/\beta}$, $\addr{1/\alpha/\beta+1}$, $...$, $\addr{2}$, $...$
to denote allocation paths. As usual, greek letters $\alpha, \beta, ...$ identify call sites.
Let $\: \addr{n} \in \cat{AllocSite}, \; \alpha_1 ... \alpha_k \in \cat{CallSite}^*, \; d \in \mathbb{Z}_{\geq 0} \:$ and $\: \addr{n/\alpha_1/.../\alpha_k+d} \in \cat{AllocPath}$.
An allocation path $\addr{n/\alpha_1/.../\alpha_k+d}$ denotes an abstract object that has been allocated at $\addr{n}$, has escaped the scope
of $k$ most recent functions on the stack by being passed to consecutive callers through call sites $\alpha_1, ..., \alpha_k$, and then has
successively entered the scope of another $d$ functions, ending up in the function that is currently being executed.
Escaping into the caller can happen in two ways: either via the return value of a function, or via mutable manipulation of other objects on the heap.
Similarly, an object can enter the scope of a callee by being passed as an argument, being pointed to by another object that enters the scope,
or by being captured in a lexical closure. We will describe this
heap abstraction more precisely by defining \textit{concretization functions} \cite{singleton} that relate abstract objects during analysis with concrete ones during
execution. Each heap abstraction has a corresponding concretization function.
\\ \\
Let's assume we are at some
point during concrete execution of a program. Let $\cat{Box}^C$ be the set of addresses of concrete objects currently allocated
on the heap, and $s \in \cat{Stack}$ be the current call stack. We start by defining a concretization function for allocation-site abstraction.
It maps addresses on the abstract heap to relevant partitions of the concrete heap:
\begin{align*}
    \widetilde{\text{all}} \;\; &\:: \;\; \cat{AllocSite} \; \rightarrow \; \P(\cat{Box}^C) \\
    \widetilde{\text{all}}(\addr{n}) \;\; &= \;\; \{b \in \cat{Box}^C \; | \; b \text{ has been allocated at $\addr{n}$}\}
\intertext{
We define a similar function for allocation paths:
}
    \text{all} \;\; &\:: \;\; \cat{AllocPath} \; \rightarrow \; \P(\cat{Box}^C) \\
    \text{all}(\addr{n/\alpha_1/.../\alpha_k+d}) \;\; &= \;\;
        \{ b \in \cat{Box}^C \; | \; b \text{ has been allocated at $\addr{n}$} \\
        & \qquad \qquad \qquad \quad \; \; \text{and } \text{alloc\_stack}(b) = \alpha_1 ... \alpha_k s_{d+1} ... s_{l} \\
        & \qquad \qquad \qquad \quad \; \; \text{and } \text{the call sites $s_{d+1} ... s_{l}$ have stayed} \\
        & \qquad \qquad \qquad \quad \; \; \;\;\;\;\;\;\:\, \text{on the stack all the time between} \\
        & \qquad \qquad \qquad \quad \; \; \;\;\;\;\;\;\:\, \text{allocation of $b$ and now} \}
\shortintertext{ where}
    \text{alloc\_stack} \;\; &\:: \;\; \cat{Box}^C \; \rightarrow \; \cat{Stack} \\
    \text{alloc\_stack}(b) \;\; &= \;\; \langle \text{call stack at the time of allocation of $b$} \rangle \\
    s_1 ... s_{d} s_{d+1} ... s_l \;=\; s \;\; &= \;\; \langle \text{current call stack} \rangle
\end{align*}
In other words, $\: \text{all}(\addr{n/\alpha_1/.../\alpha_k+d}) \:$ is a subset of $\: \widetilde{\text{all}}(\addr{n}), \:$ where
at the moment of allocation the most $k$ recent call sites on the stack were $\alpha_1, ..., \alpha_k$, and program entered
the current function by exiting these call sites and entering call sites $s_1, ..., s_d$.

\subsection*{Singular and repeating allocation paths}

We say that an allocation path $\addr{n/\alpha_1/.../\alpha_k+d}$ is \textit{singular} if none of $\addr{n}$ and $\alpha_1, ..., \alpha_k$ is syntactically
contained inside a loop. Otherwise it is a \textit{repeating} allocation path.

\subsection*{Heap specialization generalizes recency abstraction}

We observe that if an allocation path $\: \addr{n/\alpha_1/.../\alpha_k+d} \:$ is singular, then
$\: |\text{all}(\addr{n/\alpha_1/.../\alpha_k+d})| = 1. \:$ At any point during program execution, a
singular allocation path identifies a single concrete object on the heap. Overall it might refer to many different
objects throughout the execution, but at any given time there will always be just one. This means that we can perform
strong updates on singular allocation paths. This does not apply to repeating allocation paths, where allocation
has been occurring inside a loop. We note that this sounds familiar. In \textit{recency abstraction} we also had
two classes of abstract objects, where one allowed strong updates and the other didn't. Indeed:
\begin{itemize}
    \item Singular allocation paths generalize the notion of the most recently allocated object.
    \item Repeating allocation paths generalize summary nodes that represent all older objects.
\end{itemize}
One difference from recency abstraction is that when the allocation site $\addr{n}$ is directly contained inside
a loop, we will have a repeating allocation path from the start. This means we won't have the possibility to perform strong updates,
even in the initialization code that directly follows allocation. We can fix this issue by extracting loop bodies
into separate functions. Let's assume that there has been an allocation $\addr{n}$ inside a loop, and loop body has been replaced
with call site $\alpha$. The allocation $\addr{n}$ is now in a separate function and is not syntactically contained inside a loop. Inside this
function we will have:
\begin{itemize}
    \item The singular allocation path $\addr{n}$ that represents the most recently allocated object.
    \item Optionally, the repeating allocation path $\addr{n/\alpha+1}$ that summarizes all older objects allocated in previous iterations of the loop. We need
    to add \mbox{``$/\alpha+1$''} because they had already escaped the scope of this function through $\alpha$ and later entered it again, also through $\alpha$, but in
    a new iteration of the loop.
\end{itemize}
We have shown that our variant of heap specialization is strictly more precise than recency abstraction. Technically, this statement is
true only if we prepare programs before analysis and extract loop bodies into functions, but it is a minor implementation detail.
One can also easily define a more complex variant of allocation paths that won't suffer from this limitation.

\section{Abstract states}

We finally have all necessary ingredients to define abstract states of our analysis:
a context-sensitive value lattice and a precise heap abstraction.
\\ \\
Since our language has first-class functions with closures that can escape their scope,
we will also represent closures using allocation paths. We introduce a new kind of allocation
sites, \cat{Env}, one for each function:
\begin{align*}
    \cat{Env} \quad &= \quad \cat{FunctionId} \quad =\;\;\;\{id \in \cat{Id} \;|\; id \text{ is a function name}\} \; \cup \; \{\syn{main}\} \\
    \cat{AllocSite} \quad &= \quad \{e \in \cat{Expr} \;|\; e \text{ is an object or array literal}\} \, \uline{\cup \, \cat{Env}}
\intertext{
We represent lexical environments of functions as ordinary objects on the heap, where local variables
are stored as fields. Therefore, we no longer need the separate domain \cat{Loc}
that has been used in Chapter 4 to identify local variables. Our abstract states simply map abstract
addresses (allocation paths) to payloads:
}
    \cat{State} \quad &= \quad \cat{AllocPath} \, \rightarrow \, \cat{Payload} \\
    \cat{AllocPath} \quad &= \quad \cat{AllocSite} \, \times \, \cat{CallSite}^* \, \times \, \mathbb{Z}_{\geq 0} \\
    \cat{Payload} \quad &= \quad \cat{Obj} \, \times \, \cat{Arr}
\intertext{
As in Chapter 4, we represent abstract objects as maps from field names to values, and we summarize
arrays as sets of possible values of elements:
}
    \cat{Obj} \quad &= \quad \cat{Id} \, \rightarrow \, \cat{Val} \\
    \cat{Arr} \quad &= \quad \cat{Val}
\intertext{
Whenever from the context it is clear that the payload is an object,
we will use the following notational shortcuts:
}
    \text{for} \quad p \, &\in \, \cat{Payload} \quad \text{and} \quad id \, \in \, \cat{Id} \quad \text{and} \quad v \, \in \, \cat{Val} \\
    p[id \, \mapsto \, v] \, &= \, (\text{obj}(p)[id \, \mapsto \, v], \bot) \\
    p \, id \, &= \, \text{obj}(p) \, id
\intertext{
We represent values using the value lattice defined earlier in this chapter. We repeat it here for convenience:
}
\cat{Val} \quad &= \quad \stackfinval \\
\cat{RawVal} \quad &= \quad \P_{\leq N}(\cat{Int}) \, \times \, \P_{\leq N}(\cat{Str})
    \, \times \, \P(\cat{Bool} \cup \{\syn{null}\} \cup \cat{AllocPath} \cup \cat{Fun})
\intertext{
And finally, we represent first-class functions using address of a heap-allocated closure paired with a function name:
}
\cat{Fun} \quad &= \quad \cat{AllocPath} \, \times \, \cat{FunctionId}
\end{align*}

\subsection*{Entering function calls}

When an abstract address enters a function call through a call site,
we increment the stack depth of its allocation path:
\begin{align*}
    \text{enter\_call} \;\; &\:: \;\; \cat{CallSite} \, \rightarrow \, \cat{AllocPath} \, \rightarrow \, \cat{AllocPath} \\
    \text{enter\_call}(\gamma)(\addr{n/\alpha_1/.../\alpha_k+d}) \;\; &= \;\; \addr{n/\alpha_1/.../\alpha_k+(d+1)}
\intertext{
When an abstract value enters a function call, we adjust its context, and if it contains
abstract addresses, we increment their stack depth:
}
    \text{enter\_call} \;\; &\:: \;\; \cat{CallSite} \, \rightarrow \, \cat{RawVal} \, \rightarrow \, \cat{RawVal} \\
    \text{enter\_call}(\gamma)(rv) \;\; &= \;\; (\text{int}(rv), \text{str}(rv), \\
        & \quad \qquad \{\text{enter\_call}(\gamma)(x) \: | \: x \in \text{alloc\_path}(rv)\} \\
        & \quad \qquad \cup \, \{(\text{enter\_call}(\gamma)(c), f) \: | \: (c, f) \in \text{fun}(rv)\} \\
        & \quad \qquad \cup \, \text{bool}(rv) \, \cup \, \text{null}(rv)) \\
    \text{enter\_call} \;\; &\:: \;\; \cat{CallSite} \, \rightarrow \, \cat{Val} \, \rightarrow \, \cat{Val} \\
    \text{enter\_call}(\gamma)(v) \;\; &= \;\; \text{enter\_ctx}(\gamma)(\lambda s. \, \text{enter\_call}(v(s)))
\intertext{
When a payload of an abstract object enters a function call, all stored values enter a function call:
}
    \text{enter\_call} \;\; &\:: \;\; \cat{CallSite} \, \rightarrow \, \cat{Payload} \, \rightarrow \, \cat{Payload} \\
    \text{enter\_call}(\gamma)((o, a)) \;\; &= \;\; (\lambda f. \text{enter\_call}(o(f)), \text{enter\_call}(a))
\intertext{
Finally, when an abstract state enters a function call, allocation paths are remapped to account for
increased stack depth, and stored object payloads enter the call:
}
    \text{enter\_call} \;\; &\:: \;\; \cat{CallSite} \, \rightarrow \, \cat{State} \, \rightarrow \, \cat{State} \\
    \text{enter\_call}(\gamma)(\sigma) \;\; &= \;\; \lambda (\addr{n/\alpha_1/.../\alpha_k+d}).\, \begin{cases}
        \text{enter\_call}(\gamma)(\sigma(\addr{n/\alpha_1/.../\alpha_k+(d-1)})) \\
        \qquad \qquad \text{ when $d \geq 1$} \\
        \bot \qquad \quad \; \text{ otherwise}
    \end{cases}
\end{align*}

\subsection*{Exiting function calls}

When an abstract address exits a function call through a call site, two things can happen. If this object has been
allocated prior to entering this call (has nonzero stack depth), we decrement its stack depth. If it
has been allocated during this function call (stack depth is zero), we specialize it for this call site:
\begin{align*}
    \text{exit\_call} \;\; &\:: \;\; \cat{CallSite} \, \rightarrow \, \cat{AllocPath} \, \rightarrow \, \cat{AllocPath} \\
    \text{exit\_call}(\gamma)(\addr{n/\alpha_1/.../\alpha_k+d}) \;\; &= \;\; \begin{cases}
        \addr{n/\alpha_1/.../\alpha_k+(d-1)} \quad \text{when $d \geq 1$} \\
        \addr{n/\alpha_1/.../\alpha_k/\gamma} \quad \quad \;\:\, \text{when $d = 0$}
    \end{cases}
    \intertext{
When an abstract value exits a function call, we adjust its context, and if it contains
abstract addresses, we decrement their stack depth or specialize them as required:
}
    \text{exit\_call} \;\; &\:: \;\; \cat{CallSite} \, \rightarrow \, \cat{RawVal} \, \rightarrow \, \cat{RawVal} \\
    \text{exit\_call}(\gamma)(rv) \;\; &= \;\; (\text{int}(rv), \text{str}(rv), \\
        & \quad \qquad \{\text{exit\_call}(\gamma)(x) \: | \: x \in \text{alloc\_path}(rv)\} \\
        & \quad \qquad \cup \, \{(\text{exit\_call}(\gamma)(c), f) \: | \: (c, f) \in \text{fun}(rv)\} \\
        & \quad \qquad \cup \, \text{bool}(rv) \, \cup \, \text{null}(rv)) \\
    \text{exit\_call} \;\; &\:: \;\; \cat{CallSite} \, \rightarrow \, \cat{Val} \, \rightarrow \, \cat{Val} \\
    \text{exit\_call}(\gamma)(v) \;\; &= \;\; \text{exit\_ctx}(\gamma)(\lambda s. \, \text{exit\_call}(v(s)))
\intertext{
When a payload of an abstract object exits a function call, all stored values exit a function call:
}
    \text{exit\_call} \;\; &\:: \;\; \cat{CallSite} \, \rightarrow \, \cat{Payload} \, \rightarrow \, \cat{Payload} \\
    \text{exit\_call}(\gamma)((o, a)) \;\; &= \;\; (\lambda f. \text{exit\_call}(o(f)), \text{exit\_call}(a))
\intertext{
Finally, when an abstract state exits a function call, allocation paths are remapped to account for
specialization, and stored object payloads exit the call:
}
    \text{exit\_call} \;\; &\:: \;\; \cat{CallSite} \, \rightarrow \, \cat{State} \, \rightarrow \, \cat{State} \\
    \text{exit\_call}(\gamma)(\sigma) \;\; &= \;\; \lambda (\addr{n/\alpha_1/.../\alpha_k+d}).\, \begin{cases}
        \text{exit\_call}(\gamma)(\sigma(\addr{n/\alpha_1/.../\alpha_{k-1}})) \\
        \qquad \qquad \text{ when $d = 0$ and $\alpha_k = \gamma$} \\
        \text{exit\_call}(\gamma)(\sigma(\addr{n/\alpha_1/.../\alpha_k+(d+1)})) \\
        \qquad \qquad \text{ otherwise}
    \end{cases}
\end{align*}

\section{Analysis algorithm}

As in Chapter 4, we analyze the program by defining and solving an equation system with variables
ranging over the lattices of abstract states and values. We solve this system using fixpoint
iteration.

\subsection*{Prerequisites}

We enumerate various syntactic elements of the input program, such that later we can easily refer to them when defining
corresponding variables and equations:
\begin{alignat*}{2}
    \{e_1, ..., e_n\} \;&=\; \cat{Expr} & \quad - \quad & \text{the set of expressions occurring in $p$} \\
    \{s_1, ..., s_m\} \;&=\; \cat{Stmt} & \quad - \quad & \text{the set of statements occurring in $p$} \\
    \{d_1, ..., d_k\} \;&=\; \cat{Decl} & \quad - \quad & \text{the set of function declarations in $p$} \\
    \{f_1, ..., f_k\} \;&=\; \cat{FunctionId} & \quad - \quad & \text{the set of function names in $p$} \\
    \{\alpha_1, ..., \alpha_l\} \;&=\; \cat{CallSite} & \quad - \quad & \text{the set of call expressions in $p$} \\
    \{\addr{1}, ..., \addr{t}\} \;&\subseteq\; \cat{AllocSite} & \quad - \quad & \text{allocation sites corresponding to array} \\
        &&& \text{and object literals in $p$} \\
    \{\addr{f_1}, ..., \addr{f_k}\} \;&\subseteq\; \cat{AllocSite} & \quad - \quad & \text{allocation sites corresponding to lexical} \\
        &&& \text{environments of functions $f_1, ..., f_k$}
\end{alignat*}

\subsection*{Variables}

\newcommand{\sigmain}{\sigma^{\text{\tiny in}}}

\newcommand{\stackfinfunctionid}{{\cat{Stack} \; \overset{\text{fin}} \longrightarrow \; \P(\cat{FunctionId})}}

\begin{alignat*}{2}
    \sigmain_{e_1}, \, ... \, , \, \sigmain_{e_n} \: &\in \: \textbf{State} & \quad \;\; \text{---} \quad \;\; & \text{execution states before} \\
        &&& \text{evaluating expressions $e_1, ..., e_n$} \\
    \sigma_{e_1}, \, ... \, , \, \sigma_{e_n} \: &\in \: \textbf{State} & \quad \;\; \text{---} \quad \;\; & \text{execution states after} \\
        &&& \text{evaluating expressions $e_1, ..., e_n$} \\
    v_{e_1}, \, ... \, , \, v_{e_n} \: &\in \: \textbf{Val} & \quad \;\; \text{---} \quad \;\; & \text{approximate sets of values that} \\
        &&& \text{expressions $e_1, ..., e_n$ might evaluate to,} \\
        &&& \text{depending on the calling context} \\ \\
    \sigmain_{s_1}, \, ... \, , \, \sigmain_{s_m} \: &\in \: \textbf{State} & \quad \;\; \text{---} \quad \;\; & \text{execution states before} \\
        &&& \text{executing statements  $s_1, ..., s_m$} \\
    \sigma_{s_1}, \, ... \, , \, \sigma_{s_m} \: &\in \: \textbf{State} & \quad \;\; \text{---} \quad \;\; & \text{execution states after} \\
        &&& \text{executing statements  $s_1, ..., s_m$} \\ \\
        \sigmain_{f_1}, \, ... \, , \, \sigmain_{f_k} \: &\in \: \textbf{State} & \quad \;\; \text{---} \quad \;\; & \text{execution states before} \\
        &&& \text{entering functions  $f_1, ..., f_k$} \\
    \sigma_{f_1}, \, ... \, , \, \sigma_{f_k} \: &\in \: \textbf{State} & \quad \;\; \text{---} \quad \;\; & \text{execution states after} \\
        &&& \text{exiting functions  $f_1, ..., f_k$} \\
    v_{f_1}, \, ... \, , \, v_{f_k} \: &\in \: \textbf{Val} & \quad \;\; \text{---} \quad \;\; & \text{approximate sets of values that} \\
        &&& \text{functions $f_1, ..., f_k$ might return,} \\
        &&& \text{depending on the calling context} \\ \\
\end{alignat*}

\subsection*{Composing context-sensitive operations}
When writing equations of the analysis, a monadic bind operation will come in handy. It generalizes the
idea of lifting operations on \cat{RawVal} to operate pointwise on \cat{Val}
by allowing results of the lifted operation to also be context-sensitive.
We define two variants, one for values and one for states:
\begin{align*}
    \text{bind\_v} \;\; &\:: \;\; \cat{Val} \, \rightarrow \, (\cat{RawVal} \, \rightarrow \, \cat{Val}) \, \rightarrow \, \cat{Val} \\
    \text{bind\_v}(v)(f) \;\; &= \;\; \lambda s. \, f(v(s))(s) \\
    \text{bind\_s} \;\; &\:: \;\; \cat{Val} \, \rightarrow \, (\cat{RawVal} \, \rightarrow \, \cat{State}) \, \rightarrow \, \cat{State} \\
    \text{bind\_s}(v)(f) \;\; &= \;\; \lambda \addr{}. \, \big(\lambda id. \, \lambda s. \, \text{obj}(f(v \, s) \, \addr{}) \, id \, s, \; \lambda s. \, \text{arr}(f(v \, s) \, \addr{}) \, s \big)
\end{align*}
An argument similar to the one about pointwise operations on \cat{Val} can be made
that results of bind\_v and bind\_s are indeed valid elements of \cat{Val} and \cat{State}.

\subsection*{Equations}

\newcommand{\case}[2]{\\[-0.5\baselineskip] \shortintertext{$\llbracket \syn{#1} \rrbracket$ \hfill}}

Let $f$ be the name of some function in the input program. For each $f$, we generate the following equations:
\begin{alignat*}{2}
    \\[-1.5\baselineskip]
    \case{function $f$($...$) \{ $s_0$; return $e_0$ \}}{function}
        &\Longrightarrow& \quad \sigmain_{s_0} &= \sigmain_f \quad \land \quad \sigmain_{e_0} = \sigma_{s_0} \quad \land \quad
        v_f = v_{e_0} \quad \land \quad \sigma_f = \sigma_{e_0} \qquad \qquad \qquad \qquad \qquad \qquad \qquad
\end{alignat*}
Let $e$ be an expression occurring inside the body of $f$, including the
result expression after the \syn{return} keyword. We also include here all expressions occurring in the main program,
and for them we set $f = \syn{main}$. For each pair of $(f, e)$, we generate the following equations, depending on the type of $e$:
\begin{alignat*}{2}
    \\[-1.5\baselineskip]
    \case{$id$}{variable}
        &\Longrightarrow& \quad \sigma_e &= \sigmain_e \quad \land \quad v_e = \text{obj}(\sigmain_e \, \addr{f}) \, id \qquad \qquad \qquad \qquad \qquad \\
    \case{$e_1$.$id$}{property}
        &\Longrightarrow& \quad \sigmain_{e_1} &= \sigmain_e \quad \land \quad \sigma_e = \sigma_{e_1} \quad \land \\
        && v_e &= \lambda s. \, \msqcup \big\lbrace \sigma_e \, \addr{} \, id \, s \;|\; \addr{} \in \text{alloc\_path} (v_{e_1} \, s) \big\rbrace \\
    \case{$e_1$[$e_2$]}{array element}
        &\Longrightarrow& \quad \sigmain_{e_1} &= \sigmain_e \quad \land \quad \sigmain_{e_2} = \sigma_{e_1} \quad \land \quad \sigma_e = \sigma_{e_2} \\
        && v_e &= \lambda s. \, \msqcup \big\lbrace \text{arr}(\sigma_e \, \addr{}) \, s \;|\; \addr{} \in \text{alloc\_path} (v_{e_1} \, s) \big\rbrace \\
    \case{$id$ = $e_1$}{variable assignment}
        &\Longrightarrow& \quad \sigmain_{e_1} &= \sigmain_e \quad \land \quad \sigma_e = \sigma_{e_1} [\addr{f} \, \mapsto \, (\sigma_{e_1} \addr{f})[id \, \mapsto \, v_{e_1}]] \\
    \case{$e_1$.$id$ = $e_2$}{property assignment}
        &\Longrightarrow& \quad \sigmain_{e_1} &= \sigmain_e \quad \land \quad \sigmain_{e_2} = \sigma_{e_1} \quad \land \\
        && \sigma_e &= \text{bind\_s}(v_{e_1})(\lambda v. \, \msqcup \{ \sigma_{e_2}[\addr{} \, \mapsto \, (\sigma_{e_2} \, \addr{})[id \, \mapsto \, v_{e_2}]] \; | \; \addr{} \in \text{alloc\_path}(v) \}) \\
    \case{$e_1$[$e_2$] = $e_3$}{array element assignment}
        &\Longrightarrow& \quad \sigmain_{e_1} &= \sigmain_e \quad \land \quad \sigmain_{e_2} = \sigma_{e_1} \quad \land \quad \sigmain_{e_3} = \sigma_{e_2} \quad \land \\
        && \sigma_e &= \text{bind\_s}(v_{e_1})(\lambda v. \, \msqcup \{ \sigma_{e_3}[\addr{} \, \mapsto \, (\bot, \text{arr}(\sigma_{e_3} \, \addr{}) \sqcup v_{e_3})] \; | \; \addr{} \in \text{alloc\_path}(v) \}) \\
    \\[-0.5\baselineskip] \shortintertext{$
    \llbracket \syn{$e_1$ + $e_2$} \rrbracket \; / \; \llbracket \syn{$e_1$ - $e_2$} \rrbracket \; / \;
    \llbracket \syn{$e_1$ * $e_2$} \rrbracket \; / \; \llbracket \syn{$e_1$ / $e_2$} \rrbracket$ \hfill}
        &\Longrightarrow& \quad \sigmain_{e_1} &= \sigmain_e \quad \land \quad \sigmain_{e_2} = \sigma_{e_1} \quad \land \quad \sigma_e = \sigma_{e_2} \\
        && v_e &= \text{lift}^{+}(v_{e_1}, v_{e_2}) \; / \; \text{lift}^{-}(v_{e_1}, v_{e_2}) \; / \; \text{lift}^{*}(v_{e_1}, v_{e_2}) \; / \; \text{lift}^{/}(v_{e_1}, v_{e_2}) \\
    \\[-0.5\baselineskip] \shortintertext{$
    \llbracket \syn{$e_0$($e_1$, $...$, $e_d$)} \rrbracket = \alpha$ \hfill}
    &\Longrightarrow& \quad \sigmain_{e_0} &= \sigmain_e \quad \land \quad \sigmain_{e_1} = \sigma_{e_0} \quad \land \quad ... \quad \land \quad \sigmain_{e_d} = \sigma_{e_{d-1}} \quad \land \\
    && \sigmain_{f_1} &= \sigmain_{f_1} \sqcup call_{f_1} \quad \land \quad ... \quad \land \quad \sigmain_{f_k} = \sigmain_{f_k} \sqcup call_{f_k} \quad \land \\
    && \sigma_e &= \text{exit\_call}(\alpha)(\text{bind\_s}(v_{e_0})(\lambda v. \, \msqcup \big\lbrace \sigma_{f} \;|\; (cl, f) \in \text{fun}(v) \big\rbrace)) \quad \land \quad \\
    && v_e &= \text{exit\_call}(\alpha)(\text{bind\_v}(v_{e_0})(\lambda v. \, \msqcup \big\lbrace v_{f} \;|\; (cl, f) \in \text{fun}(v) \big\rbrace)) \\
    &\text{where} \\
    &&& call_{f_i} = \text{bind\_s}(v_{e_0})(\lambda v. \, \msqcup \big\lbrace \text{raw\_call}(f_i, cl) \;|\; (cl, f) \in \text{fun}(v) \, \land \, f = f_i \big\rbrace) \\
    &&& \text{raw\_call} \; : \; \cat{FunctionId} \, \times \, \cat{AllocPath} \, \rightarrow \, \cat{State} \\
    &&& \text{pass\_arguments} \; : \; \cat{FunctionId} \, \times \, \cat{State} \, \rightarrow \, \cat{Val}^* \, \rightarrow \, \cat{State} \\
    &&& \text{raw\_call}(f_i, cl) = \begin{cases}
        \text{pass\_arguments}(f_i, \text{enter\_call}(\alpha)(\sigma_{e_d}))( \\
        \qquad \lambda s. \, \text{enter\_call}(\alpha)(cl), \\
        \qquad \text{enter\_call}(\alpha)(v_{e_1}), \\
        \qquad ..., \\
        \qquad \text{enter\_call}(\alpha)(v_{e_d}), \\
        ) \end{cases} \\
    &&& \text{pass\_arguments}(f_i, \sigma)(v_0, ..., v_d) = \begin{cases}
        \sigma[\addr{f_i} \, \mapsto \, \bot_{\cat{Payload}}[\\
            \qquad id_0^{\,i} \, \mapsto \, v_0, \\
            \qquad ..., \\
            \qquad id_d^{\,i} \, \mapsto \, v_d, \\
        ]]\qquad\!\!\!\text{when } \text{\#args}(f_i) = d \\
        \bot \quad \text{ otherwise}
    \end{cases} \\
    &&& id_0^{\,i}, id_1^{\,i}, ..., id_{\text{\#args}(f_i)}^{\,i} \quad \text{--} \quad \text{formal parameters of function $f_i$} \\
    &&& \text{note: in programs generated by closure conversion,}\\
    &&& \text{$id_0^{\,i}$ is always equal to \syn{closure}} \\
    \\[-0.5\baselineskip] \shortintertext{$
    \llbracket \syn{\{$id_1$:\;$e_1$, $...$, $id_d$:\;$e_d$\}} \rrbracket = \addr{e}$ \hfill}
        &\Longrightarrow& \quad \sigmain_{e_1} &= \sigmain_e \quad \land \quad \sigmain_{e_2} = \sigma_{e_1} \quad \land \quad ... \quad \land \quad \sigmain_{e_d} = \sigma_{e_{d-1}} \quad \land \\
        && v_{e} &= \lambda s. \, \addr{e} \quad \land \quad \sigma_e = \sigma_{e_d}[\addr{e} \, \mapsto \, \bot_{\cat{Payload}}[
            id_1 \, \mapsto \, v_{e_1}, ..., id_d \, \mapsto \, v_{e_d}
        ]] \\
    \\[-0.5\baselineskip] \shortintertext{$
    \llbracket \syn{[$e_1$, $...$, $e_d$]} \rrbracket = \addr{e}$ \hfill}
        &\Longrightarrow& \quad \sigmain_{e_1} &= \sigmain_e \quad \land \quad \sigmain_{e_2} = \sigma_{e_1} \quad \land \quad ... \quad \land \quad \sigmain_{e_d} = \sigma_{e_{d-1}} \quad \land \\
        && v_{e} &= \lambda s. \, \addr{e} \quad \land \quad \sigma_e = \sigma_{e_d}[\addr{e} \, \mapsto \, (\bot, v_{e_1} \, \sqcup \, ... \, \sqcup \, v_{e_d})] \\
    \case{bind-closure $id$}{closure capture}
    &\Longrightarrow& \quad \sigma_e &= \sigmain_e \quad \land \quad v_e = (\addr{f}, id)
\end{alignat*}
Let $s$ be a statement occurring inside the body of $f$.
For each pair of $(f, s)$, we generate the following equations, depending on the type of $s$:
\begin{alignat*}{2}
    \\[-1.5\baselineskip]
    \case{$\langle empty \rangle$}{}
        &\Longrightarrow& \quad \sigma_s &= \sigmain_s \qquad \qquad \qquad \qquad \qquad \qquad \qquad \qquad \qquad \qquad \qquad \qquad \qquad \\
    \case{$e$}{}
        &\Longrightarrow& \quad \sigmain_e &= \sigmain_s \quad \land \quad \sigma_s = \sigma_e \\
    \case{$s_1$; $s_2$}{}
        &\Longrightarrow& \quad \sigmain_{s_1} &= \sigmain_s \quad \land \quad \sigmain_{s_2} = \sigma_{s_1} \quad \land \quad \sigma_s = \sigma_{s_2} \\
    \case{var $id$ = $e$}{}
        &\Longrightarrow& \quad \sigmain_{e} &= \sigmain_s \quad \land \quad
        \sigma_s = \sigma_{e} [\addr{f} \, \mapsto \, (\sigma_{e} \addr{f})[id \, \mapsto \, v_{e}]] \\
    \case{if ($e$) \{ $s_1$ \} else \{ $s_2$ \}}{}
        &\Longrightarrow& \quad \sigmain_e &= \sigmain_s \quad \land \\
        && \sigmain_{s_1} &= \text{bind\_s}(v_e)(\lambda v. \, \sigma_e \text{ if } \syn{true} \in \text{bool}(v) \text{ else } \bot_{\cat{State}}) \quad \land \\
        && \sigmain_{s_2} &= \text{bind\_s}(v_e)(\lambda v. \, \sigma_e \text{ if } \syn{false} \in \text{bool}(v) \text{ else } \bot_{\cat{State}}) \quad \land \\
        && \sigma_s &= \text{bind\_s}(v_e)(\lambda v.\\
        &&& \qquad (\sigma_{s_1} \text{ if } \syn{true} \:\; \in \text{bool}(v) \text{ else } \bot_{\cat{State}}) \; \sqcup \\
        &&& \qquad (\sigma_{s_2} \text{ if } \syn{false} \in \text{bool}(v) \text{ else } \bot_{\cat{State}}) \\
        &&& ) \\
    \case{for (var $id$ in $e$) \{ $s_0$ \}}{}
        &\Longrightarrow& \quad \sigmain_e &= \sigmain_s \quad \land \quad \sigma_s = \sigma_{s_0} \, \sqcup \, \sigma_e \quad \land \\
        && \sigmain_{s_0} &= \text{bind\_s}(v_e)(\lambda v. \, (\sigma_e \, \sqcup \, \sigma_{s_0})[
            \addr{f} \, \mapsto \, (\sigma_e \, \sqcup \, \sigma_{s_0}) \addr{f} [ \\
        &&& \qquad id \, \mapsto \, \text{arr}(\sigma_e(\text{alloc\_path}(v))) \\
        &&& ]]) \\
    \case{while ($e$) \{ $s_0$ \}}{}
        &\Longrightarrow& \quad \sigmain_e &= \sigmain_s \, \sqcup \, \sigma_{s_0} \quad \land \quad
        \sigmain_{s_0} = \sigma_e \quad \land \quad \sigma_s = \sigma_{s_0} \, \sqcup \, \sigma_e
\end{alignat*}
}

\cleardoublepage

\#\chapter{Implementation}

Over the course of this thesis, the context-sensitive analysis
described in Chapter 5 has been implemented in a prototype analyzer.

\subsection*{Screenshots}
Below are some screenshots of running our analyzer on the examples from Chapter 5:
\includegraphics[width=15cm]{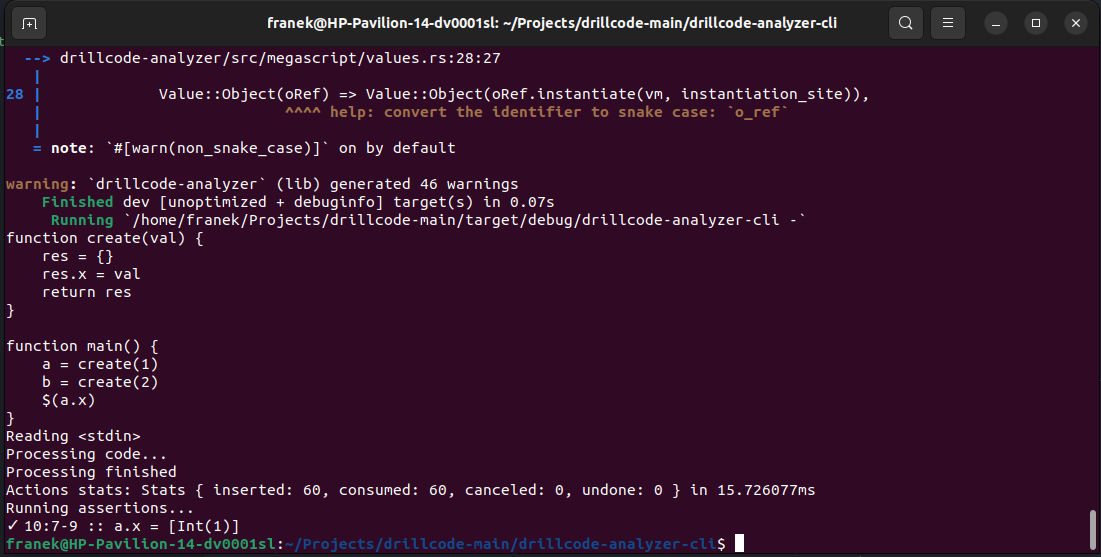}
\includegraphics[width=15cm]{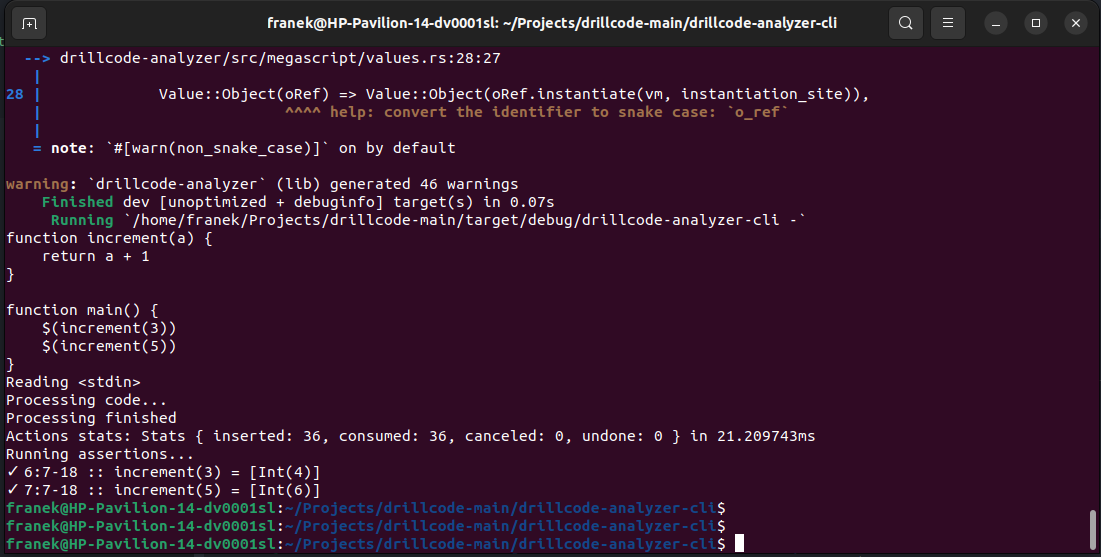}
\\ \\
\includegraphics[width=15cm]{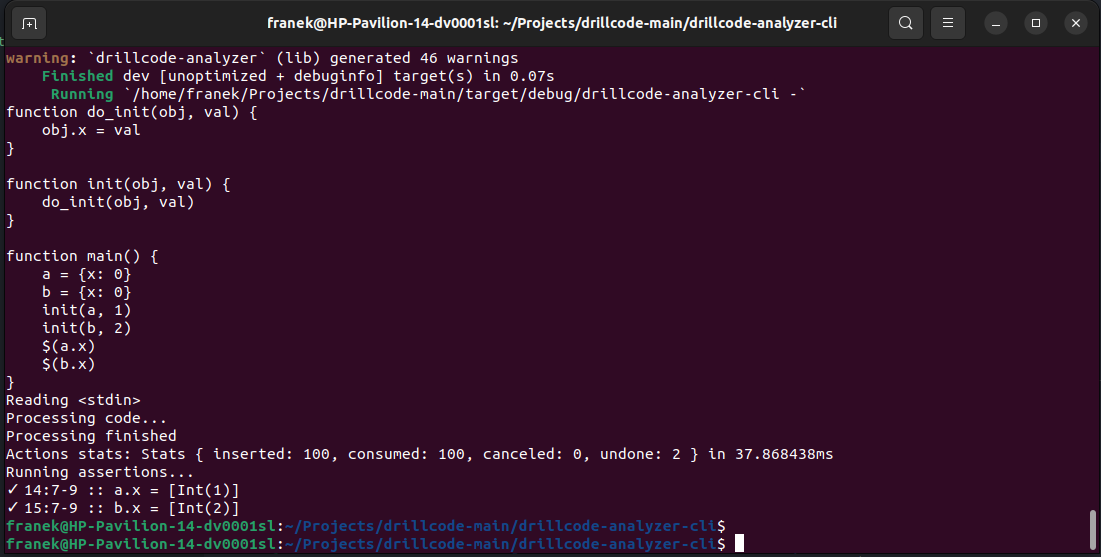}

\subsection*{Monadic DSL}

The analyzer is implemented in a mix of Rust and a custom DSL implemented using
procedural macros in Rust.
The DSL is similar to the \textit{do notation} in Haskell, and provides a composable way to
express computations in our custom monad. This monad it is simlar to a combination of the
State and List monads from Haskell. The State part abstracts away the fact that instead
of operating on values and memory locations, we are composing constraints on the lattice 
of abstract states. The List part provides nondeterminism and context-sensitivity - the
fact that we operate on the value lattice. Thanks to this
monad we can write code in a direct way, as if we were manipulating single concrete values.
It automatically translates it to equivalent lattice-valued operations on abstract
values, and takes care of correct propagation of calling contexts.

\subsection*{Implicit representation of abstract states}

In fact, we don't store entire abstract states in memory at
any point. Instead, for each memory location, we store a slice of the control flow graph that is
spanned by all read and write operations that touch this location, in a similar spirit to \cite{sparse1,sparse2,sparse3}.
We store one abstract value at each vertex of this slice of the CFG. This way we drastically reduce memory
consumption of the analyzer - instead of storing $O\left(\#\textit{states} \cdot \#\textit{locations}\right)$
elements of the value lattice, we only need
$O\left(\sum_{l \in \textit{locations}} \#\textit{reads}(l) + \#\textit{writes}(l)\right)$.

\subsection*{Incremental updates}

This monad described above also handles the bookkeeping that is necessary for incremental updates.
All memory writes and reads are instrumented with precise tracking. Any operation
can be rolled back, with the effect of undoing all traces of it and all subsequent data flow
that was caused by this operation. Moreover, such a rollback will not mess up any other data
that did not depend on it, even if it has been computed later than the operation being undone.

\subsection*{Extensions}

The prototype analyzer implements a few additional features that have not been described in Chapter 5:
\begin{itemize}
    \item Loop unrolling for a small number of iterations.
    \item Adaptive representation of arrays, where arrays that are statically known
        are represented with full precision by enumerating elements, and arrays
        that are results of complex computations are represented as a \textit{bag of possible elements}.
        Precision degrades gradually and it can also express things in-between such as
        ``an array of ints that begins with \texttt{[2, 3, 5]}'' or ``one or more repetitions of \texttt{['a', 'b']}''.
    \item Flow-insensitive global variables. This is a less precise representation, but in the future it
        will be essential for good performance on larger codebases. In particular, it can
        prevent a lot of unnecessary recomputations of unrelated pieces of code when preforming incremental updates.
\end{itemize}

\cleardoublepage

\chapter{Conclusions and future work}

\subsection*{Conclusions}

We have developed an advanced context-sensitive static analysis algorithm that can partially execute
programs, obtaining information that is commonly believed to be available only at runtime. We have introduced it in
a gradual way, starting with a simple analysis in the framework of abstract interpretation. Later we
used multiple figures and metaphors to make our description intuitive and accessible. We hope that this
will spark more interest in advanced techniques from the field of static analysis, and that they finally will
make their way to everyday developer tools such as IDEs, language servers and linters. Over the course of
this thesis, our algorithm has been implemented in a prototype that can analyze simple programs.

\subsection*{Future work}

Further development is needed to extend our prototype to understand a substantial subset of JavaScript or Python and test how it
performs on larger, real-world codebases. We speculate that a future version of our analyzer might have
to lazily load large codebases piece by piece, because our analysis is very advanced and might consume significant resources.
This can be addressed by equipping it with a controller that predicts which files
and components are relevant to the problem the programmer is currently working on, and loads or unloads them
as needed. This vision is plausible because our system is designed to be incremental and decoupled from the start, and even in its
current state it is able to incrementally load and unload parts of programs.

\cleardoublepage

\appendix

\setcounter{secnumdepth}{-1}
\chapter{Appendix: Closure Conversion}

{
\newcommand{\denotation}[1]{\\[-0.8\baselineskip] \shortintertext{$#1$}}
\newcommand{\cat}[1]{\textbf{#1}}
\newcommand{\syn}[1]{(\texttt{#1})}
\renewcommand{\P}{\mathcal{P}}
\newcommand{\CC}[1]{\mathcal{CC} \llbracket \texttt{#1} \rrbracket}
\newcommand{\CCs}[1]{\mathcal{CC_S} \llbracket \texttt{#1} \rrbracket}
\newcommand{\CCe}[1]{\mathcal{CC_E} \llbracket \texttt{#1} \rrbracket}
\newcommand{\BV}[1]{\operatorname{BV} \llbracket \texttt{#1} \rrbracket}
\newcommand{\Rs}[1]{\mathcal{R_S} \llbracket \texttt{#1} \rrbracket}
\renewcommand{\Re}[1]{\mathcal{R_E} \llbracket \texttt{#1} \rrbracket}

Here we define the process of closure conversion. $\CC{$\cdot$}$ converts surface-level programs (TinyScript$^+$)
to the intermediate representation (TinyScript). $\BV{$\cdot$}$, as in \textit{bound variables}, is a
function that takes a statement, scans it for the occurrences of \syn{var $id$}, and gathers identifiers
$id$ into a set. This scanning only traverses compound statements, ifs and loops, and ignores nested
function definitions. Functions $\Rs{$\cdot$}$ and $\Re{$\cdot$}$ take $bv$, the set of bound variables
\textit{in the outer scope}, as input, and rewrite their occurrences $a$ into $\texttt{closure.}a$. Functions
$\CCs{$\cdot$}$ and $\CCe{$\cdot$}$ perform closure conversion respectively for statements and expressions.

\begin{align*}
\CC{$\cdot$} \;&:\; \cat{Prog}^+ \,\longrightarrow\, \cat{Prog} \\
\\
\BV{$\cdot$} \;&:\; \cat{Stmt}^+ \,\longrightarrow\, \P(\cat{Id}) \\
\Rs{$\cdot$} \;&:\; \cat{Stmt}^+ \,\longrightarrow\, \P(\cat{Id}) \rightarrow \cat{Stmt}^+ \\
\Re{$\cdot$} \;&:\; \cat{Expr}^+ \,\longrightarrow\, \P(\cat{Id}) \rightarrow \cat{Expr}^+ \\
\\
\CCs{$\cdot$} \;&:\; \cat{Stmt}^+ \,\longrightarrow\, \P(\cat{Id}) \rightarrow \cat{Stmt} \times \cat{Decl}^* \\
\CCe{$\cdot$} \;&:\; \cat{Expr}^+ \,\longrightarrow\, \P(\cat{Id}) \rightarrow \cat{Expr} \times \cat{Decl}^* \\
\end{align*}

\begin{alignat*}{2}
    \denotation{\CC{$s$} \;=\; \CCs{$s$} \, bv \:\text{ where}}
    bv &:\; \P(\cat{Id}) \;\;=\;\; \BV{$s$} \cup \{\texttt{input}\} && \\
    \\
    \denotation{\Rs{S($s_1$, $...$, $s_n$; $e_1$, $...$, $e_n$)} \, bv \;=\; \syn{S($s_1'$, $...$, $s_n'$; $e_1'$, $...$, $e_n'$)} \:\text{where}}
        s_1' &:\; \cat{Stmt}^+ \;\;=\;\; \Rs{$s_1$} \, bv && \\
        &... \qquad \qquad \: ... && \\
        s_n' &:\; \cat{Stmt}^+ \;\;=\;\; \Rs{$s_n$} \, bv && \\
        e_1' &:\; \cat{Expr}^+ \;\;=\;\; \Re{$e_1$} \, bv && \\
        &... \qquad \qquad \: ... && \\
        e_n' &:\; \cat{Expr}^+ \;\;=\;\; \Re{$e_n$} \, bv && \\
    \denotation{\Rs{function $id_0$($id_1$, $...$, $id_n$) \{ $s$; return $e$ \}} \, bv}
    \shortintertext{$\quad\;\; = \syn{function $id_0$($id_1$, $...$, $id_n$) \{ $s_1$; return $e_1$ \}} \:\text{ where}$}
        bv_1 &:\; \P(\cat{Id}) \;\;=\;\; bv \, \setminus \, (\{id_1, ..., id_n\} \cup \BV{$s$}) && \\
        s_1 &:\; \cat{Stmt}^+ \;\;=\;\; \Rs{$s$} \, bv_1 && \\
        e_1 &:\; \cat{Stmt}^+ \;\;=\;\; \Re{$e$} \, bv_1 && \\
    \\
    \denotation{\Re{E($e_1$, $...$, $e_n$)} \, bv \;=\; \syn{E($e_1'$, $...$, $e_n'$)} \:\text{where}}
        e_1' &:\; \cat{Expr}^+ \;\;=\;\; \Re{$e_1$} \, bv && \\
        &... \qquad \qquad \: ... && \\
        e_n' &:\; \cat{Expr}^+ \;\;=\;\; \Re{$e_n$} \, bv && \\
    \denotation{\Re{$id$} \, bv \;=\; \begin{cases} \syn{closure.$id$} \:\text{ when }\: id \in bv \\ id \:\text{ otherwise} \end{cases}}
    \denotation{\Re{$id$ = $e$} \, bv \;=\; \begin{cases} \syn{closure.$id$ = $\Re{$e$} \, bv$} \:\text{ when }\: id \in bv \\ \syn{$id$ = $\Re{$e$} \, bv$} \:\text{ otherwise} \end{cases}}
    \denotation{\Re{function($id_1$, $...$, $id_n$) \{ $s$; return $e$ \}} \, bv}
    \shortintertext{$\quad\;\; = \syn{function($id_1$, $...$, $id_n$) \{ $s_1$; return $e_1$ \}} \:\text{ where}$}
        bv_1 &:\; \P(\cat{Id}) \;\;=\;\; bv \, \setminus \, (\{id_1, ..., id_n\} \cup \BV{$s$}) && \\
        s_1 &:\; \cat{Stmt}^+ \;\;=\;\; \Rs{$s$} \, bv_1 && \\
        e_1 &:\; \cat{Stmt}^+ \;\;=\;\; \Re{$e$} \, bv_1 && \\
    \\
    \denotation{\CCs{S($s_1$, $...$, $s_n$; $e_1$, $...$, $e_n$)} \, bv \;=\; (\texttt{S($s_1'$, $...$, $s_n'$; $e_1'$, $...$, $e_n'$)}, d) \:\text{ where}}
        (s_1', d^s_1) &:\; \cat{Stmt} \times \cat{Decl}^* \;\;=\;\; \CCs{$s_1$} \, bv && \\
        &... \qquad \qquad \qquad \quad \: \: ... && \\
        (s_n', d^s_n) &:\; \cat{Stmt} \times \cat{Decl}^* \;\;=\;\; \CCs{$s_n$} \, bv && \\
        (e_1', d^e_1) &:\; \cat{Expr} \times \cat{Decl}^* \;\;=\;\; \CCe{$e_1$} \, bv && \\
        &... \qquad \qquad \qquad \quad \: \: ... && \\
        (e_n', d^e_n) &:\; \cat{Expr} \times \cat{Decl}^* \;\;=\;\; \CCe{$e_n$} \, bv && \\
        d &:\; \cat{Decl}^* \;\;=\;\; d^s_1 ... d^s_n \, d^e_1 ... d^e_n && \\
    \denotation{\CCs{function $id_0$($id_1$, $...$, $id_n$) \{ $s$; return $e$ \}} \, bv \;=\; (f, d_1 d_2 d_3) \:\text{ where}}
        bv_1 &:\; \P(\cat{Id}) \;\;=\;\; bv \, \setminus \, (\{id_1, ..., id_n\} \cup \BV{$s$}) && \\
        s' &:\; \cat{Stmt}^+ \;\;=\;\; \Rs{$s$} \, bv_1 && \\
        e' &:\; \cat{Expr}^+ \;\;=\;\; \Re{$e$} \, bv_1 && \\
        bv_2 &:\; \P(\cat{Id}) \;\;=\;\; bv \cup \{id_1, ..., id_n\} \cup \BV{$s$} && \\
        (s'', d_1) &:\; \cat{Stmt} \times \cat{Decl}^* \;\;=\;\; \CCs{$s'$} \, bv_2 && \\
        (e'', d_2) &:\; \cat{Expr} \times \cat{Decl}^* \;\;=\;\; \CCs{$e'$} \, bv_2 && \\
        d_3 &:\; \cat{Decl} \;\;=\;\; (\texttt{function $id_0$(closure, $id_1$, $...$, $id_n$) \{} \\
        & \qquad \qquad \qquad \qquad \texttt{$s''$; return $e''$ \}} && \\
        & \qquad \qquad \qquad \texttt{\}}) && \\
        f &:\; \cat{Stmt} \;\;=\;\; \syn{var $id_0$ = bind-closure $id_0$} && \\
    \\
    \denotation{\CCe{E($e_1$, $...$, $e_n$)} \, bv \;=\; (\texttt{E($e_1'$, $...$, $e_n'$)}, d) \:\text{where}}
        (e_1', d_1) &:\; \cat{Expr} \times \cat{Decl}^* \;\;=\;\; \CCe{$e_1$} \, bv && \\
        &... \qquad \qquad \: ... && \\
        (e_n', d_n) &:\; \cat{Expr} \times \cat{Decl}^* \;\;=\;\; \CCe{$e_n$} \, bv && \\
        d &:\; \cat{Decl}^* \;\;=\;\; d_1 ... d_n && \\
    \denotation{\CCe{function($id_1$, $...$, $id_n$) \{ $s$; return $e$ \}} \, bv \;=\; (f, d_1 d_2 d_3) \:\text{ where}}
        id_0 &:\; \cat{Id} \;\;=\;\; \langle \text{generate a fresh identifier} \rangle && \\
        bv_1 &:\; \P(\cat{Id}) \;\;=\;\; bv \, \setminus \, (\{id_1, ..., id_n\} \cup \BV{$s$}) && \\
        s' &:\; \cat{Stmt}^+ \;\;=\;\; \Rs{$s$} \, bv_1 && \\
        e' &:\; \cat{Expr}^+ \;\;=\;\; \Re{$e$} \, bv_1 && \\
        bv_2 &:\; \P(\cat{Id}) \;\;=\;\; bv \cup \{id_1, ..., id_n\} \cup \BV{$s$} && \\
        (s'', d_1) &:\; \cat{Stmt} \times \cat{Decl}^* \;\;=\;\; \CCs{$s'$} \, bv_2 && \\
        (e'', d_2) &:\; \cat{Expr} \times \cat{Decl}^* \;\;=\;\; \CCs{$e'$} \, bv_2 && \\
        d_3 &:\; \cat{Decl} \;\;=\;\; (\texttt{function $id_0$(closure, $id_1$, $...$, $id_n$) \{} \\
        & \qquad \qquad \qquad \qquad \texttt{$s''$; return $e''$ \}} && \\
        & \qquad \qquad \qquad \texttt{\}}) && \\
        f &:\; \cat{Expr} \;\;=\;\; \syn{bind-closure $id_0$} && \\
\end{alignat*}
}

\cleardoublepage

\addcontentsline{toc}{chapter}{Bibliography}

\bibliographystyle{alpha}
\bibliography{mylit}

\cleardoublepage
\thispagestyle{empty}
\section*{Selbst\"andigkeitserkl\"arung}

Hiermit versichere ich, dass ich die vorliegende Masterarbeit 
selbst\"andig und nur mit den angegebenen Hilfsmitteln angefertigt habe und dass alle Stellen, die dem Wortlaut oder dem 
Sinne nach anderen Werken entnommen sind, durch Angaben von Quellen als 
Entlehnung kenntlich gemacht worden sind. 
Diese Masterarbeit wurde in gleicher oder \"ahnlicher Form in keinem anderen 
Studiengang als Pr\"ufungsleistung vorgelegt.

\vskip 3cm

Ort, Datum	\hfill Unterschrift \hfill 

\end{document}